\documentclass[amsmath,amssymb,11pt]{article}
\usepackage{jheppub2}
\pdfoutput=1
\usepackage{graphicx,epsfig}
\usepackage{float}
\usepackage{subfloat}
\usepackage[utf8]{inputenc}
\usepackage{hyperref}
\usepackage{caption}
\usepackage{color}
\usepackage{cleveref}
\usepackage{subcaption}
\usepackage{booktabs}

\usepackage{tikz}

\usepackage{soul}
\usepackage{cancel}

\newcommand*{\affmark}[1][*]{\textsuperscript{#1}}

\newcommand{\beq}{\begin{equation}}

\newcommand{\eeq}{\end{equation}}

\newcommand{\total}{\text{total}}
\newcommand{\corner}{\text{corner}}

\title{Liouville gravity at the end of the world:\\Deformed defects in AdS/BCFT}

\author{Dominik Neuenfeld\affmark[a],}
\emailAdd{dominik.neuenfeld@uni-wuerzburg.de}
\author{Andrew Svesko\affmark[b], and}
\emailAdd{andrew.svesko@kcl.ac.uk}
\author{Watse Sybesma\affmark[c,d,e]}
\emailAdd{zhws2@cam.ac.uk}

  \affiliation{\affmark[a]Institute for Theoretical Physics and Astrophysics, Julius-Maximilians-Universit\"{a}t W\"{u}rzburg,\\ Am Hubland, 97074 W\"{u}rzburg, Germany}
\affiliation{\affmark[b]Department of Mathematics, King’s College London,
Strand, London, WC2R 2LS, United Kingdom}
\affiliation{\affmark[c]Department of Applied Mathematics and Theoretical Physics, University of Cambridge, \\Cambridge CB3 0WA, United Kingdom}
\affiliation{\affmark[d]Isaac Newton Institute for Mathematical Sciences,
University of Cambridge, \\ Cambridge CB3 0EH, United Kingdom}
\affiliation{\affmark[e]Science Institute, University of Iceland, Dunhaga 3, 107 Reykjav\'{i}k, Iceland}

\abstract{We study shape deformations of two-dimensional end-of-the-world (ETW) branes, such as those in bottom-up models of two-dimensional holographic boundary conformal field theories (BCFT), and derive an action for the theory of brane deformations in any bulk three-dimensional maximally symmetric spacetime. In the case of a bulk anti-de Sitter (AdS) spacetime, at leading order in the ultraviolet cutoff, the induced theory on the brane controlling its shape 
is Liouville gravity coupled to quantum matter.
We show in certain limits the theory reduces to semi-classical AdS, dS or flat Jackiw-Teitelboim (JT) gravity, thus providing the first doubly-holographic derivation  of two-dimensional models of dilaton gravity minimally coupled to a large number of conformal fields.
Specializing to the AdS JT gravity limit, 
we discuss the dual BCFT interpretation and provide evidence that changing the boundary conditions of JT gravity on the brane is equivalent to a deformation of the dual BCFT with the displacement operator. This establishes a doubly-holographic triality between (i) brane deformations in the bulk, (ii) JT gravity in the brane description, and (iii) irrelevant deformations of the CFT boundary. Lastly, in the presence of a non-trivial dilaton profile, we prove that the Ryu-Takayanagi formula for holographic BCFTs receives a contact term whenever the minimal surface ends on the brane.}

\begin{document}

\maketitle
\section{Introduction} \label{sec:intro}

Semi-classical gravity serves as a useful proxy to study quantum effects in gravity. In this context, quantum fields backreact on a classical dynamical spacetime governed by semi-classical field equations. If we add a conformal field theory (CFT) with sufficiently large central charge to two-dimensional dilaton gravity, specifically  Jackiw-Teitelboim (JT) gravity \cite{Jackiw:1984je,Teitelboim:1983ux,Almheiri:2014cka} or the Callan-Harvey-Giddings-Strominger (CGHS) model \cite{Callan:1992rs,Russo:1992ax}, the semi-classical field equations can be solved consistently.
This is because in a large-$c$ approximation backreaction effects due to the CFT are encoded in the conformal anomaly \cite{Christensen:1977jc}, entering via the Polyakov effective action \cite{Polyakov:1981rd}, and are minimally coupled to the dilaton.
Such a context thus enables analytic studies of quantum corrections to black hole physics, \emph{e.g.},
\cite{Susskind:1993if,Fiola:1994ir,Almheiri:2014cka,Penington:2019npb,Almheiri:2019psf,Gautason:2020tmk,Pedraza:2021cvx,Pedraza:2021ssc,Svesko:2022txo,Hirano:2023ebw, Hartman:2020swn}. 
Models of pure dilaton gravity can be obtained as low-energy effective descriptions of certain ultraviolet (UV) theories \cite{Kitaev_SYK, Maldacena:2016upp,Saad:2019lba}, or low-energy descriptions of UV complete higher-dimensional physics \cite{Achucarro:1993fd,Fabbri:2000xh,Nayak:2018qej,Sachdev:2019bjn,Castro:2018ffi,Moitra:2019bub,Castro:2021fhc,Maldacena:2019cbz,Cotler:2019nbi,Sybesma:2020fxg,Kames-King:2021etp,Svesko:2022txo,Castro:2022cuo}.
Consistent semi-classical JT gravity, however, is thus far an effective toy model: a large number of quantum fields whose stress-tensor sources the gravitational field equations are added \emph{by hand} at the level of the two-dimensional action.

\begin{figure}[t!]
\centering
\begin{center}
\includegraphics[scale=.65]{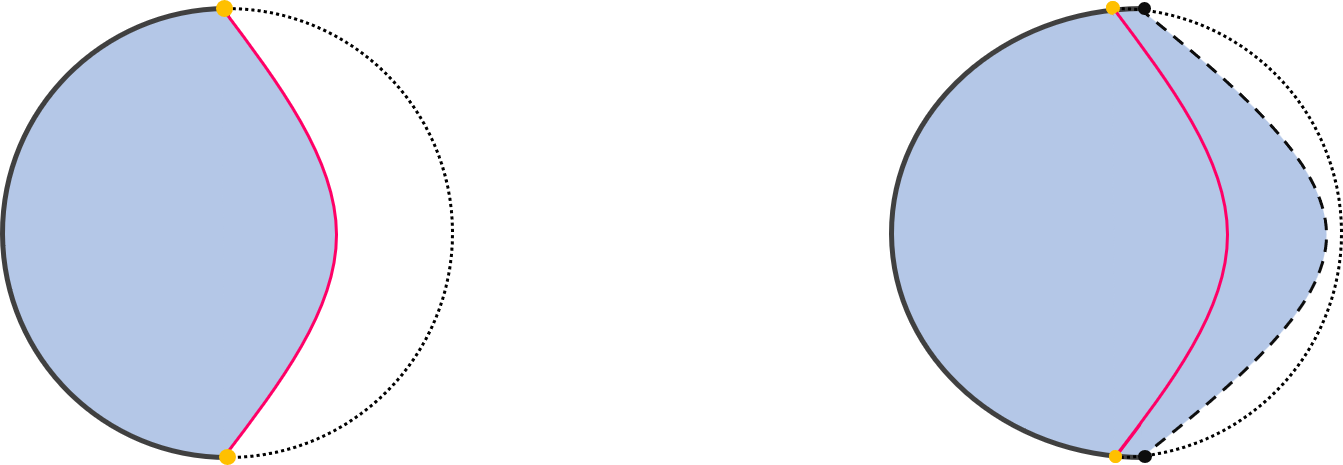}
\put(-95,65){$\text{AdS}_{d+1}$}
\put(-375,65){$\text{AdS}_{d+1}$}
\put(-421,6){$\text{BCFT}_{d}$}
\put(-140,6){$\text{BCFT}_{d}$}
\put(-60,95){$\mathcal{Q}_{0}$}
\put(-340,95){$\mathcal{Q}_{0}$}
\put(-20,65){$\mathcal{Q}$}
\caption{ \small Spatial slice of holographic braneworld. \textbf{Left:} Standard set-up with no brane deformations. An end-of-the-world (Karch-Randall) brane $\mathcal{Q}_{0}$ fixed at $\theta=\phi_{0}$ (magenta line) inside bulk $\text{AdS}_{d+1}$ (shaded blue region) with a holographic $\text{BCFT}_{d}$ at its boundary (thick black curve). The brane perspective is that of induced $\text{AdS}_{d}$ gravity coupled to the $\text{CFT}_{d}$. The boundary perspective uses holography of the brane gravity and is described by the $\text{CFT}_{d}$ coupled to a defect $\text{CFT}_{d-1}$ where the brane and $\text{AdS}_{d+1}$ boundary intersect (gold points). \textbf{Right:} A deformed brane $\mathcal{Q}$ (dashed, black line) sits at $\theta=\phi_{0}+\phi(\xi)$. The deformation has changed the location of the boundary of the $\text{BCFT}_{d}$.} \vspace{-7mm}
\label{fig:branefluc}
\end{center}
\end{figure}
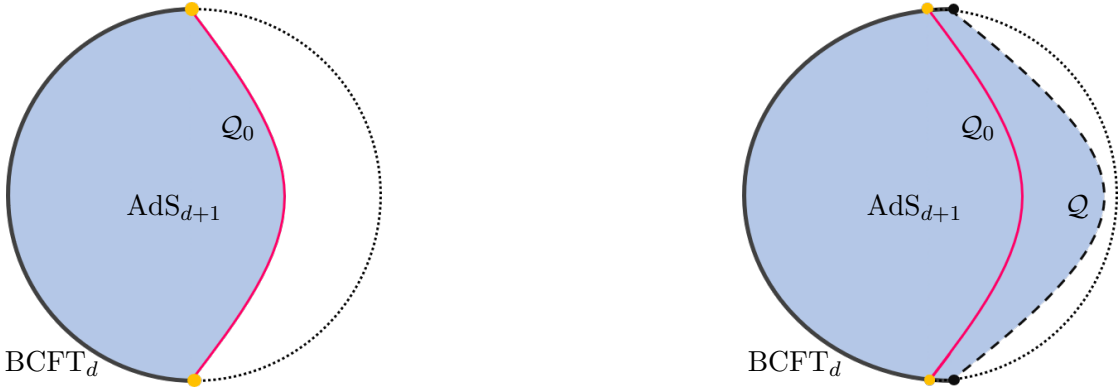

On the other hand, so-called `doubly-holographic' models \cite{Randall:1999vf,Karch:2000ct,deHaro:2000wj,Takayanagi:2011zk} naturally describe dynamical gravity minimally coupled to a large number of matter fields. Such models are constructed by coupling a $d$-dimensional end-of-the-world (ETW) brane to Einstein gravity in an asymptotically $d+1$-dimensional Anti-de Sitter (AdS) background, usually referred to as the `bulk picture'. Upon integrating out the bulk between the brane and the asymptotic boundary (i.e., the blue shaded volume in \cref{fig:branefluc}), the `brane description' of the system emerges in which quantum matter lives on a dynamical background $d$-dimensional spacetime and semi-classical gravity is induced by matter loops. Further, the bulk or brane perspectives have a UV description in terms of a $d$-dimensional holographic conformal field theory with a boundary ($\text{BCFT}_{d}$) coupled to a $(d-1)$-dimensional conformal defect. In two dimensions, however, the gravitational theory in the brane description -- at least naively -- is non-local Polyakov gravity \cite{Polyakov:1981rd,Chen:2020hmv} and does not capture many interesting gravitational solutions. In the study of black hole physics one thus usually resorts to adding a dilaton field in an \emph{ad hoc} manner to the brane action \cite{Almheiri:2019hni,Chen:2020uac,Chen:2020hmv,Grimaldi:2022suv,Lee:2022efh}.

This paper bridges two-dimensional semi-classical dilaton gravity and double holography by providing a doubly-holographic derivation of
two-dimensional models of dilaton gravity that are minimally coupled to a large number of CFTs and have a dual UV description. 
As we will see, fluctuating modes of ETW branes become physical and are controlled at low energies to leading order by Liouville gravity coupled to a two-dimensional conformal field theory,\footnote{While finishing our work we became aware of \cite{Kawamoto:2023wzj}, which states the brane theory for a brane near the AdS$_{3}$ boundary is equal to Liouville field theory. We thank Shan-Ming Ruan for pointing this out to us. Similar ideas, without ETW branes, were also discussed in, e.g., \cite{Carlip:2005tz,Nguyen:2021pdz}.}
\begin{align}
    \label{eq:liouvilleIntro}
    I = \frac 1 {4 \pi} \int d^2\xi \sqrt{-g} (Q \varphi \mathcal{R} + \nabla_i \varphi \nabla^i \varphi + \mu e^{2 b \varphi}) + W_\text{CFT}[g] + \mathcal O(\epsilon),
\end{align}
where $\mathcal{R}$ is the Ricci scalar and $Q$, $\mu$, and $b$ parameters.
The subleading terms schematically indicated by $\mathcal O(\epsilon)$ depend on a UV cutoff. By `Liouville gravity' we mean a theory where both the metric $g_{ij}$ and the dilaton $\varphi$ are dynamical. 
In exact Liouville gravity (i.e., a matter CFT coupled to gravity) one can absorb the dilaton into the conformal scale factor of the metric and only the conformal factor is dynamical. Here, however, the explicit breaking of Weyl invariance due to $\mathcal O(\epsilon)$ terms prevents this absorption. 
To distinguish those situations we will call the former, i.e., a theory with only a dynamical scale factor `Liouville Theory', whereas we reserve the name `Liouville gravity' for our theory where both a scalar field and the metric are integrated over.
The fact that both $\phi$ and $g_{ij}$ are dynamical fields makes it possible that, in appropriate limits, JT gravity in flat, AdS and dS spacetimes coupled to quantum matter can be obtained. As we will show, in those cases the renormalized dilaton value near the boundary is dual to a source for the boundary displacement operator in the dual BCFT description. 

Describing brane fluctuations in terms of a dilaton is not new. Historically, induced models of \emph{classical} dilaton-gravity on a braneworld arise by including a second brane \cite{Garriga:1999yh,Gen:2000nu,Kudoh:2001wb}. In such contexts, the dilaton, also known as the `radion',  is the (unique) scalar mode of the vacuum bulk gravity theory which describes the relative displacement of the two branes. Similarly, via wedge holography \cite{Akal:2020wfl,Miao:2020oey,Geng:2020fxl,Geng:2021hlu}, AdS JT gravity arises when two $\text{AdS}_{2}$ Karch-Randall \cite{Karch:2000ct} branes undergo small fluctuations characterized by the dilaton \cite{Geng:2022slq,Geng:2022tfc, Bhattacharjee:2022pcb,Deng:2022yll,Aguilar-Gutierrez:2023tic}. Notably, however, the dilaton theory considered in those works has some undesirable features. In particular, to ensure small brane fluctuations, the renormalized dilaton value at the asymptotic boundary has to scale to zero as the bulk IR cutoff is removed. Thus, in the limit of vanishing cutoff, the dilaton cannot diverge as it approaches the $\text{AdS}_{2}$ boundary, which means this theory cannot be consistently coupled to quantum matter \cite{Maldacena:1998uz,Almheiri:2014cka,Sarosi:2017ykf}. In contrast, in the present work we recover more standard models of dilaton gravity where the dilaton diverges near the boundary.

More precisely, we consider a `reference' ETW brane with constant tension $T_0$ coupled to three-dimensional general relativity which solves the brane equations of motion. We then take the off-shell action of a brane of constant tension $T$, which might be different than $T_0$, located at $\theta=\phi_0 + \phi(\xi)$, where $\phi(\xi)$ is the displacement in normal coordinates around the reference brane, and expand it in $\phi$. For a sketch of this setup we refer to figure \ref{fig:branefluc}.
After integrating out the bulk, we find the induced action on the brane is a two-dimensional dilaton theory coupled to an effective CFT, cf.\ \eqref{eq:res_final},

\begin{align}
    \label{eq:fullactionintro}
\begin{split}
I_{\total}[g, \phi] ={}&\frac{1}{16\pi G_{N}}\int d^{2}\xi\sqrt{-g}(\mu_1/2+\phi_{0}\mathcal{R})+W_{\text{CFT}}[g]+ \mathcal O(\epsilon)\\
 &+\frac{1}{16\pi G_{N}}\int d^{2}\xi\sqrt{-g}\left(\phi(\mathcal{R}+\mu_{1})+T_{0}(\partial_{i}\phi)^{2}+T_{0}\mu_{1}\phi^{2}\right)\\
&-\frac{\Delta T}{16\pi G_{N}}\int d^{2}\xi\sqrt{-g}\left(2 + 4T_{0}\phi +(\partial_{i}\phi)^{2}+\mu_{2}\phi^{2}\right)\;\\
&-\frac{1}{8\pi G_{N}}\int d\xi \sqrt{-g^\mathcal{I}}  \phi (\mathcal K + \sqrt{1 - ( T_0 + \Delta T)^2}) + \mathcal O(\phi^3) \;,\\
\end{split}
\end{align}
which, for $\Delta T = 0$ and after a field redefinition, reproduces the first few orders of an expansion of \eqref{eq:liouvilleIntro} in $b \phi$.
Here $\mu_{1}=2(1-T_{0}^{2})$, $\mu_{2}=2(1+T_{0}^{2})$, $\Delta T =  T - T_0$, and $G_{N}$ denotes the bulk Newton's constant. This action is valid up to second order in $\phi$ for bulk terms and at linear order in $\phi$ at the boundary.
The first line arises from the undeformed brane and captures a (cutoff) CFT living on the world volume of the reference brane, together with a cosmological constant term and a topological term. The $\mathcal O(\epsilon)$ term are suppressed in the UV cutoff and consist of terms higher order in $\mu_1 \sim (1 - T_0)$ and higher derivative corrections.

In other situations where dilaton gravity arises from dimensional reduction, the coefficient of the topological term typically measures an entropy. For example, JT gravity follows from a spherical reduction of higher-dimensional near-extremal black holes, where the dilaton encodes small deviations away from extremality and its constant background value is proportional to the ground state entropy of the black hole. Remarkably, a similar interpretation is also possible here. In our case $\phi_0/ (4G_N)$ is precisely the boundary entropy in the dual BCFT description, i.e., the Affleck-Ludwig  `ground state degeneracy' of the BCFT$_{2}$ \cite{Affleck:1991tk}.

The dilaton contribution in the second and third lines, meanwhile, captures the dynamics of braneworld deformations. As can easily be seen from \eqref{eq:fullactionintro}, for $\Delta T = 0$, the cosmological constant terms together with the term linear and quadratic in $\phi$ are the first few orders of an exponential potential $\sim \mu_1 e^{2 \phi}$, suggesting that the theory is indeed Liouville gravity at leading order in $\mu_1 = \mathcal O(\epsilon)$. The final line are the familiar Gibbons-Hawking-York (GHY) \cite{Gibbons:1976ue, York:1972sj} term and counterterm for dilaton gravity, which appear as a result of a delicate cancellation between contributions from the brane action, the GHY and counterterm action at the asymptotic boundary of the higher dimensional spacetime \cite{Balasubramanian:1999re} and the Hayward corner term \cite{Hayward:1993my} where the brane intersects the bulk AdS boundary. 

At linear order in $\phi$ and $\Delta T = 0$, we recover either AdS or dS JT gravity coupled to quantum matter, depending on whether $T_0$ is smaller or larger than one. 
For $T_0 = 1$ but $\Delta T \neq 0$, the equations of motion for this model are also compatible with a flat background, $\mathcal R = 0$, if $\Box \phi = - 2 \Delta T$ and $\phi \sim \mathcal O(\Delta T)$. Under such assumptions, the equation of motion reduce to those of flat-JT with action $\int d^2\xi \sqrt{-h} (\phi \mathcal R + \lambda^2)$, whose relation to other models for two-dimensional flat space gravity, specifically the CGHS model, we discuss.
See Table \ref{tab:JTtaxonomy} for a summary. Our work thus provides the first higher-dimensional origin of dilaton gravity coupled to a large-$c$ conformal field theory.

As mentioned above, previously, dilaton gravity has been introduced into doubly holographic models by adding JT gravity by hand in the brane perspective. This was reflected in the bulk by replacing the constant tension term of the ETW brane by the JT gravity action. While the implication for the boundary picture are not clear, it is believed that there should be a relation to boundary CFTs coupled to an SYK model \cite{Almheiri:2019hni}. Moreover, it was proposed that the Ryu-Takayanagi \cite{Ryu:2006bv,Hubeny:2007xt} formula in the presence of a JT term on the brane needs to be modified with a contact term whenever the RT surface $\sigma_A$ touches the ETW brane $\mathcal Q$ \cite{Almheiri:2019hni},
\begin{align}
    \label{eq:rt_intro}
    S_\text{RT} =\underset{\sigma_A}{\min \operatorname{ext}} \left( \frac{\text{Area}(\sigma_A)}{4 G_N}\right)  \to  \underset{\sigma_A}{\min \operatorname{ext}} \left( \frac{\text{Area}(\sigma_A)}{4 G_N} + \frac{\phi(\sigma_A \cap \mathcal Q)}{4 G_N} \right),
\end{align}
in order to correctly reproduce generalized entropies in the brane description of the system.

In the present paper we show that dilaton gravity, and for small fluctuations JT gravity, naturally arises in the brane perspective. Thus one may wonder if our construction sheds light on the boundary interpretation of JT gravity on the brane and whether the modification of the RT formula, \cref{eq:rt_intro}, can be derived. We answer the latter question in the affirmative and explicitly show how the RT formula acquires a contact term whenever the minimal surface ends on the ETW brane in the limit of small brane fluctuations. 
While a detailed discussion of the boundary interpretation of the dilaton is beyond the scope of the present work, we demonstrate that in the JT limit the location of the BCFT$_{2}$ boundary, constituting the location of a one-dimensional defect, is determined by the boundary value of the dilaton. Consequently, the deformations codified by the dilaton act as a source for a boundary operator of conformal dimension $\Delta=2$; namely, the $\text{BCFT}_{2}$ displacement operator. This provides another entry in the doubly-holographic dictionary in addition to the already mentioned fact that the topological contribution to the action (\ref{eq:fullactionintro}) precisely equals the boundary entropy. 

\begin{table}[t]
    \centering
    {\bfseries\strut Semi-classical dilaton gravity theories at leading order in the dilaton}
    \begin{tabular}{@{}llll@{}}
    \toprule
        tension & $\Delta T$ & theory & comment\\
        \midrule
         $T_0 < 1$ & $\Delta T = 0$ & AdS JT  + matter  & \\
         $T_0 = 1$ & $\Delta T < 0$ & Flat-JT  + matter & gravity becomes weak at spacelike infinity \\
         $T_0 = 1$ & $\Delta T = 0$ & Flat-JT  + matter & no black hole solutions \\
         $T_0 = 1$ & $\Delta T > 0$ & Flat-JT  + matter & gravity becomes strong at spacelike infinity \\
         $T_0 > 1$ & $\Delta T = 0$ & dS JT  + matter & \\
         \bottomrule
    \end{tabular}
    \caption{    Taxonomy of 2D dilaton gravity coupled to a large-$c$ CFT via deformed braneworld holography at leading order in deformations.  For AdS- and dS-JT gravity the cosmological constant is given by $\Lambda = 2(T_{0}-1)$. The cosmological constant term in flat-JT corresponds to $2\lambda^2 =- \Delta T$.    }
    \label{tab:JTtaxonomy}
\end{table}

The remainder of this article is organized as follows. In Section \ref{sec:dilagravaction} we present a covariant, perturbative derivation of the induced action (\ref{eq:fullactionintro}) that governs deformations about an ETW brane up to second order. We argue in Section \ref{sec:whyLiouv} the theory describing brane deformations is Liouville gravity
with a topological term and coupled to a large-$c$ CFT with UV-cutoff corrections. Section \ref{sec:emergeJTbrane} demonstrates how to recover various semi-classical models of JT gravity. 
We present a boundary conformal field theory interpretation of the (Karch-Randall) brane gravity theory in Section \ref{sec:BCFTpov}. In Section \ref{sec:RTderivation} we apply our set-up to derive the Ryu-Takayanagi formula in AdS JT gravity. We conclude in Section \ref{sec:disc} with a summary and outlook of future directions. To be self-contained and pedagogical, we include a number of appendices. Appendix \ref{app:conventions} describes our conventions. In Appendix \ref{app:perturbative_computations} we detail the derivation of the second order action for a fluctuating braneworld in arbitrary dimensions. Appendix \ref{app:branedefsJT} characterizes deformations parallel to an AdS brane. 

\section{Perturbative action in the brane perspective} \label{sec:dilagravaction}
In this section we derive the action that governs deformations of an ETW brane in a three-dimensional maximally symmetric bulk spacetime up to second order and the corresponding boundary terms to first order.\footnote{To clarify, the induced two-dimensional action receives cubic order corrections in deformations, however, the Gibbons-Hawking-York-like boundary term derived in Section \ref{sec:boundary_term} receives quadratic order corrections.} We will also consider the case where the location of a brane of tension $T$ is expressed as deformations of a brane of tension $T_0$ with $T \neq T_0$. In principle our approach can straightforwardly be extended to higher orders in fluctuations, although it becomes increasingly cumbersome to do so.  

After establishing the general form of the action, we specialize to the case of an AdS$_3$ bulk and argue that, in the brane description, the dilaton theory of gravity is coupled to quantum matter and discuss the interpretation of the coefficient of the topological term. A higher-dimensional analog is briefly discussed in Section \ref{sec:disc}, a complete treatment of which is left for future work.

\subsection{Set-up}

Let $\mathcal{M}$ be a $(2+1)$-dimensional spacetime endowed with metric $G_{\mu\nu}$. We consider the situation where the boundary of $\mathcal{M}$ consists of a contribution $\mathcal Q$ at a finite distance, a so-called end-of-the-world (ETW) brane, and a contribution $\mathcal N$ at asymptotic infinity, see Figure \ref{fig:branefluc}. For now we do not require an AdS bulk and $\mathcal N$ is allowed to be the empty set, i.e., there might not be an asymptotic boundary. In its most general form, the system is governed by the action
\begin{align}
\label{eq:Ibulkgen}
     \hspace{-4mm} I_\total = I_\text{bulk} + I_\text{brane} + I_\text{bdry} + I_\corner.
\end{align}
The first contribution,
\begin{align}\label{eq:bulkaction}
    I_\text{bulk} = \frac{1}{16\pi G_{N}} \int_{\mathcal M} d^3 x \sqrt{-G}\left(R[G] - 2 \Lambda\right),
\end{align}
is the usual Einstein-Hilbert action for the three-dimensional ambient geometry $\mathcal{M}$. The action of the ETW brane $\mathcal {Q}$, given by the second contribution, is composed of a Gibbons--Hawking--York (GHY) boundary term and a tension-term with constant tension $T$,
\begin{align}
    \label{eq:Ibrane}
    I_\text{brane} = - \frac{1}{8\pi G_{N}} \int_{\mathcal Q} \hspace{-1mm} d^2 \xi \sqrt{-\gamma} (K + T).
\end{align}
If the bulk spacetime has a remaining asymptotic boundary $\mathcal{N} = \partial \mathcal{M} - \mathcal{Q}$, the third term,
\begin{align}
    \label{eq:Iboundary}
    I_\text{bdry} =  - \frac{1}{8\pi G_{N}} \int_{\mathcal N} d^2 \xi \sqrt{-\gamma^\mathcal{N}} (K + 1),
\end{align}
needs to be added, which consists of a GHY term for the asymptotic boundary plus a counterterm such that the on-shell value of the action is finite. Lastly, since in this case the $\text{AdS}_{2}$ brane intersects the asymptotic boundary at an angle, the theory must be supplemented by a codimension-2 Hayward corner term \cite{Hayward:1993my},
\begin{align}
    \label{eq:Icorner}
    I_\corner = -  \frac{1}{8\pi G_{N}} \int_\mathcal{I} d \xi \sqrt{-\gamma^\mathcal I} (\Theta - \pi).
\end{align}
Here $\mathcal I = \overline {\mathcal N} \cap \overline{\mathcal Q}$ is the location where the brane intersects the asymptotic boundary with induced metric $\gamma^\mathcal{I}$ and $\Theta$ denotes the angle between the $\mathcal N$ and $\mathcal Q$.
Above and in the following we have set the bulk length scale $L$ to unity. The ETW brane is taken to be dynamical, i.e., we impose Neumann boundary conditions on $\mathcal Q$. The boundary conditions at the asymptotic boundary $\mathcal N$ are taken to be Dirichlet. 

Consider now an ETW brane $\mathcal Q_0$ with constant tension $T_0$. We can choose Gaussian normal coordinates $\{\theta, \xi^1, \xi^2\}$ such that $\mathcal Q_0$ is located at constant $\theta=\phi_0$ and the line element near $\mathcal Q_0$ has the expansion
\begin{align}
\begin{split}
\label{eq:metric_ansatz}
ds^2  &= d\theta^2 + \tilde g_{ij}(\theta, \xi) d\xi^i d\xi^j \\
&= d\theta^2 + \left( \tilde g_{ij}(\phi_0, \xi) + \delta \theta \; \partial_\theta \tilde g_{ij}(\phi_0, \xi) + \frac 1 2 \delta \theta^2\; \partial^2_\theta \tilde g_{ij}(\phi_0, \xi) + \dots \right)  d\xi^i d\xi^j,
\end{split}
\end{align}
where in the second line we expanded the metric up to second order in $\delta \theta < 0$, $|\delta \theta| \ll 1$. Using \eqref{eq:metric_ansatz}, the geometry can be extended behind the brane by allowing $\delta \theta$ to take on positive values. The coordinates $\xi^{i}$ parametrize the two-dimensional brane.

Intuitively, we want to think of the brane $\mathcal Q_0$ with tension $T_0$ as a `reference' or `undeformed' brane. The original, `deformed' brane $\mathcal Q$ with tension $T$ that appears in \cref{eq:Ibulkgen} is 
located at $\theta = \phi_0 + \phi(\xi)$ for small $\phi(\xi)$. The tension $T$ may, but does not have to be equal to the tension of the undeformed brane $T_0$. The goal of this section is to reexpress the action \cref{eq:Ibulkgen} involving $\mathcal Q$ as a perturbative series in $\phi(\xi)$ around an action involving $\mathcal Q_0$.  Our notation already reflects the expected result: The action of $\mathcal Q$ can be recast as an action for a dilaton-like field $\phi(\xi)$ living on the undeformed brane $\mathcal Q_0$. 

Before we continue, let us fix our conventions (see Appendix \ref{app:conventions} for further details). We denote the two-dimensional induced metric on the reference brane $\mathcal Q_0$ by $g_{ij}(\xi)$ and the metric on the deformed brane $\mathcal Q$ by $\gamma_{ij}(\xi)$. Both are generally different from $\tilde g_{ij}(\theta, \xi)$ in (\ref{eq:metric_ansatz}).
The projector onto the tangent space of the brane is denoted $h_{\mu\nu} = G_{\mu\nu} - n_\mu n_\nu$, with outwards-pointing spacelike unit normal $n_{\mu}$ obeying $G_{\mu\nu}n^{\mu}n^{\nu}=+1$. The induced metric on the asymptotic boundary is $\gamma^\mathcal N_{ij} = g^\mathcal N_{ij}$. The induced metric on the intersection of the branes $Q_0$ and $Q$ with $\mathcal N$ are written as $g^\mathcal I_{ab}$ and $\gamma^\mathcal I_{ab}$, respectively. See Table \ref{tab:metric_conventions} for a summary.

\begin{table}[t]
    \centering
    \begin{tabular}{@{}ll@{}}
    \toprule
        notation & description\\
        \midrule
         $G_{\mu\nu}(\theta, \xi)$ & bulk metric \\
         $h_{\mu\nu}(\theta, \xi)$ & projector onto the hypersurfaces perpendicular to $n_\mu$\\
         $\tilde g_{ij}(\theta, \xi)$ & induced metric on the hypersurface at constant $\theta$\\
         $g_{ij}(\xi)$ & induced metric on the brane at $\theta = \phi_0$\\
         $\gamma_{ij}(\xi)$ & induced metric on the brane at $\theta = \phi_0 + \phi(\xi)$\\
         $\mathfrak g_{ij}(\xi)$ & coefficients in a Fefferman-Graham-expansion\\
         $g^\mathcal{N}_{ij} = \gamma^\mathcal{N}_{ij}$ & induced metric at the asymptotic boundary\\
         $g^\mathcal{I}_{ab}$ & induced metric at the intersection of the asymp.\ boundary and undeformed brane\\
         $\gamma^\mathcal{I}_{ab}$ & induced metric at the intersection of the asymp.\ boundary and deformed brane\\
         \bottomrule
    \end{tabular}
    \caption{Summary of metrics and related tensors used in the main text. Superscripts $\mathcal N$ and $\mathcal I$ are sometimes omitted, when it is clear from the context which induced metric is being referred to.}
    \label{tab:metric_conventions}
\end{table}

Below, in Section \ref{sec:bulk_term}, we make no specific assumptions about the bulk geometry, except that it is maximally symmetric. In particular, we imagine embedding both branes in a background satisfying the vacuum Einstein equations. In \cref{sec:boundary_term} and the following we will specialize to the case where the background is asymptotically AdS and has a dual holographic description in terms of a CFT living on the remaining portion of the AdS boundary.  
Presently, we only make two essential assumptions. First, the undeformed brane $\mathcal Q_{0}$ satisfies the brane equations of motion which follow from \eqref{eq:Ibulkgen} (c.f.~Appendix \ref{app:conventions}),
\begin{align}
    \label{eq:brane_eom}
    K^{(0)}_{\mu\nu} = - T_{0} h^{(0)}_{\mu\nu},
\end{align}
where $K^{(0)}_{\mu\nu}$ is the brane's extrinsic curvature.
Second, we further assume the branes are embedded in a bulk spacetime that is a solution to the $\theta\theta$ component of Einstein's vacuum field equations with an arbitrary cosmological constant, such that part of the bulk metric is fixed, but the induced metric on the brane is not.

Incidentally, these assumptions are sufficient to considerably simplify the metric ansatz (\ref{eq:metric_ansatz}).
Namely, the first two orders in $\delta \theta = \phi(\xi)$ are respectively proportional to $g_{ij}(\xi) = \tilde g_{ij}(\phi_0, \xi)$, the induced metric on the undeformed brane, and $\partial_\theta \tilde g_{ij}(\phi_0, \xi)$, which is proportional to the extrinsic curvature $K^{(0)}_{ij}$ of the undeformed brane. Via the brane equation of motion (\ref{eq:brane_eom}), we learn that the latter expression must also be proportional to $h_{ij}^{(0)} = g_{ij}(\xi)$. Further, the bulk vacuum Einstein equations may be understood as a second order differential equation in $\theta$ with both initial conditions proportional to $g_{ij}(\xi)$. Since no other tensor structures enters, all higher order terms in the ansatz (\ref{eq:metric_ansatz}) must also be proportional to $g_{ij}(\xi)$. Therefore, we can always cast the bulk metric as
\begin{align}
\begin{split}
\label{eq:metric_ansatz_simple}
ds^2  &=  d\theta^2 + f^2(\theta) g_{ij}(\xi) d\xi^i d\xi^j\;,
\end{split}
\end{align}
where $f(\phi_0) = 1$. The form of $f(\theta)$ depends on the choice of bulk cosmological constant and is otherwise completely determined by its first two terms in a series expansion around $\phi_0$. We have that $f(\theta) \sim \cosh \theta, e^\theta, \sinh \theta$, for bulk AdS, flat space, and dS, respectively. In Appendix \ref{app:perturbative_computations} we give a more explicit argument for the form of (\ref{eq:metric_ansatz_simple}) using perturbation theory.
The line element on the deformed brane at $\theta=\phi_{0}+\phi(\xi)$ is 
\beq 
ds^{2}_{\mathcal Q}=\gamma_{ij}d\xi^{i} d\xi^{j}
=
\left( f^{2}(\phi_{0}+\phi)g_{ij}(\xi)+\partial_{i}\phi \, \partial_{j}\phi\right)d\xi^{i}d\xi^{j}\;,\label{eq:inducedmetdefB}
\eeq
with $\partial_{i}\phi\equiv\partial_{\xi^{i}}\phi$.

To continue, we split the full action (\ref{eq:Ibulkgen}) into three terms,
\beq I_\total = I^{(0)}_\total + \Delta I^\text{bulk}_\total  + \Delta I^\text{bdry}_\total \;,\eeq
The first one, denoted $I^{(0)}_{\total}$, is independent of deformations $\phi(\xi)$ and describes the theory in the presence of the undeformed brane $\mathcal Q_0$.
The second contribution $\Delta I^\text{bulk}_{\total}$, meanwhile, is the action for the brane deformations $\phi(\xi)$ living on $\mathcal Q_0$ which arises from $I_\text{bulk} + I_\text{brane}$. The last term $\Delta I^\text{bdry}_{\total}$ are $\phi$-depedent terms that stem from contributions at $\mathcal I$ and are not captured by $\Delta I^\text{bulk}_{\total}$. In the next three sections we analyze all three contributions, starting with the bulk contributions to $\Delta I_{\total}$ before we focus on $\Delta I^\text{bdry}_{\total}$ and $I^{(0)}_\total$ in Sections \ref{sec:boundary_term} and \ref{sec:topmatcoup}, respectively.

\subsection{Action for deformations: The bulk term} \label{sec:bulk_term}

The action $\Delta I^\text{bulk}_{\total}$ is obtained by expanding the action $I_\text{bdry}$ and $I_\text{brane}$ in $\phi(\xi)$ and keeping terms at linear or higher order in $\phi$. Since the main result of this subsection, \cref{eq:defactfinal}, has been obtained before \cite{Geng:2022slq,Geng:2022tfc} (with the exception of a total derivative term which will be discussed in more detail in the next subsection), our presentation will be rather brief and only summarize the most important intermediate results to go to quadratic order in $\phi(\xi)$, leaving details to Appendix \ref{app:perturbative_computations} (including when the brane is $d$-dimensional). 

However, it should be noted that our derivation, unlike previous expositions, is fully covariant. This is made possible by the key input in this section: the Gauss-Codazzi relation 
\begin{align}
    \label{eq:gauss_codazzi}
    R = \mathcal{R} - K^2 - K^{\mu\nu} K_{\mu\nu} + 2 n^\mu \partial_\mu K + 2 D_\mu a^\mu - 2 a^2,
\end{align}
which relates the three-dimensional Ricci scalar  $R$ to the two-dimensional Ricci scalar $\mathcal R$ on the brane.
Upon substitution of \cref{eq:gauss_codazzi} into $I_\text{bulk} + I_\text{brane}$, the contribution $\Delta I_{\total}$ becomes
 \beq
\begin{split}
\Delta I^\text{bulk}_{\total}&=\frac{1}{16\pi G_{N}}\int d^{2}\xi\biggr[\sqrt{-\tilde{g}}\phi(\xi)\left(\mathcal{R}[\tilde{g}]-K^{2}-K^{\mu\nu}K_{\mu\nu}+2n^{\mu}\partial_{\mu}K-2\Lambda\right)\big|_{\theta=\phi_{0}+\frac{1}{2}\phi}\\
&-2\sqrt{-\gamma}(K+T)\big|_{\theta=\phi_{0}+\phi}+2 \sqrt{-g} (K + T)\big|_{\theta = \phi_0}\biggr]+\mathcal{O}(\phi^{3})\;,
\end{split}
\label{eq:DeltaIbulkint}\eeq
where we used $\sqrt{-G}=\sqrt{-\tilde{g}}$, and dropped terms involving the `acceleration' $a^{\mu}=n^{\nu}\nabla_{\nu}n^{\mu}$, since they will ultimately contribute terms of order $\mathcal{O}(\phi^{3})$ (see Appendix \ref{app:perturbative_computations}, around (\ref{eq:athetaak})). We used the fundamental theorem of calculus to expand the integral over $\theta$, which also prompts the factor of $\phi(\xi)$ in the first line and the evaluation at $\theta=\phi_{0}+\frac{1}{2}\phi$.

To proceed, we need only expand $\tilde g_{ij}(\theta,\xi)$ and $n^\mu \partial_\mu K$ to first order in deformations, and $\gamma$ and $K$ to second order. Further, we parametrize the cosmological constant as $\Lambda=\kappa$
where $\kappa=-1,0,1$ respectively corresponds to AdS, flat, dS geometry in the bulk.
The bulk equations of motion then imply 
\beq f''(\theta) = - \kappa f(\theta)\;,\eeq
together with boundary condition $f(\phi_{0})=1$.
Remaining agnostic to the choice of $\kappa$ and the slicing $f(\theta)$, we find
\begin{align}
    \tilde g_{ij} &= g_{ij} (1 + 2 T_{0} \phi) + \mathcal O(\phi^2),    \label{eq:gtilde_expansion}\\
    n^\mu \partial_\mu K 
    &= 
    2 \left(T_{0}^2 + \kappa \right) 
    - 
    2  \left( 
    	2T_{0}^{3}+3 T_{0}\kappa +f'''(\phi_{0})
    \right)\phi
    - 
    2 T_{0} \Box \phi  + \mathcal O(\phi^2),\\
    K 
    &= - 2 T_{0} + \Box \phi 
    +
    2  \left(T_{0}^2+\kappa \right)  \phi
    -
    2T_{0}\phi\box\phi \nonumber \\ & \qquad
    +
    \left(
    	2 T_{0}^{3}+3T_{0}\kappa+f'''(\phi_{0})
    \right)\phi^{2}
    + \mathcal O(\phi^3), \\
    \sqrt{-\det \gamma} 
    &  =  
    \sqrt{-\det g} \left(1 + 2 T_{0} \phi + \frac {1} 2 g^{ij} \partial_i \phi \partial_j \phi + \left(T^{2}_{0} -\kappa \right) \phi^2   \right)  + \mathcal O(\phi^3),
\end{align}
leaving the computational details for Appendix \ref{app:perturbative_computations}.

The resulting total action for fluctuations up to second order is
\beq 
\begin{split}
\label{eq:defactfinal}
\Delta I^\text{bulk}_{\total}={}&\frac{1}{16\pi G_{N}}\int d^{2}\xi\sqrt{-g}(\phi\mathcal{R}+\mu_{1}\phi+T_{0}(\partial_{i}\phi)^{2}+T_{0}\mu_{1}\phi^{2})\\
&-\frac{\Delta T}{16\pi G_{N}}\int d^{2}\xi\sqrt{-g}(4\phi T_{0}+(\partial_{i}\phi)^{2}+\mu_{2}\phi^{2})
\\&
-\frac{1}{8\pi G_{N}}\int d^{2}\xi \sqrt{-g}\Box\phi\;,
\end{split}
\eeq
where  $(\partial_{i}\phi)^{2}=g^{ij}(\partial_{i}\phi)(\partial_{j}\phi)$ and we have introduced parameters 
\begin{align}
\label{eq:mus}
    \mu_{1}\equiv -2(\kappa+T_{0}^{2}), \qquad \mu_{2}\equiv 2(T_{0}^{2}-\kappa), \qquad \Delta T = T - T_0.
\end{align}
\Cref{eq:defactfinal} gives the general action for fluctuations where we have allowed for the tension of the deformed brane $T$ to be different than that of the undeformed brane $T_{0}$. Note that all terms of the form $\phi \Box \phi$ cancel and so no integration by parts is necessary. However, the final line results in a boundary contribution. This term seems rather unusual and its fate will be discussed momentarily.

The linear terms in \eqref{eq:defactfinal} make up the usual action of JT gravity with cosmological constant $\Lambda_{2}=2(\kappa+TT_{0})$.
When $T_{0}=T$, i.e., $\Lambda_{2} = 2(\kappa+T_{0}^{2})$, the bulk cosmological constant imposes a lower bound on the brane cosmological constant. This implies for bulk AdS $(\kappa=-1)$, (A)dS$_{\text{2}}$ brane slicings are allowed, while for de Sitter $(\kappa=+1)$ or flat ($\kappa=0$) bulk, only a dS brane slicing is allowed. 
Though not clear at this stage, when $\kappa=-1$, recovering flat-JT gravity requires $\Delta T \neq 0$ and $T_{0} = 1$, which is the reason to allow for $T \neq T_0$. Thus, when the bulk is $\text{AdS}_{3}$, we can recover three types of brane slicings. This, together with the fact that the full boundary theory in the case of AdS$_3$ is well-understood, motivates us to only consider the case of $\kappa = -1$ from now on.

\subsection{Action for deformations: The boundary term}
\label{sec:boundary_term}
Seemingly missing from the deformed action \eqref{eq:defactfinal} is the familiar GHY boundary term for dilaton gravity, proportional to $\int d\xi\sqrt{-g^\mathcal I}\phi \mathcal K$. Instead, the action has an unusual $\Box\phi$ piece, which contributes as a boundary term. Such an action would clearly be inconsistent, since the absence of the GHY term causes \eqref{eq:defactfinal} to result in an ill-posed variational problem in the presence of a boundary.
However, as we will now see, the contribution $\Delta I^\text{bdry}_\total$ which arise if the brane is deformed in the presence of an asymptotic boundary $\mathcal N$ solves this problem. To be more precise, instead of the asymptotic boundary we have to consider a regulator surface which sits at a small but finite value of the holographic coordinate and only take it to infinity at the end of our computations. The choice of this regulator surface and the limiting procedure determines the Weyl frame for the background on which the dual CFT is defined.

For concreteness, we consider an $\text{AdS}_{2}$ brane in an $\text{AdS}_{3}$ bulk and express the bulk in Poincar\'e coordinates
\begin{align}
    \label{eq:poincare_coords}
    ds^2_{\text{AdS}_{3}} = \frac{-dt^2 + dx^2 + dz^2}{z^2},
\end{align}
which is cut off by our ETW brane $\mathcal{Q}_0$ of tension $T_0$ that solves the (undeformed) brane equations of motion (\ref{eq:brane_eom}). Here we take the dual CFT to live on the half-space $x\leq0$ with some boundary conditions imposed at $x=0$.
In these coordinates, we choose the regulator surface at $z = z_c$. In an abuse of notation, we will refer to this surface in the following as the `asymptotic boundary' and still denote it by $\mathcal N$, although it should be understood that the asymptotic boundary is only reached as $z_c\to0$.

The coordinate choice \eqref{eq:poincare_coords} is different from the one made in the rest of \cref{sec:dilagravaction}. While the present choice of coordinates does not alter the relevant physics, it is much better suited to understand the effect of the brane deformation at the asymptotic boundary. The metric \eqref{eq:poincare_coords} is related to a metric of the form \eqref{eq:metric_ansatz_simple} through the coordinate transformation
\begin{align}
    \label{eq:coord_trans}
    z=\frac{y}{\cosh \theta }, \qquad x=y \tanh \theta,
\end{align}
from which it immediately follows that
\begin{align}
\label{eq:x_vs_z}
    x = z \sinh \theta.
\end{align}
Applying this change of coordinates to \eqref{eq:poincare_coords}, we recover $ds^{2}_{\text{AdS}_{3}}=d\theta^{2}+\cosh^{2}(\theta) g_{ij}^{\text{AdS}_{2}}d\xi^{i}d\xi^{j}$, with $g_{ij}^{\text{AdS}_{2}}d\xi^{i}d\xi^{j}=\frac{1}{y^{2}}(-dt^{2}+dy^{2})$. The ETW brane with tension $T_0$ is located at 
\begin{align}
    T_0 = \tanh \theta_0
\end{align}
and the AdS length on the brane is given by
\begin{align}
    \label{eq:ads_length_brane}
    L_\text{brane} = \frac{L}{\sqrt{1-T_0^2}}.
\end{align}
The location of the brane in $(t,x,z)$ coordinates is given to zeroth order by $x = X(z)$ with
\begin{align}
    \label{eq:standard_solution_poincare}
    X(z) = \sinh(\theta_0) z =  \frac{T_0}{\sqrt{1 - T_0^2}}\; z,
\end{align}
where we have capitalized the function $X(z)$ to distinguish it from the coordinate $x$.

As indicated in the complete action \eqref{eq:Ibulkgen}, the asymptotic boundary is equipped with GHY and Hayward terms. 
Deformations of the brane will in general change the location and shape of the intersection of the brane with the asymptotic boundary, and thus we expect additional $\phi$-dependent terms to arise from such boundary terms. Moreover, upon deforming, parts of the brane will move behind or in front of the cutoff surface, i.e., the range of integration along the brane will change. This also contributes a correction such that the overall boundary contribution to the action for $\phi$ in \eqref{eq:defactfinal} is
\begin{align}
    \label{eq:remaining_boundary}
    \Delta I^\text{bdry}_\total = \delta_{\partial} \left( I_\text{bdry} + I_\corner + I_\text{brane} \right)
\end{align}
where $\delta_\partial$ denotes deformations of the intersection of the brane with the asymptotic boundary (i.e., regulator surface) and the actions were defined in \eqref{eq:Iboundary} -- \eqref{eq:Ibrane}.

As we now demonstrate, the effect of the boundary term \eqref{eq:remaining_boundary} is to exactly replace the $\Box \phi$ term in the deformed action \eqref{eq:defactfinal} by the usual boundary term of dilaton gravity, including the correct counterterm. To show this, we study \eqref{eq:remaining_boundary} to linear order in $\phi$. The higher order contributions are left for future work. A crucial aspect in the following is the relation between deformations $\delta \theta = \phi$ and deformations in $(t,x,z)$ coordinates. Recall that $\phi$ was defined to be the brane deformation at a constant value of $y$. Thus, to linear order, we have under a displacement $\phi$ of the brane, 
\begin{align}
    \delta x = \sqrt{1-T_0^2} z \phi \qquad \text{and} \qquad \delta z = - T_0 z \phi,
\end{align}
from the coordinate transformation \eqref{eq:coord_trans}.

\begin{figure}[t!]
    \centering
    \begin{subfigure}{0.49\textwidth}
    \centering
    \begin{tikzpicture}
        \draw (-2,0) -- (0,0);
        \draw[dotted] (-2,0.2) node[left] {$z_c$} -- (0.5,0.2);
        \draw[thick] (0,0)  -- node [above left] {$\mathcal Q$} (45:2);
        \draw[->, dashed, gray] (0,0) ++ (45:1) arc (45:15:1);
        \draw[->, dashed, gray] (0,0) ++ (45:0.5) arc (45:15:0.5);
        \draw[->, dashed, gray] (0,0) ++ (45:1.5) arc  (45:15:1.5) node[above right] {\textcolor{black}{$\delta \theta = \phi$}};
        \draw[gray, ->] (-2,1.5) -- ++(0.75,0);
        \draw[gray, ->] (-2,1.5) -- ++(45:0.75) node[above right] {$y$};
        \draw[gray, ->] (-2,1.5) -- ++(90:0.75);
        \draw[gray, ->] (-2,1.5) -- ++(135:0.75);
        \draw[gray, ->] (-2,1.5) -- ++(180:0.75);
        \draw[gray, ->] (-2,1.5) ++(0:0.5) arc (0:180:0.5) node[below] {$\theta$};
        \draw (0,0) ++ (-0.5,0) arc (180:45:0.5);
        \draw (0,0) ++ (-0.1,0.25)  node {$\Theta$};
    \end{tikzpicture}
    \caption{}
    \end{subfigure}
    \begin{subfigure}{0.49\textwidth}
    \centering
    \begin{tikzpicture}
        \draw (-2,0) -- (0,0);
        \draw[dotted] (-2,0.2) node[left] {$z_c$} -- (0.5,0.2);
        \draw[thick] (0,0) -- node [above left] {$\mathcal Q$}(45:2);
        \draw[->, dashed, gray] (0,0) ++ (45:1) -- ++(0.7,0) node[right] {\textcolor{black}{$\delta X$}};
        \draw[->, dashed, gray] (0,0) ++ (45:0.5) -- ++(0.7,0) ;
        \draw[->, dashed, gray] (0,0) ++ (45:1.5) -- ++(0.7,0) ;
        \draw[gray, ->] (-2,1.5) -- ++(0.75,0) node[right] {$x$};
        \draw[gray, ->] (-2,1.5) -- ++(0,0.75) node[above] {$z$};
        \draw (0,0) ++ (-0.5,0) arc (180:45:0.5);
        \draw (0,0) ++ (-0.1,0.25)  node {$\Theta$};
    \end{tikzpicture}
    \caption{}
    \label{fig:angle}
    \end{subfigure}
    \caption{Two different expressions for deformations of the brane $\mathcal Q_0$, shown on a constant time-slice. (a) Direction of deformations is perpendicular to the brane. (b) Deformations of the brane are parametrized parallel to the boundary.}
    \label{fig:fluctuations}
\end{figure}
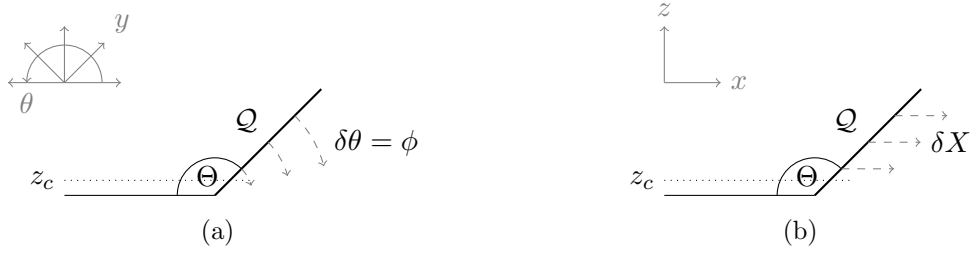

To evaluate the boundary terms, we are interested in deformations of the intersection locus of the brane with the asymptotic boundary \emph{at constant $z = z_c$}. Under a deformation with $\phi$ the brane trajectory, \eqref{eq:standard_solution_poincare} changes to
\begin{align}
    \label{eq:new_traj}
    X(z) \mapsto X(z) + \delta X(z) = X(z - \delta z) + \delta x = X(z) + \frac{\phi}{\sqrt{1-T_0^2}} z.
\end{align}
\Cref{fig:fluctuations} illustrates the difference of the parametrization of the fluctuations in the case of an $\text{AdS}_{2}$ brane in an $\text{AdS}_{3}$ bulk. 
The integration limit of the $x$ integral in the GHY boundary-term in \eqref{eq:remaining_boundary} shifts by $\delta X(z_c)$, which produces
\begin{align}
    \delta I_{\text{bdry}} = - \frac{1}{8 \pi G_N} \int dt \sqrt{-g^\mathcal{I}} \frac{\delta X(z_c)}{z_c} (K^{\mathcal N} + 1) = - \frac{1}{8 \pi G_N} \int dt \sqrt{-g^\mathcal{I}}  \phi \frac{K^{\mathcal N} + 1}{\sqrt{1-T_0^2}},
\end{align}
where we have added a superscript to the extrinsic curvature scalar $K^\mathcal N$ to indicate that it captures the extrinsic curvature of the asymptotic boundary $\mathcal N$. 
We now need to express $K^\mathcal N$ in terms of the extrinsic curvature of the boundary of the brane $\mathcal Q$, which we will denote by $\mathcal K$. To this end, it is useful to give explicit expressions for the normal covectors $N_{\mathcal N}$, $N_{\mathcal Q}$ and $n_{\mathcal Q}$ at the intersection $\mathcal I$, namely, the bulk normal to the asymptotic boundary $\mathcal N$, the bulk normal to the brane $\mathcal Q$ and the brane normal to the intersection $\mathcal I$, respectively. Explicitly,
\begin{align}
N_{\mathcal N} = -\frac{dz}{z_c}, && N_{\mathcal Q} = \frac{\sqrt{1-T_0^2} dx - T_0 dz}{z_c}, && n_{\mathcal Q} = \frac{- T_0 dx - \sqrt{1-T_0^2} dz}{z_c}.
\end{align}
Thus, we can decompose
\begin{align}
\begin{split}
    K^{\mathcal N} &= G^{tt} K^{\mathcal N}_{tt} + G^{xx} K^{\mathcal N}_{xx} 
    = - G^{tt} \nabla_t (N_{\mathcal N})_t - 1\\
    &= - G^{tt} \nabla_t \left( \sqrt{1-T_0^2} n_{\mathcal Q} + T_0 N_{\mathcal Q} \right)_t - 1 \\
    &= \sqrt{1-T_0^2} \mathcal K - 1 - T_0^2,
\end{split}
\end{align}
where $\mathcal K$ denotes the brane extrinsic curvature of the intersection surface $\mathcal I$. Ultimately,
\begin{align}
    \label{eq:corr1}
    \delta_\partial I_{\text{bdry}} = - \frac{1}{8 \pi G_N}\int dt \sqrt{-g^\mathcal{I}}   \phi \left( \mathcal K   - \frac{T_0^2}{\sqrt{1-T_0^2}}\right).
\end{align}

To obtain the contribution from the Hayward corner term, note that since we are only interested in terms linear in $\phi$, we can ignore the variation of the induced metric $\sqrt{-\gamma^\mathcal{I}}$. The only contribution then comes from $\Theta$, the angle between the brane and the asymptotic boundary. The angle $\Theta$ can be computed, e.g., from the inner product of the normal vectors of the asymptotic boundary and the brane. Alternatively, from Figure \ref{fig:angle} it follows that $\tan (\pi - \Theta) = \partial_z X(z)$. Thus, 
\begin{align}
    \cot \Theta = - \partial_z X(z),
\end{align}
which at linear order translates to
\begin{align}
    \delta \Theta = (1 - T_0^2) \partial_z \delta X(z) = \sqrt{1 - T_0^2} (\phi + z \partial_z \phi) = \sqrt{1 - T_0^2} \phi - \partial_n \phi,
\end{align}
where in the last step we have used that $n_{\mathcal Q}$ can be written as $n_{\mathcal Q} = -\frac{dy}{z_c}$. This yields the a correction coming from the Hayward term which takes the form
\begin{align}
    \label{eq:corr2}
    \delta_\partial I_\corner = - \frac 1 {8 \pi G_N} \int dt \sqrt{-g^\mathcal{I}} \left(\sqrt{1-T_0^2} \phi - \partial_n \phi\right)\;.
\end{align}

Lastly, we need to correct for the fact that under deformations parts of the brane will have crossed the regulator surface. This contributes the on-shell value of the brane action over a range of $\Delta y = \frac{\delta z}{\sqrt{1 - T_0^2} }$,
\begin{align}
    \label{eq:corr3}
    \delta_\partial I_\text{brane} = - \int dt \sqrt{-g^\mathcal{I}} \frac{\Delta y}{z_c} (K^{(0)} + T) = - \int dt \sqrt{-g^\mathcal{I}} \frac{T_0^2 + T_0 \Delta T}{\sqrt{1-T_0^2}} \phi,
\end{align}
where we used $K^{(0)} = - 2 T_0$. 
Altogether, adding \cref{eq:corr1,eq:corr2,eq:corr3} yields
\begin{align}
    \Delta I^\text{bdry}_\total[\phi] = -\frac{1}{8 \pi G_N}\int dt \sqrt{-g^{\mathcal I}} \left( \phi (\mathcal K + \sqrt{1 - T_0^2} - \frac{\Delta T T_0}{\sqrt{1 - T_0^2}}) - \partial_n \phi \right).
\end{align}
Note that in the limit of $T_0 \to 1$ this expression stops making sense. The reason is that in our choice of coordinates this limit forces the brane behind the cutoff surface and our analysis breaks down. However, one can easily recognise the constant in the integrand to be the first terms of a series expansion of $\sqrt{1 - (T_0 + \Delta T)^2}$. We thus propose that the correct expression for all $T_0$ is given by  
\begin{align}
    \Delta I^\text{bdry}_\total[\phi] = -\frac{1}{8 \pi G_N}\int dt \sqrt{-g^{\mathcal I}} \left( \phi (\mathcal K + \sqrt{1 - (T_0 + \Delta T)^2}) - \partial_n \phi \right) + \mathcal O(\Delta T).
\end{align}

We can recognize the contribution proportional to $\phi$ as the boundary term for dilaton gravity, including the counterterm proportional to $1/L_\text{brane}$, \eqref{eq:ads_length_brane}. Moreover, the normal derivative contribution cancels the corresponding $\Box\phi$ in \eqref{eq:defactfinal}. Lastly, note that one could repeat the computation for a dS brane in AdS. In this case we would need to take the absolute value of the expression under the square root.

\subsection{Matter coupling and the topological term} \label{sec:topmatcoup}
So far we have only focused on the action for brane deformations. Let us now turn to the remaining piece of the action that is independent of brane fluctuations,
\begin{align}
    &I^{(0)}_\total =  \frac{1}{16\pi G_N} \int d^2\xi \int^{\phi_0 } d\theta \sqrt{-G}\left(R[G] - 2 \Lambda\right) - \frac{1}{8 \pi G_N} \int d^2 \xi \sqrt{- g} (K^{(0)} + T)\big|_{\theta = \phi_0}\;,\label{eq:I1}
\end{align}
with $K^{(0)}=G^{\mu\nu}K^{(0)}_{\mu\nu}=-2T_{0}$ via the brane equations of motion, \eqref{eq:brane_eom}. It is comprised of three contributions: an Einstein-Hilbert term $I_{\text{bulk}}$, a GHY term $I_{\text{GHY}}$, and the purely tensional brane action $I_{\text{T}}$ of tension $T$ (the latter two assemble into $I_\text{brane}$ in \cref{eq:Ibrane}), i.e.,
\begin{align}
    \label{eq:Izero}
    I^{(0)}_{\total}=I_{\text{bulk}}+I_{\text{GHY}}+I_{\text{T}}\;,
\end{align}
with
\begin{equation}\label{eq:Ibrane2}
    I_{\text{T}}=-\frac{T}{8\pi G_{N}}\int d^{2}\xi\sqrt{-g}
    \,.
\end{equation}
We will now show these terms give rise to an effective large-$c$ CFT with UV-cutoff corrections, a topological term which encodes the boundary entropy of the dual BCFT, and a cosmological constant term.

\subsubsection*{The matter sector}
Let us start by recalling how \eqref{eq:Izero} is evaluated in standard AdS/CFT.
In the absence of an ETW brane, the on-shell bulk gravitational action suffers from long distance infrared (IR) divergences since the metric grows to infinity at the AdS boundary. Under the standard AdS/CFT dictionary, these bulk IR divergences coincide with ultraviolet (UV) divergences in the dual CFT.
These divergences can be dealt with by using holographic renormalization \cite{deHaro:2000vlm}, a prescription that removes the bulk IR divergences by adding appropriate local counterterms \cite{deHaro:2000vlm,Kraus:1999di,Emparan:1999pm,Papadimitriou:2004ap} and using a minimal subtraction scheme.

For holographic renormalization, it is standard to express the AdS bulk geometry in a Fefferman-Graham expansion,
\beq ds^{2}=\frac{1}{4\rho^{2}}d\rho^{2}+\frac{1}{\rho}\mathfrak{g}_{ij}(\rho,\xi)d\xi^{i}d\xi^{j},\label{eq:FGexpangen}\eeq
near the asymptotic boundary $\rho=0$.
The IR divergences are regulated by introducing an IR cutoff surface at $\rho=\epsilon$ for $\epsilon \ll 1$ and evaluating the on-shell value of the bulk action, restricted to the integration region $\epsilon \leq \rho$. The asymptotic boundary also lives at $\rho\to\infty$. However, for us it is sufficient to focus on the region near $\rho=0$.
The regulated action $I^{\epsilon}_{\text{reg}}$ is then given by
\beq I^{\epsilon}_{\text{reg}}\equiv I^{\epsilon}_{\text{bulk}}+I_{\text{GHY}}^{\epsilon}\big|_{\text{on-shell}}\;. \label{eq:regact}\eeq

To evaluate the right hand side on-shell, one expands the metric $\mathfrak{g}_{ij}(\rho,\xi)=\mathfrak{g}_{ij}^{(0)}(\xi)+\rho\mathfrak{g}^{(2)}_{ij}(\xi)+...$, and determines coefficients $\mathfrak{g}_{ij}^{(i\geq2)}$ by perturbatively solving the bulk Einstein equations \cite{deHaro:2000vlm}. 
Schematically, the regulated action (\ref{eq:regact}) is composed of a divergent and finite contribution in the limit $\epsilon\to0$.

Specifically, for $d=2$, using $\operatorname{tr}(\mathfrak{g}^{(2)}_{ij})=\mathfrak{g}_{(0)}^{ij}\mathfrak{g}_{ij}^{(2)}=-\frac{1}{2}\mathcal{R}[\mathfrak{g}^{(0)}]$ as well as $g_{ij}(\xi)=\frac{1}{\epsilon}\mathfrak{g}_{ij}(\epsilon,\xi)$ the on-shell action can be cast as
\begin{align}
I^{\epsilon}_\text{reg} = \frac{1}{16\pi G_{N}}\int d^{2}\xi\sqrt{-g} \big(2-\log \sqrt{\epsilon} \; \mathcal{R}[g] \big)+ (\text{finite})\;,\label{eq:covariant2don}
\end{align}
with a finite contribution that survives the $\epsilon \to 0$ limit.

From here, the renormalized action $I_{\text{ren}}$ is defined via a simple minimal subtraction,
\begin{align}
    \label{eq:def_ren}
    I_{\text{ren}}\equiv \lim_{\epsilon\to0}(I^{\epsilon}_{\text{reg}}+I^{\epsilon}_{\text{ct}})\;,
\end{align} 
with $I^{\epsilon}_{\text{ct}}$ being a local counterterm equal to (minus) the divergent contribution to the regularized on-shell action, i.e., 
\beq I^{\epsilon}_{\text{ct}}\equiv-\frac{1}{16\pi G_{N}}\int d^{2}\xi\sqrt{-g}\big(2-\log \sqrt{\epsilon} \; \mathcal{R}[g] \big)\;.\label{eq:ctact2d}\eeq
Via the AdS/CFT dictionary, metric $g_{ij}$ variations of the renormalized action yields the quantum expectation value of the stress-tensor of the holographic CFT, $\langle T^{\text{CFT}}_{ij}\rangle$. Hence,
\beq I_{\text{ren}} \equiv W_{\text{CFT}}[g]\;,\label{eq:CFTanomact}\eeq
with $W_{\text{CFT}}$ being the quantum effective action of the CFT, such that 
\begin{equation}\label{eq:CFTtrace}
    g^{ij}\langle T^{\text{CFT}}_{ij}\rangle=-\frac{2}{\sqrt{-g}}
    g^{ij}
    \frac{\delta W_{\text{CFT}}[g]}{\delta g^{ij}}
    =
    \frac{c}{24\pi}\mathcal{R}
    \,.
\end{equation}

Now consider the situation in the presence of an ETW brane. In this context, the IR bulk regulator surface is instead replaced with an ETW brane of tension $T$, excising the region between $0 < \rho < \epsilon$ (we will see shortly how the cutoff $\epsilon$  relates to $T$). Consequently, the limit $\epsilon\to0$ is not taken, the on-shell action (\ref{eq:covariant2don}) is not divergent, and a counterterm as in (\ref{eq:def_ren}) is not introduced.

Taking $\epsilon\neq0$ (but still small), the effective theory in the brane description is
\beq I^{(0)}_{\total}= I^{\epsilon}_{\text{ren}} +I_{\text{T}}\;,\eeq
with $I_{\text{T}}$ given in \eqref{eq:Ibrane2}. Since $I^{(0)}_\total$ implicitly depends on the tension $T_0$ and the value of $\epsilon$ is determined by $T_0$ (see next subsection), we drop the $\epsilon$ superscript on $I^{(0)}_\total$.
Now add and subtract the local counterterm action (\ref{eq:ctact2d}). Then,
\begin{align}
    I^{(0)}_{\total} = I^\epsilon_\text{grav} + I^\epsilon_\text{matter},
\end{align}
with 
\begin{align}
    I^\epsilon_\text{grav} = - I^\epsilon_\text{ct} + I_T, \qquad I^\epsilon_\text{matter} = I^{\epsilon}_{\text{reg}} + I_\text{ct}^\epsilon,
\end{align}
where $I^\epsilon_\text{matter}$ is the term that appears in the limit in \eqref{eq:def_ren}, i.e., $W_\text{CFT}[g] = \lim_{\epsilon \to 0} I^\epsilon_\text{matter}$. Thus, up to $\mathcal O(\epsilon)$-corrections,
the effective brane theory is 
\begin{align}
\label{eq:expansionI0}
I^{(0)}_{\total} = I^\epsilon_\text{grav} + W_{\text{CFT}}[g] + \mathcal{O}(\epsilon)\;.
\end{align}
The effective action of a CFT living on the undeformed brane, given by the Polyakov effective action \cite{Polyakov:1981rd}, 
\beq W_{\text{CFT}}
 =-\frac{c}{24\pi}\int d^{2}\xi\sqrt{-g}[\chi \mathcal{R}+(\nabla\chi)^{2}]
 +
 \frac{c}{12\pi}\int d\xi\sqrt{-\gamma^{\mathcal{N}}}\chi \mathcal K\;,\label{eq:polyact2}\eeq
 cast in its localized form by introducing an auxiliary scalar field $\chi$, where we have also included its GHY boundary term (and may be further supplemented with a local counterterm). Substituting the formal solution of $\chi$ equation of motion, $2\Box\chi=\mathcal{R}$, into the two-dimensional contribution returns the familiar non-local Polyakov term, $I_{\text{Polyakov}}=-\frac{c}{96\pi}\int d^{2}x\sqrt{-g}\mathcal{R}\Box^{-1}\mathcal{R}$, while the boundary term becomes more complicated.

\subsubsection*{The topological term}
The gravitational part of the action \eqref{eq:expansionI0} is, explicitly, 
\beq I^{\epsilon}_{\text{grav}}=\frac{1}{16\pi G_{N}}\int d^{2}\xi\sqrt{-g}\left(2(1-T) - \log \sqrt{\epsilon} \; \mathcal{R}[g]\right)
+
\frac{1}{8\pi G_{N}}\int d\xi\sqrt{-\gamma^{\mathcal{N}}}\log\sqrt{\epsilon}\,\mathcal K
\;.\eeq
In examples such as JT gravity coming from the near horizon limit of higher-dimensional near-extremal black holes, the constant coefficient of the Ricci scalar equals $4G_N$ times the ground state entropy of the black hole. In the present set-up there is no higher-dimensional black hole to appeal to, and it is therefore
 natural to ask about the interpretation of $\log\sqrt{\epsilon}$.

While we have no answer for a general ETW brane, we can make an interesting observation if the brane geometry is AdS$_2$. Using the Fefferman-Graham expansion in the bulk which in the case of AdS$_3$ takes the form (cf. Eq. (11) of \cite{Skenderis:1999nb})
\begin{align}
   ds^2 =  \frac{d\rho^2}{4 \rho^2} + \left(\frac{1}{\sqrt \rho} + \sqrt{\rho}\right) ^ 2 \mathfrak g^{(0)}_{ij} d\xi^i d\xi^j
\end{align}
we can obtain a relation between the extrinsic curvature scalar of the brane at $\epsilon$ and its location
\begin{align}
    K^{(0)} = - 2 T_0 = -2 \left( \frac{1 - \epsilon}{1 + \epsilon}\right).
\end{align}
where the first equality follows from the brane equations of motion. We can solve this for $\epsilon$ and find
\begin{align}\label{eq:phi0eps1}
    - \log \sqrt{\epsilon} = \operatorname{arctanh} T_0.
\end{align}
Moreover, in the case of an AdS$_3$ brane the warp factor in \eqref{eq:metric_ansatz_simple} is given by $f(\theta) = \frac{\cosh \theta}{\cosh \phi_0}$, such that $K^{(0)}=-2f'(\phi_{0})=-2\tanh(\phi_{0})$ (see above \eqref{eq:Kabove}). It immediately follows then
\begin{align}\label{eq:phi0eps2}
\phi_0 = - \log \sqrt\epsilon.
\end{align}
Therefore, it is reasonable to expect the system has a ground state entropy of
\begin{align}
    \label{eq:bdry_entropy}
    S_0 = \frac {\phi_0} {4 G_N}  = \frac c 6 \operatorname{arctanh} T_0.
\end{align}
For $\epsilon\ll1$, we see the ground state entropy is large, $S_{0}\gg1$.

To understand what this ground state entropy counts, recall that an AdS$_3$ bulk with an AdS$_2$ ETW brane provides a bottom-up model for a BCFT$_2$. The boundary dual possesses a \emph{boundary entropy} \cite{Calabrese:2004eu}, defined as the logarithm of the $g$-factor $g_{\alpha}\equiv\langle 0|\alpha\rangle$, where $|\alpha\rangle$ is the BCFT state realizing boundary condition $\alpha$ and $|0\rangle$ is the BCFT vacuum state, and measures the ground state degeneracy of the BCFT \cite{Affleck:1991tk}. Computing the holographic entanglement entropy of a system, the authors of \cite{Takayanagi:2011zk,Fujita:2011fp} (see also \cite{Suzuki:2022xwv}) have shown that in this case the boundary entropy is precisely given by the right hand side of \eqref{eq:bdry_entropy}, i.e., the ground state entropy $S_{0}$ in our case is exactly the boundary entropy.\footnote{While finishing our work, the article \cite{Myers:2024zhb} appeared which also showed (\ref{eq:bdry_entropy}) (up to a normalization factor of two because they are considering a two-sided defect; see their Eq. (A.69)). See also Eq. (2.18) of \cite{Grimaldi:2022suv}, where $S_{0}$ is interpreted as the central charge characterizing the conformal degrees of freedom on the defect.}

Altogether, the zeroth order bulk action $I^{(0)}_{\total}$ is equivalent to the semi-classical brane theory,
\begin{align} 
I^{(0)}_{\total}& = \frac{1}{16\pi G_{N}}\int d^{2}\xi\sqrt{-g}(2(1-T_0)+\phi_{0}\mathcal{R}[g])+W_{\text{CFT}}[g] 
-
\frac{1}{8\pi G_{N}}\int d\xi \sqrt{-\gamma^{\mathcal{N}}}\phi_{0}\mathcal{K}
+ \mathcal O(\epsilon) \;, \nonumber \\
& = \frac{1}{16\pi G_{N}}\int d^{2}\xi\sqrt{-g}(\frac {\mu_1} 2+\phi_{0}\mathcal{R}[g])+W_{\text{CFT}}[g] - \frac{\Delta T}{8\pi G_{N}}\int d^{2}\xi\sqrt{-g} \label{eq:zeroorderactcom}
\\&\quad-
\frac{1}{8\pi G_{N}}\int d\xi \sqrt{-\gamma^{\mathcal{N}}}\phi_{0}\mathcal{K}
+ \mathcal O(\epsilon) \;. \nonumber
\end{align}
In the above we included terms of higher order in $(1-T_0)$ into $\mathcal O(\epsilon)$, such that up to the given order,
\begin{align}
    \label{eq:quickref}
    2 ( 1-T) \simeq (1-T_0^2) + 2 (T_0 - T) \simeq \frac {\mu_1} 2 - 2 \Delta T, 
\end{align}
with $\Delta T\equiv T-T_0$. In summary, at leading order action \eqref{eq:zeroorderactcom} is comprised of a cosmological constant and topological term, together with an effective action for a CFT.
Since $W_{\text{CFT}}$ takes the form of the Polyakov action, in the absence of brane fluctuations, the induced theory of gravity on a two-dimensional holographic braneworld is Polyakov gravity together with a cosmological constant and a topological term.

The higher order terms in $(1 - T_0)$ are suppressed as $\phi_{0} \gg 1$. Note that in this limit
\beq \phi_{0}/G_{N}\gg c\gg1\;,\label{eq:scalesbrane}\eeq
where we can use the Brown-Henneaux relation $c= 3/(2G_{N}) \gg 1$ \cite{Brown:1986nw}. This is the semi-classical limit of the brane theory. Away from this critical tension limit, the order of scales \eqref{eq:scalesbrane} need not be the case. Indeed $\phi_{0}$ and $c$ are independent parameters, and in the event they are comparable the semi-classical description breaks down.

\subsection{Full action to second order}
\label{sec:fullresult}

Finally, let us collect the full action for an AdS$_2$ ETW brane in the brane description. By adding the action for deformations \eqref{eq:defactfinal} and topological/matter coupling \eqref{eq:zeroorderactcom}, we find for an $\text{AdS}_{3}$ bulk ($\kappa=-1$),
\begin{align}
    \label{eq:res_final}
\begin{split}
 I_{\total} ={}&\frac{1}{16\pi G_{N}}\int d^{2}\xi\sqrt{-g}(2(1-T_0)+\phi_{0}\mathcal{R})+W_{\text{CFT}}[g]+ \mathcal O(\epsilon)\\
 &+\frac{1}{16\pi G_{N}}\int d^{2}\xi\sqrt{-g}\left(\phi(\mathcal{R}+\mu_{1})+T_{0}(\partial_{i}\phi)^{2}+T_{0}\mu_{1}\phi^{2}\right)\\
&-\frac{\Delta T}{16\pi G_{N}}\int d^{2}\xi\sqrt{-g}\left(2 + 4T_{0}\phi +(\partial_{i}\phi)^{2}+\mu_{2}\phi^{2}\right)+ \mathcal O(\phi^3)\;\\
&-\frac{1}{8\pi G_{N}}\int d\xi \sqrt{-g^\mathcal{I}}  \left( \phi_0 \mathcal K + \phi (\mathcal K + \sqrt{1 - ( T_0 + \Delta T)^2})\right)+\mathcal{O}(\phi^{2}) \;\\
\end{split}
\end{align}
with 
\begin{align}
\mu_{1}=2(1-T_{0}^{2}), \qquad \mu_{2}=2(1+T_{0}^{2}),
\end{align}
and $\phi_{0}$ is the boundary entropy of the dual BCFT.

Below, we consider the two cases for which $T_{0}=1$ but $\Delta T\neq0$, and $T_{0}\neq1$ but $\Delta T=0$. For the latter, we can consistently replace $T_{0}$ by $1 + \mathcal O(\epsilon)$, and set $\mu_{1}=4(1-T_{0})$ and $\mu_{2}=4$ up to the order indicated in \eqref{eq:res_final}. Either way, as indicated, the corrections to the two-dimensional action are cubic in $\phi$, while corrections to the boundary term are quadratic in deformations. The $\phi$ independent action receives subleading corrections in $(1-T_0)$, which contribute at $\mathcal O(\epsilon)$. 

The power counting in $(1-T_0)$ deserves a further comment. In view of the cosmological constant term the action \eqref{eq:res_final} seems to be of order $\mathcal O(\epsilon)$. This, however, is only because we chose to set the bulk AdS length $L=1$. From the brane point of view, the natural length scale is set by $(1-T_0^2)^{-\frac 1 2}$, c.f. \eqref{eq:ads_length_brane}.\footnote{For a dS brane the corresponding length scale is set by $(T_0^2-1)^{-\frac 1 2}$.} 
Thus, to correctly count powers of $1-T_0$ we need to remember $\mathcal O(\sqrt{-g}) \sim \mathcal O(\mathcal R^{-1}) \sim \mathcal O((1-T_0)^{-1})$. Alternatively, as long as $T_0 \neq 1$ we can Weyl-rescale the metric such that lengths are measured in units of $L_\text{brane}$. Such a rescaling would remove all explicit occurrences of $1-T_0$ in \eqref{eq:res_final} except possibly for terms of $\mathcal O(\epsilon)$ and higher. In the case of flat space, i.e. $T_0 = 1$, we have no explicit factor of $1 - T_0$, but higher derivative corrections still appear. The expression $\mathcal O(\epsilon)$ should of course be understood to capture those.

\section{Liouville gravity on the brane} \label{sec:whyLiouv}

We now come to the main observation of this paper: We argue that, to leading order in $\epsilon$ and at $\Delta T = 0$, the theory controlling brane fluctuations, \eqref{eq:res_final}, is given by Liouville gravity with a topological term and coupled to a large-$c$ CFT. Here, by `Liouville gravity' we mean a theory governed by the Liouville action,
\begin{align} 
\label{eq:liouville_action}
\begin{split}
I_L &=\frac{1}{4\pi}\int d^{2}\xi\sqrt{-g}\left[\partial_{i}\varphi \partial^{i} \varphi + Q \mathcal R \varphi + 4 \pi \mu e^{2 b \varphi}\right]\;,
\end{split}
\end{align}
where both the Liouville field $\varphi$ and metric $g_{ij}$ are dynamical. We use this terminology, which emphasizes the dynamical metric, to contrast to `Liouville theory', i.e., the effective action of a two-dimensional CFT together with a cosmological constant, in which only the Liouville field is dynamical while the metric is fixed to some fiducial value (see, e.g., \cite{Erbin:2015}). 

Indeed, considering the action \eqref{eq:res_final} of the previous section with $\Delta T = 0$ and dropping terms of $\mathcal O(\epsilon)$, we find
\begin{align}
\label{eq:Liouvgravbrane}
I_{\total}&=\frac{1}{16\pi G_{N}}\int d^{2}\xi\sqrt{-g}\left[(\partial_{i}\phi)^{2}+(\phi_0 + \phi) \mathcal{R}+4\pi\tilde{\mu} (1 + 2 \phi + 2 \phi^2)\right]+W_{\text{CFT}}[g] + \mathcal O(\phi^3) \nonumber \\
&= \frac{1}{16\pi G_{N}}\int d^{2}\xi\sqrt{-g}\left[(\partial_{i}\phi)^{2}+(\phi_0 + \phi)\mathcal{R}+4\pi\tilde{\mu} e^{2\phi}\right]+W_{\text{CFT}}[g] + \mathcal O(\phi^3)\;,
\end{align}
where $4\pi\tilde{\mu}\equiv \frac 1 2 \mu_1$, and we have neglected writing the boundary terms. To bring this into a more canonical from of the (classical) Liouville action, we use the holographic identification $G_{N}^{-1}=\frac{2c}{3}$ with $\frac{c}{6}=b^{-2}$ and define $\phi = b \varphi$ and $\tilde{\mu}= b^2 \mu$ to arrive at\footnote{We could have chose to absorb $\phi_{0}$ in $\phi$ and $\mu$, resulting into $\mu' = \mu e^{-2b\varphi_0}$.
}
\begin{align} 
\begin{split}
I_{\total}&=\frac{1}{4\pi}\int d^{2}\xi\sqrt{-g}\left[\partial_{i}\varphi \partial^{i} \varphi + Q \mathcal R (\varphi_0 + \varphi) + 4 \pi \mu e^{2 b \varphi}\right]+W_{\text{CFT}}[g] + \mathcal O(\varphi^3),
\end{split}
\end{align}
where $Q = \frac 1 b$ as usual for classical Liouville theory.\footnote{Taking into account the path integral measure, which we will not discuss here in detail, shifts $Q \to Q + b$.} The $\varphi$ dependent terms reproduce the form of \eqref{eq:liouville_action}. 

This argument establishes part of our claim, namely that we find the correct classical action at the first few orders in the dilaton $\phi$. In the following we will gather more evidence by arguing why we expect the agreement to hold to higher orders in $\phi$. Furthermore, so far it is not clear that both the dilaton and the metric are independent and dynamical fields. We will address this issue by showing from the bulk that the path integral over the two-dimensional brane metric as well as the dilaton follow from the path integral over all bulk metrics. In this language, we can understand the dilaton as a Goldstone associated with partial breaking of diffeomorphism symmetry in the presence of the brane.

Similarly, let us examine the situation for $T_{0}=1$ and $\Delta T\neq0$. The action \eqref{eq:res_final} takes the form
\begin{align}
\label{eq:Liouvgravbraneflat}
I_{\total}=&~\frac{1}{16\pi G_{N}}\int d^{2}\xi\sqrt{-g}\left[(\partial_{i}\phi)^{2}+(\phi_0 + \phi) \mathcal{R}
-
2\Delta T (1 + 2 \phi)\right]
+W_{\text{CFT}}[g] + \mathcal O(\phi^3) 
\;.
\end{align}
Up to quadratic order, this action can be trusted, as is reflected by $\mathcal{O}(\phi^3)$. 
Due to the fact that $\Delta T$ is of order $\phi$, we lose a full order of $\phi$ of information in the action as compared to the $T_{0}\neq 1$ and $\Delta T=0$ case.
Meanwhile, we point out that contrasting the $\Delta T=0$ case, it is suggestive that the prefactor of the kinetic term gets modified to $(1-\Delta T)$ if one would allow for higher order corrections, which could point to a departure from a potential match to a Liouville type of action. 
Unfortunately, we lack information to make a concrete statement about higher order corrections in this case.

\subsection{Why Liouville?}
\label{sec:why}

From a holographic point of view the ETW brane plays two roles. First, in the bulk and brane description it defines a UV regulator which makes the matter effective action well-defined even without renormalization. Second, ETW branes also define the background geometry of the brane description in the semi-classical limit.

The large-but-finite contributions to the regulated effective matter action can be written as a derivative expansion in intrinsic geometric quantities of the background geometry. In this way, the gravitational action is entirely induced by matter loops. We have seen an explicit example in \eqref{eq:zeroorderactcom} above, where the leading contribution of the bulk action could be written as a cosmological constant together with a Ricci term. In higher dimensions, higher derivative corrections appear \cite{Chen:2020uac}.

The fact gravity is induced from only the matter effective action immediately suggests a mechanism by which a dilaton governed by the Liouville action arises. Consider a brane with induced metric $h_{ij}$ which under a small deformation $\delta$ is taken to a brane with induced metric $h'_{ij}$. In two dimensions we can always apply a deformation dependent diffeomorphism such that the deformed metric takes the form of the original metric up to a Weyl transformation,
\begin{align}
    h'_{ij} = e^{2 \phi} h_{ij},
\end{align}
where now $\phi = \phi[\delta]$ is a function of the deformation. Integrating out the bulk up to the location of the deformed brane gives a topological term, a cosmological constant term, and a matter effective action. In other words, the result will be \eqref{eq:zeroorderactcom} with metric $h'_{ij}$. The matter effective action approximately looks like that of a CFT at low energies; it is not quite that of a conformal theory, since the location of the brane acts as a UV cutoff and thus breaks the conformal symmetry. This will lead to cutoff-dependent corrections in the effective action. Such terms, however, are suppressed as the brane approaches the asymptotic boundary, i.e., in the limit where the UV cutoff is removed. Since our claim is only concerned with the leading order in the cutoff, we can ignore these cutoff dependent corrections and treat the matter on the brane as a large-c CFT.

The quantum effective action of a two-dimensional CFT together with a cosmological constant is diffeomorphism invariant, but transforms under Weyl rescaling as \cite{Polyakov:1981rd}
\begin{align}
\label{eq:liouville_deriv}
\begin{split}
 W_\text{CFT}[h'_{ij}] + \frac{\tilde \mu}{4 G_N} \int d^2x \sqrt{-h'} &= W_\text{CFT}[e^{2 \phi} h_{ij}] + \frac{\tilde \mu}{4 G_N} \int d^2x \sqrt{-h} e^{2 \phi} \\
& = W_\text{CFT}[h_{ij}] + \frac c 6 I_\text{Polyakov}[\phi] + \frac{\tilde \mu}{4 G_N} \int d^2x \sqrt{-h} e^{2 \phi},
\end{split}
\end{align}
where $I_\text{Polyakov}$ is the Polyakov term with boundary contribution \cite{Herzog:2015ioa},  
\begin{align}
    I_\text{Polyakov} = \frac 1 {4\pi} \int d^2x \sqrt{-h} \left(\phi \mathcal{R} + \partial_i \phi \partial^i \phi  \right) - \frac 1 {2 \pi} \int ds \sqrt{-\gamma^\mathcal{I}}\phi K
\end{align}
The last two terms in \eqref{eq:liouville_deriv} combine to the dilaton-dependent expression in \eqref{eq:Liouvgravbrane}, which, as previously shown, is related to the Liouville action \eqref{eq:liouville_action}  by a field redefinition.
Thus, if we decide to express the action which depends on $h'_{ij}$ as an action depending on the `reference metric' $h_{ij}$, the action takes the form proposed in \eqref{eq:Liouvgravbrane}, even to higher orders in $\phi$.

Of course, the cutoff-dependent terms in the matter effective action will yield corrections to the Liouville action. We therefore expect that the dilaton gravity theory induced on the brane from deformations takes the form of Liouville theory plus corrections in $\epsilon$. The Liouville field $\phi$ can be a complicated function of the deformation $\delta$. However, as Section \ref{sec:dilagravaction} has shown explicitly, $\phi \propto \delta  + \mathcal O(\delta^3)$.

\subsection{Higher order check}\label{Section:3.2}

In Section \ref{sec:dilagravaction} we derived the action for brane deformations to quadratic order in fluctuations. As observed above, in a particular limit, the brane theory is Liouville gravity, at second order in deformations. Here we investigate corrections to the equations of motion that are higher order in $\phi$ to confirm Liouville gravity continues to describe brane deformations at leading order in $1 - T_0$. We accomplish this explicitly to third order in deformations in the equations of motion, i.e., fourth order in the action, using the bulk Einstein equations and Israel junction conditions, though in principle our analysis can be extended to arbitrary order.

To this end, it is useful to express the Israel junction conditions as
\begin{equation}\label{eq:israeljuncLiou}
	K_{ij}
	+
	(T_{0}+\Delta T)\gamma_{ij}
	=
	0
	\,,
\end{equation}
where $K_{ij}$ is the extrinsic curvature and $\gamma_{ij}$ is the induced metric, both evaluated at the deformed brane $\theta = \phi_0 + \phi(\xi)$. Furthermore, $\Delta T \sim \mathcal{O}(\phi)$ when $\Delta T\neq0$. The bulk Einstein equations read
\begin{equation}
	R_{\mu\nu}
	-
	\frac{1}{2}R G_{\mu\nu}
	+
	\Lambda G_{\mu\nu}
	=
	0
	\,,
\end{equation}
where we choose a negative cosmological constant.
 Substituting in the bulk metric ansatz \eqref{eq:metric_ansatz_simple}, at zeroth order in $\phi$ both the junction conditions and bulk Einstein equations are solved when 
\begin{equation}
	f'(\phi_{0})
	=
	T_{0}f(\phi_{0})
	\,,
	\quad
	f''(\phi_{0})
	=
	f(\phi_{0})
	\,,
	\quad
	\mathcal{R}
	=
	-2f(\phi_{0})^{2}
	+
	2f'(\phi_{0})^{2}
	\,.
\end{equation}
From the first two equations we conclude, using our choice of normalization $f(\phi_{0})=1$, 
\begin{equation}
    \left.
	f^{(n=\text{odd})}(\theta)\right|_{\theta=\phi_{0}}
	=
	T_{0}
	\,,
	\quad
    \left.
	f^{(n=\text{even})}(\theta)\right|_{\theta=\phi_{0}}
	=
	1
	\,,
\end{equation}
where $n>0$ denotes the number of derivatives. Consequently,
\begin{equation}\label{eq:RiccscalLiou}
	\mathcal{R}
	=
	-2(1-T_{0}^{2})
	\,.
\end{equation}
With this information, we now examine the Israel junction conditions \eqref{eq:israeljuncLiou} at higher order in $\phi$ and show that they agree with the equations following from Liouville gravity, up to corrections in $\epsilon$, i.e., corrections in $(1 - T_0)$ and higher derivatives.

Let us from this point on restrict to $\Delta T=0$ and $T_{0}\neq 1$.
For reference, first note the equations of motion which follow from the Liouville action as presented in \eqref{eq:Liouvgravbrane}. Varying with respect to the dilaton yields the Liouville equation,
\begin{equation}\label{eq:liouville3}
	\Box\phi
	=
	(1-T_{0}^{2})
		e^{2\phi}
		+
		\frac{\mathcal{R}}{2}
	\,.
\end{equation}
The metric equations are
 \beq\label{eq:einsteinlouie} g_{ij}\Box\phi-\nabla_{i}\nabla_{j}\phi-\frac{1}{2}g_{ij}[(\partial_{k}\phi)^{2}+(1-T_{0}^{2})) e^{2\phi}]+(\nabla_{i}\phi)(\nabla_{j}\phi)=8\pi G_{N}\langle T^{\text{CFT}}_{ij}\rangle\;,\eeq
 where
 \beq \langle T^{\text{CFT}}_{ij}\rangle=\frac{c}{12\pi}\left[(g_{ij}\Box-\nabla_{i}\nabla_{j})\chi+(\partial_{i}\chi)(\partial_{j}\chi)-\frac{1}{2}g_{ij}(\partial_{k}\chi)^{2}\right]\;,
 \eeq
 with $\chi$ being the auxiliary field localizing the Polyakov effective action \eqref{eq:polyact2}.  Taking the trace of (\ref{eq:einsteinlouie}) returns the Liouville equation \eqref{eq:liouville3} upon invoking the CFT anomaly \eqref{eq:CFTtrace}. This is a direct result of gravity being holographically induced on the brane with $c G_{N}=\frac{3}{2}$; otherwise the right hand side would have a factor of $c G_{N} \neq \frac 3 2$ from which the Liouville equation (\ref{eq:liouville3}) does not follow.

Since the bulk is in equilibrium we expect solutions to the brane theory to be in equilibrium as well. Working in conformal gauge $ds^{2}=-e^{2\rho}dx^{+}dx^{-}$, this implies $\langle T_{\pm\pm}^{\text{CFT}}\rangle=0$. 
Requiring regularity, we further impose that the normal ordered matter stress-energy tensor vanishes in $x^{\pm}$-coordinates,
 from which \cite{Christensen:1977jc}
\begin{equation}\label{eq:Tppmm}
   0= \langle T_{\pm\pm}^{\text{CFT}}\rangle
    =
    -\frac{c}{12}\left(
        \partial_{\pm}\rho\partial_{\pm}\rho
        -
        \partial_{\pm}^{2}\rho
    \right)
    \,.
\end{equation}
This can be derived using the explicit form of the quantum effective action \eqref{eq:polyact2} and invoking the auxiliary field solution $\chi=-\rho$ (see Eq. (2.26) of \cite{Pedraza:2021cvx}).  
Solving for $\rho$ yields 
\begin{align}
    \rho = - \log \left[ \frac{a + b x^+ + c x^- + x^+ x^-}{d(a - bc)} \right]
    \,,
\end{align}
with constants $a$, $b$, $c$, and $d$. This solution implies a constant two-dimensional Ricci scalar $\mathcal{R}=8/((bc-a)d^{2})$.
Meanwhile,  the $ij=\pm\pm$ components of the metric equations \eqref{eq:einsteinlouie} are
\begin{equation}\label{eq:braneeq2}
    \partial_{\pm}\phi\partial_{\pm}\phi
    -
    2\partial^{2}_{\pm}\phi
    +
    \partial_{\pm}\rho\partial_{\pm}\phi
    =
    8\pi G_{N}
    \langle T_{\pm\pm}^{\text{CFT}}\rangle
    =
    0
    \,.
\end{equation}
The content of the remaining  $ij=\pm\mp$ components are encoded in the trace, and therefore they need not be considered separately.

We can now verify that the bulk equations for a deformed brane, i.e., the Israel junction conditions \eqref{eq:israeljuncLiou}, are consistent with the equations from Liouville gravity presented above. Let us start by considering the trace of the junction condition
\begin{equation}
	\gamma^{ij}
	\left(
	K_{ij}
	+
	T_{0}\gamma_{ij}
	\right)
	=
	0\;.
\end{equation}
Notice that the junction condition is evaluated at the location of the deformed brane (see Appendix \ref{app:conventions} for conventions). Expanding to third order in $\phi$, we find 
\begin{equation}\begin{aligned}\label{eq:israel2}
    0
    =
    &
    -2(1-T_{0}^{2})\phi
    +
    \Box\phi
    +
    2T_{0}\phi
    \left(
        (1-T_{0}^{2})\phi
        -
        \Box\phi
    \right)
    +
    (1-3T_{0}^{2})\phi^{2}
    \left(
    \frac{2}{3}(1-T_{0}^{2})\phi
        -
    \Box\phi
    \right)
    \\
    &
    -
    \nabla^{j}\phi(\nabla_{j}\nabla_{i}\phi)\nabla^{i}\phi
    -
    \frac{1}{2}
    (\nabla_{i}\phi)(\nabla^{i}\phi)\Box\phi
    +
    \mathcal{O}(\phi^{4})
    \,.
\end{aligned}\end{equation}
Recall that we only expect Liouville gravity to describe the equations up to leading order in the cutoff. In particular, we do not expect to be able to reproduce the partial derivative terms which only appear at higher order as discussed in Section \ref{sec:fullresult}. In fact, this is already suggested by \cref{eq:israel2}, since at leading order in $\phi$ we see that $\Box \phi \sim (1 - T_0)\phi$. We can thus collect terms and replace $T_{0}$ with $1$ up to subleading terms which together with the higher derivative equations can be summarized as $\mathcal{O}((1-T_{0})^2)$ and are thus also part of the higher order corrections. The remaining parts of the equation then are

\beq 0=e^{-2\phi}\Box\phi +2(1-T_{0})\left(	e^{-2\phi}-1\right)+\mathcal{O}(\phi^{4})+\mathcal{O}(\epsilon)\;,\eeq
which we recognize to be the Liouville equation \eqref{eq:liouville3} together with the Ricci scalar (\ref{eq:RiccscalLiou}).

The other two equations (\ref{eq:braneeq2}) are directly encoded in the $ij=\pm\pm$ components of the junction conditions. To wit,
\begin{equation}\begin{aligned}
   0= K_{\pm\pm}+T_{0}g_{\pm\pm}
    =&~
    -
    T_{0}\partial_{\pm}\phi\partial_{\pm}\phi
    +
    2\partial^{2}_{\pm}\phi
    -
    \partial_{\pm}\rho\partial_{\pm}\phi
    \\&
    -
    \frac{1}{2}\partial_{i}\phi\partial^{i}\phi
    \left(
         \partial_{\pm}\rho\partial_{\pm}\phi
         -
         2\partial^{2}_{\pm}\phi
    \right)
    -2(1-T_{0}^{2})(\partial_{\pm}\phi)^{2}\phi+...
    \\
    =&~
    -
    \partial_{\pm}\phi\partial_{\pm}\phi
    +
    2\partial^{2}_{\pm}\phi
    -
    \partial_{\pm}\rho\partial_{\pm}\phi
    +
    \mathcal{O}(\phi^{4})
    +
    \mathcal{O}(\epsilon)
    \,,
\end{aligned}\end{equation}
where $+...$ denotes quartic order corrections in $\phi$. Here we have again dropped higher derivative terms.  
We have thus demonstrated that at leading order in $\epsilon$ and to cubic order in $\phi$ the equations coming from Liouville gravity agree with equations governing the brane deformations. This implies that the actions agree to quartic order in $\phi$.

\subsection{The dynamical dilaton}
\label{sec:dynamical_dilaton}
So far the discussion has purely been at the level of the classical action. To claim the theory on the brane is Liouville gravity, we need to argue that both the metric and dilaton need to be integrated over in the path integral. In fact, above we saw the dilaton is merely the Weyl factor of the induced metric on the brane. This might suggest the dilaton could be absorbed into the metric degrees of freedom and is not an independent field. However, as we will now show, the dilaton is indeed a dynamical degree of freedom and its presence is intimately linked to spontaneous breaking of local diffeomorphism invariance at the ETW brane.

From the bulk point of view, the bulk path integral instructs us to consider a three-dimensional topological manifold with boundary and path-integrate over all metrics, weighted by the action. As is well known, the diffeomorphism symmetry of bulk Einstein gravity introduces flat directions in the path integral which need to be removed by fixing a gauge. In the absence of the brane, the gauge symmetry 
\begin{align}
    \theta \to \theta' = \theta'(\theta, \xi^i), \qquad {\xi}^i \to {\xi'}^i = {\xi'}^i(\theta, \xi^i),
\end{align}
is sufficient to bring the metric into a Fefferman-Graham-like gauge, \eqref{eq:metric_ansatz}. This removes three out of the six independent metric degrees of freedom, leaving us with three metric degrees of freedom, the correct number for a two-dimensional metric. Moreover, the residual gauge symmetry
\begin{align}
    \label{eq:res_diff}
    {\xi}^i \to {\xi'}^i = {\xi'}^i(\xi^i),
\end{align}
leaves us in Fefferman-Graham gauge and agrees with the symmetry of infinitesimal diffeomorphisms on a two-dimensional spacetime.\footnote{Note that there might be obstructions in the higher dimensional spacetime for \eqref{eq:res_diff} to exist, so these diffeomorphisms might also be broken. This is responsible for the fact that the dilaton is not a Goldstone but a pseudo-Goldstone.}

The presence of a boundary $\theta = \phi_0$, however, breaks the diffeomorphism symmetry to those diffeomorphisms that leave the location of the boundary, i.e., the ETW brane, unchanged. 
The broken diffeomorphisms, which move the location of the brane, are of the form $\theta \to \theta' = \theta'(\theta, \xi^i)$ with $\theta'(\phi_0, t, x) \neq \phi_0$.
Explicit expressions can be derived by expanding $\theta'$ around $\phi_0$, 
\begin{align}
    \theta'_\text{broken} &= \theta + \vartheta^{(0)}(\xi^i), \\
    \theta'_\text{unbroken} &= \phi_0 + \sum_{n=1}^\infty \;\vartheta^{(n)}(\xi^i)\; (\theta - \phi_0)^n,
\end{align}
where $\theta^{(n)}(\xi^i)$ for $n \in \mathbb N_0$ are $\xi^i$-dependent functions.
By acting with diffeomorphisms $\xi'^i$ and $\theta'_\text{unbroken}$ we can only bring the bulk metric $G_{\mu\nu}$ into the form\footnote{A closely related gauge has been discussed in \cite{Ciambelli:2023ott}, where \eqref{eq:WeylFG} was treated as a partially-gauge fixed metric. There the gauge was called \emph{Weyl-Fefferman-Graham gauge}.}
\begin{align}
    \label{eq:WeylFG}
    ds_{\text{AdS}_3}^2 = (d\theta + \partial_i \vartheta^{(0)}(\xi^i)\; d\xi^i)^2 + \tilde g_{ij} d\xi^i d\xi^j,
\end{align}
where the defect is still located at $\phi_0$.
This can be seen, for example, by starting in Fefferman-Graham gauge and showing that the broken diffeomorphisms change the metric to \eqref{eq:WeylFG}. 

Let us now consider what happens if we attempt to bring the metric into the form \eqref{eq:metric_ansatz} by using the broken diffeomorphisms. Although the diffeomorphisms are broken, it is still useful to implement the Faddeev-Popov method. Decompose unity into
\begin{align}
    \label{eq:FPone}
    1 = \int \mathcal D \phi \Delta_{FP}(F[G^\phi]) \delta(G^\phi),
\end{align}
where $\mathcal D \phi$ is an integral over all functions in two variables, $\phi = \vartheta^{(0)}(\xi)$, and we assume we have gauge fixed the metric using the unbroken diffeomorphisms. The gauge fixing condition $F[ G]$ could be chosen as $F[G^\phi] = G_{t\theta} + \partial_t \phi$, but the precise choice is immaterial in the following. Inserting \eqref{eq:FPone} into the bulk path integral, we schematically obtain
\begin{align}
    Z = \int \mathcal D G\, \mathcal D\phi \, \Delta_{FP}(F[ G^\phi]) \, \delta(G^\phi) \, e^{i I_{\total}^{(0)}[G]}.
\end{align}
We have used a superscript on $I_\total^{(0)}$ to make explicit that the domain of integration of $I_{\total}^{(0)}[G]$ extends to $\phi_0$.

Now change $G \to G^\phi$ by a diffeomorphism. The action $I_{\total}^{(0)}[G]$, however, is not dif\-feo\-mor\-phism-invariant but rather transforms as
\begin{align}
    I_\total^{(0)}[G^\phi] \to I_\total[G^\phi, \phi] = I_\total^{(0)}[G^\phi] + \Delta I^\text{bulk}_{\total}[G^\phi, \phi]+ \Delta I^\text{bdry}_{\total}[G^\phi, \phi].
\end{align}
After renaming $G^\phi \to G$ and performing the integration of the delta-function, the path integral thus turns into
\begin{align}
Z = \int \widetilde{\mathcal D G_{\mu\nu}} \mathcal D\phi \; \Delta_{FP}(F[G^\phi]) \; e^{iI_\total[G^\phi, \phi]},
\label{eq:bulkPIZ}\end{align}
where we integrate only over the unfixed components of the metric. Unlike in the case where we fixed a proper gauge symmetry, we cannot factor out the $\mathcal D\phi$ integral due to the presence of the extra $\phi$-dependent term. Instead $\phi$ becomes a dynamical field with a bulk action given by $\Delta I_\total$ given in \eqref{eq:defactfinal}.

\section{Emergence of JT gravity on the brane} \label{sec:emergeJTbrane}

Consider our complete induced action \eqref{eq:res_final} with $\Delta T = 0$ and perform the Weyl rescaling $\tilde g_{ij} = e^{2 \phi} g_{ij}$. Taking into account the contributions from the cosmological constant term as well as $W_\text{CFT}$, we find (ignoring boundary terms and terms subleading in $(1-T_0)$) that the dependence on the dilaton field completely drops out:
\begin{align}
\begin{split}
    I_{\total} &= \frac{1}{16\pi G_{N}}\int d^{2}\xi\sqrt{-\tilde g}\left((\phi_{0} + \phi) \mathcal{R}[\tilde g] + 2 (\phi_0 + \phi) \tilde \Box \phi + (\partial_{i}\phi)^{2} + \frac{\mu_1}{2} \right) \\
    &\qquad +W_{\text{CFT}}[\tilde g] - \frac c {24\pi} \int d^2 \xi \sqrt{-\tilde g} \left( \phi \mathcal{R}[\tilde g] + \phi \tilde \Box \phi \right) \\
    &= \frac{1}{16\pi G_{N}}\int d^{2}\xi\sqrt{-\tilde g}\left( \phi_{0} \mathcal{R}[\tilde g] + 2 \phi_0 \tilde \Box \phi  + \frac{\mu_1}{2} \right)+W_{\text{CFT}}[\tilde g].
\end{split}
\end{align}
This, of course, is only the reverse of the computation of Section \ref{sec:why}, where we argued, to the given order, the dilaton dependence can be explained by a Weyl rescaling of only the effective matter action. Thus, from our point of view, JT gravity on the brane cannot arise by choosing a particular Weyl frame. This observation stands in contrast with the conclusions of \cite{Geng:2022tfc,Geng:2022slq,Bhattacharjee:2022pcb,Aguilar-Gutierrez:2023tic}, where the $\mathcal{O}(\phi^{2})$ fluctuations are Weyl-rescaled away leaving only JT gravity on the brane. While our $\phi$-corrections agree in the induced action \eqref{eq:defactfinal}, in those the works the leading order contribution \eqref{eq:zeroorderactcom}, and in particular the fact that a matter theory lives on the brane, was not considered. 

Nonetheless, in this section we explain how JT gravity and other related theories arise on the brane. We do this by studying consistent limits of the equations of motion of the complete induced theory. Moreover, in this section we show that, as an induced gravity scenario, our braneworld set-up offers an explicit realization of the Susskind-Uglum proposal that generalized gravitational entropy equals von Neumann entropy \cite{Susskind:1994sm}.

\subsection{A lightning summary of 2d dilaton gravities}\label{sec:taxonomy}
We are primarily interested in reproducing theories of dilaton gravity of the type
\begin{equation}\label{eq:sphi}
	I_{\phi}
	=
	\frac{\phi_{0}}{16\pi G_{N}}
	\int  d^{2}x\sqrt{-g}\mathcal{R}
	+
	\frac{1}{16\pi G_{N}}
	\int  d^{2}x\sqrt{-g}\left[\phi( \mathcal{R}-2\Lambda)+4\lambda^{2}\right]
	\,,
\end{equation}
with $\Lambda$ being a cosmological constant and $\lambda$ some scale.
To have a well-posed variational problem, the action should be supplemented with a GHY-like boundary term, which we ignore for simplicity.
In such models the dilaton $\phi$ acts as a Lagrange multiplier, such that the dilaton equations of motion force the Ricci scalar $\mathcal{R}=2\Lambda$. The class of models described by (\ref{eq:sphi}) are widely studied because they provide exactly solvable models of quantum gravity \cite{Saad:2019lba,Johnson:2019eik,Kar:2022sdc,Rosso:2022tsv}, holography \cite{Sachdev:1992fk,Kitaev_SYK}, and, when coupled to matter, can be used to analytically study black hole formation and evaporation \cite{Susskind:1993if,Fiola:1994ir,Almheiri:2014cka,Penington:2019npb,Almheiri:2019psf,Gautason:2020tmk,Pedraza:2021cvx,Pedraza:2021ssc,Svesko:2022txo}. Cases worth highlighting include:

\paragraph{JT ($\Lambda<0$, $\lambda=0$):}
The Jackiw-Teitelboim (JT) model \cite{Jackiw:1984je,Teitelboim:1983ux} yields (nearly) AdS$_{2}$ solutions and describes the low-energy dynamics of a wide class of higher-dimensional nearly extremal black holes, e.g., \cite{Achucarro:1993fd,Nayak:2018qej,Sachdev:2019bjn,Castro:2018ffi,Moitra:2019bub,Castro:2021fhc}.

\paragraph{de Sitter-JT ($\Lambda>0$, $\lambda=0$):}
This model yields (nearly) asymptotically $\text{dS}_{2}$ solutions and arises, e.g., from spherical reduction of higher-dimensional de Sitter black holes in their near-Nariai limit \cite{Maldacena:2019cbz,Cotler:2019nbi,Sybesma:2020fxg,Kames-King:2021etp,Svesko:2022txo,Castro:2022cuo}.

\paragraph{flat-JT and CGHS ($\Lambda=0$, $\lambda\neq0$):}
Flat-JT gravity \cite{Dubovsky:2017cnj}, where the dilaton equation of motion, $\mathcal{R}=0$, yields two-dimensional Minkowski geometries and arises from, e.g., the spherical reduction of higher-dimensional dS black holes in their ultracold limit \cite{Castro:2022cuo}. Further, one requires $\lambda^{2}>0$ for flat-JT to become positive near spacelike infinity, signalling gravity turning off; $\lambda^{2}<0$ causes the dilaton to become negative near spacelike infinity, which implies regions of strong gravity where this is not reasonably expected. The case $\lambda=0$, meanwhile, yields a trivial theory without black holes due to the lack of a length scale.

Notably, flat-JT gravity can be related to the classical part of Callan-Giddings-Harvey-Strominger (\text{CGHS}) model \cite{Callan:1992rs}  through a Weyl rescaling $g_{\mu\nu}\to e^{-2\varphi}g_{\mu\nu}$ and field redefinition $\phi_{0}+\phi\to e^{-2\varphi}$:
\begin{equation}\label{eq:cghs1}
    I_{\text{flat-JT}} 
    \quad
	\overset{\text{Weyl}}{\Longrightarrow}
	\quad
	I^{\text{(classical)}}_{\text{CGHS}}
	=
	\frac{1}{16\pi G_{N}}
	\int d^{2}x\sqrt{-g}
	e^{-2\varphi}
	\left[\mathcal{R}+ 4(\partial\varphi)^{2}+4\lambda^{2}\right]\,.
\end{equation}
The dilaton gravity contribution arises as a low-energy limit of some string/brane constructions, e.g., \cite{Maldacena:1997cg}, and has Witten's stringy black hole as a solution \cite{Mandal:1991tz,Witten:1991yr}.

Thus far we have only described \emph{classical} dilaton theories. A benefit of working in two-dimensions is that one has 
 analytical control over the problem of semi-classical backreaction, where the gravitational field equations are sourced by a CFT with central charge $c\gg1$.\footnote{The total central charge receives a contribution from the conformal factor, the ghosts of lightcone gauge, and the dilaton. 
 These contributions do not scale with $c$, the central charge of the CFT, and for $c\gg1$ we can neglect those contributions.} Backreaction effects are then captured by the Polyakov effective action (\ref{eq:polyact2}). Adding the quantum effective action (\ref{eq:polyact2}) to any of the classical JT actions yields a soluble model for which one can find analytic semi-classical black hole solutions. Alternatively, the classical CGHS action \eqref{eq:cghs1} plus the quantum effective action $W_{\text{CFT}}$, dubbed the CGHS model, $I_{\text{CGHS}}=I^{\text{(classical)}}_{\text{CGHS}}+W_{\text{CFT}}$, does not have any known analytical black hole solutions. This is because the classical CGHS theory enjoys a symmetry which is broken when $W_{\text{CFT}}$ is introduced. This shortcoming can be ameliorated by adding a purely semi-classical term, resulting in the Russo-Susskind-Thorlacius (\text{RST}) model \cite{Russo:1992ax}
\begin{align}\label{eq:RST}
	I_{\text{RST}}
	=
    I_{\text{CGHS}}
	-
	\frac{c}{96\pi}
	\int \hspace{-1mm} d^{2}x\sqrt{-g}
		2\varphi \mathcal{R}
	\,,
\end{align}
where the $2\varphi R$ is a (local) RST term that restores the classical CGHS symmetry responsible for analytic solutions.

It is instructive to try to relate the RST model (\ref{eq:RST}) and semi-classical flat-JT gravity via a Weyl transformation plus field redefinition. Consider $g_{ij}\to e^{2\varphi}g_{ij}$, such that $\mathcal{R}\to e^{-2\varphi}\mathcal{R}-2e^{-2\varphi}\Box\varphi$. Consequently, under this rescaling the classical CGHS action (\ref{eq:cghs1}), quantum effective action (\ref{eq:polyact2}), and RST term, respectively transform as, 
\beq
\begin{split}
I^{(\text{classical})}_{\text{CGHS}} \quad &  \overset{\text{Weyl}}{\Longrightarrow}  \quad \frac{1}{16\pi G_{N}}\int d^{2}x\sqrt{-g}e^{-2\varphi}[\mathcal{R}+4\lambda^{2}e^{2\varphi}]\;,\\
W_{\text{CFT}}\quad& \overset{\text{Weyl}}{\Longrightarrow}\quad W_{\text{CFT}} - \frac{c}{96\pi} \int d^{2}x \sqrt{-g}\left[- 4\varphi \mathcal{R}  -  4(\nabla\varphi)^{2}\right]\;,\\
-\frac{c}{96\pi}\int d^{2}x\sqrt{-g}2\varphi\mathcal{R} \quad & \overset{\text{Weyl}}{\Longrightarrow}  \quad -\frac{c}{96\pi}\int d^{2}x\sqrt{-g}[2\varphi\mathcal{R}+4(\nabla\varphi)^{2}]\;,
\end{split}
\eeq
where we have dropped boundary terms arising from integrations by parts. Altogether,
\beq I_{\text{RST}}\quad  \overset{\text{Weyl}}{\Longrightarrow}  \quad \frac{1}{16\pi G_{N}}\int d^{2}x\sqrt{-g}\left[\left(e^{-2\varphi}+\frac{G_{N}c}{3}\varphi\right)\mathcal{R}+4\lambda^{2}\right]+W_{\text{CFT}}\;. \label{eq:rst2}\eeq
We recognize the transformed action as flat-JT gravity coupled to $W_{\text{CFT}}$ upon the field redefinition $\left(e^{-2\varphi}+\frac{G_{N}c}{3}\varphi\right)\to(\phi+\phi_{0})$. 
In a holographic setting, as in our case, we can use the Brown-Henneaux relation to set $G_{N}c/3=1/2$. 
Notice, however, $e^{-2\varphi}+\varphi/2$ has two branches of solutions, with a minimal value $\varphi_{\text{min}}=-\log(4)/2$, whereas the field redefinition $\phi_{0}+\phi\to e^{-2\varphi}\geq0$ for CGHS \eqref{eq:cghs1}, is single valued. 
It is unknown if there is any way to amend the path integral measure such that $\varphi\geq \varphi_{\text{min}}$ \cite{Thorlacius:1994ip}. 
 
Lastly, we emphasize that, unlike the aforementioned classical dilaton theories, their semi-classical extensions do not have a higher-dimensional pedigree; it is not known how to obtain a minimally coupled CFT with $c\gg1$ via a dimensional reduction. Rather, two-dimensional semi-classical dilaton-gravity is historically treated as an effective toy model to study the problem of semi-classical backreaction and black hole evaporation. One of the upshots of the current work is that we provide a derivation of the JT models using braneworld holography (and implicitly RST by the above argument) as we will exemplify in the next paragraphs.

\subsection{Semi-classical dilaton gravity on the brane}

Let us now see how each of the JT models are embedded in the induced brane theory (\ref{eq:res_final}). While in principle this can be done at the level of the action, it is not \emph{a priori} obvious which terms can be consistently ignored. We will thus first show the equations of motion of our setup consistently reduce to the equations of motion for the host of JT dilaton gravities described above, from which we determine which terms in the action may be safely discarded.

A first apparent puzzle is the number of equations of motion. JT gravity has four equations of motion in total: three components of the metric equations,  and the dilaton equation of motion. On the other hand, as argued in Section \ref{Section:3.2}, the trace of the equations of motion coming from varying the metric in Liouville gravity coupled to the quantum effective action agrees in our case with the equation of motion coming from varying the Liouville field, and thus we seem to only have three equations. The resolution of this discrepancy is that in deriving the action we have made a choice about the reference brane. To be precise, we have that the intrinsic curvature of the background is given by
\begin{align}
    \label{eq:constraint}
    \mathcal R = - 2 (1 - T_0^2) + \mathcal O ((1 - T_0)^2).
\end{align}
It is well known that the equation for the curvature is a constraint equation in JT gravity, and as such belongs to the four equations of motion. However, when considering the equations of motion for brane deformations we need to impose this equation by hand. This accounts for the missing equation.

Let us now start by considering the case where $T_0 \neq 1$ and $\Delta T = 0$. The dilaton equation of motion for brane fluctuations \eqref{eq:liouville3} becomes
\begin{equation}
	\mathcal{R}
	=
	2\Box\phi
	-
	2 (1 - T_0^2)
		e^{2\phi}
	  +  \mathcal{O}((1-T_{0})^{2}).
\end{equation}
Imposing our constraint \eqref{eq:constraint} and expanding in powers of $\phi$ we find 
\begin{equation}
	\Box\phi
	=
	2 (1 - T_0^2) \phi + \mathcal O(\phi^2)
	  +  \mathcal{O}((1-T_{0})^{2}).
\end{equation}
As expected, this equation is the trace of the metric equations of motion \eqref{eq:einsteinlouie} to the same order,
\begin{equation}
\label{eq:dileqJT}
\begin{aligned}	
	0=&~
	g_{ij}\Box\phi
	-
	\nabla_{i}\nabla_{j}\phi
	-
	\frac{1}{2}g_{ij}[(\partial_{k}\phi)^{2}
	+
	4\pi\tilde{\mu} e^{2\phi}]
	+
	(\nabla_{i}\phi)(\nabla_{j}\phi)
	-
	8\pi G_{N}\langle T^{\text{CFT}}_{ij}\rangle
	\\
	=&~
	g_{ij}\Box\phi
	-
	\nabla_{i}\nabla_{j}\phi
	+
	g_{ij}(T_{0}-1)
	(
		1
		+
		2\phi)
	-
	8\pi G_{N}\langle T^{\text{CFT}}_{ij}\rangle	
	+
	\mathcal{O}(\phi^{2}) +  \mathcal{O}((1-T_{0})^{2})
	\;.
\end{aligned}
\end{equation}
This shows the Liouville equation \eqref{eq:dileqJT} together with the constraint \eqref{eq:constraint} form a consistent set of equations to $\mathcal O(\phi^2)$. These equations of motion can also be obtained from an action of the form  
\begin{align}
\begin{split}
    \label{eq:JTfromLiouville}
 I_{\text{(A)dS JT}} ={}&\frac{1}{16\pi G_{N}}\int d^{2}\xi\sqrt{-g}(2(1-T_0)+\phi_{0}\mathcal{R})+W_{\text{CFT}}[g] \\
 &+\frac{1}{16\pi G_{N}}\int d^{2}\xi\sqrt{-g}\left(\phi(\mathcal{R}+\mu_{1}) \right)\\
&-\frac{1}{8\pi G_{N}}\int d\xi \sqrt{-g^\mathcal{I}}  \phi (\mathcal K + \sqrt{|1 - T_0^2|}) \;,
\end{split}
\end{align}
which precisely is \eqref{eq:res_final} truncated to linear order in $\phi$.
We thus conclude truncating the full dilaton action at linear order in $\phi$ is consistent and yields JT gravity on a geometry with cosmological constant $- 2 (1 - T_0^2)$. This clearly captures JT in AdS ($T_0 < 1$) and dS ($T_0 > 1$). 

To obtain JT gravity in flat space we could naively take the limit $T_0 \to 1$. However, as can be easily seen from \eqref{eq:JTfromLiouville}, in this case the theory would reduce to a term $\int d^2 x \sqrt{-g }\phi \mathcal R$ coupled to matter. Such a theory is trivial, as discussed in Section \ref{sec:taxonomy}, and does not allow for black hole solutions. 
To capture an interesting flat space dilaton theory we must keep the cosmological constant term in the first line of \eqref{eq:JTfromLiouville} whilst removing $\mu_1$ in the second line. 

This is precisely why in \eqref{eq:Liouvgravbraneflat} we retained the parameter $\Delta T$ which facilitates taking such a limit. Let us therefore now focus on the case where $T_0 = 1$ and $\Delta T \sim \mathcal O(\phi) \neq 0$  and follow the same logic as above. Our constraint \eqref{eq:constraint} now imposes that the brane we expand about has a flat intrinsic geometry. Up to higher derivative terms, the brane equation of motion coming from varying with respect to brane fluctuations is
\begin{equation}
	\mathcal{R}
	=
	2\Box\phi
	+ 4 \Delta T
	+
	\mathcal{O}(\phi^2) +
	\mathcal{O}(\text{higher derivatives})		
	\,,
\end{equation} 
which, in view of our constraint, is solved by 
\begin{align}
\label{eq:dilaton_sourced}
	\Box\phi
	= 
 - 2 \Delta T +
	\mathcal{O}(\phi^2).
\end{align}
As before, this is the trace of the equation of motion obtained by varying with respect to the metric,
 \beq
 g_{ij}\Box\phi-\nabla_{i}\nabla_{j}\phi
 +
 g_{ij}\Delta T 
 =8\pi G_{N}\langle T^{\text{CFT}}_{ij}\rangle  + \mathcal O(\phi^2)\;,\eeq
 where we recall the trace anomaly of the stress-energy tensor is proportional to the Ricci tensor of the background, i.e., the flat metric. 
 The above equation, together with \eqref{eq:constraint} follow from the action of flat-JT gravity coupled to a CFT,
\begin{align}
\begin{split}
 I_{\text{flat-JT}} ={}&\frac{1}{16\pi G_{N}}\int d^{2}\xi\sqrt{-g}\phi_{0}\mathcal{R}+W_{\text{CFT}}[g]\\
 &+\frac{1}{16\pi G_{N}}\int d^{2}\xi\sqrt{-g}\left(\phi\mathcal{R} - 2 \Delta T\right)\\
&-\frac{1}{8\pi G_{N}}\int d\xi \sqrt{-g^\mathcal{I}}  \phi (\mathcal K + \sqrt{|2\Delta T|})  + \mathcal O(\epsilon) \;.
\end{split}
\end{align}

This action can be obtained from \eqref{eq:res_final} by dropping all terms of $\mathcal O(\phi^2)$ and higher while treating $\Delta T = \mathcal O(\phi)$ and setting $T_0 = 1$. Here, we emphasize again, that the correction $\mathcal O(\epsilon)$ includes higher derivative corrections which we likewise ignore. Comparing to the canonical form of the flat-JT action, $\int d^2x \sqrt{-g} ( \phi \mathcal R + 4\lambda^2)$, we identify the cosmological constant to be
\begin{align}
    4\lambda^2 = - 2 \Delta T.
\end{align}
Thus, we conclude, from a doubly-holographic point of view, flat-JT gravity is nothing but fluctuations of a brane with tension $T \gtrless 1$ expanded around a flat brane. The fact that the geometry of the deformed brane is not flat is captured in the inhomogenous solution for the dilaton \eqref{eq:dilaton_sourced}, which describes how the distance of the deformed brane with tension $T = 1 + \Delta T$ to the reference brane with tension $T_0 = 1$ changes as a function of location. Moreover, the usual requirement of $\lambda^2 > 0$ in order to have gravity turn off at infinity translates to the requirement we choose the deformed brane to have subcritical tension $\Delta T < 1$, for the brane to be AdS.

\subsection{The relation between von Neumann and generalized entropy}
\label{sec:entropy}

For two-dimensional models of the type (\ref{eq:sphi}) plus the Polyakov action (\ref{eq:polyact2}), the analog of the Bekenstein-Hawking entropy \cite{Bekenstein:1972tm,Bekenstein:1973ur,Hawking:1975vcx} $S_\text{BH}$ is 
\beq S_\text{BH}=\frac{1}{4G_{N}}(\phi_{0}+\phi)|_{\partial\Sigma}-\frac{c}{6}\chi|_{\partial\Sigma}\;,\label{eq:entSgen}\eeq
where $\chi$ is the auxiliary scalar field introduced to localize the Polyakov action (\ref{eq:polyact2}), and $\partial\Sigma$ is a codimension-2 surface, e.g., a black hole bifurcation surface. A formal derivation of this entropy formula follows from applying the Iyer-Wald formalism \cite{Wald:1993nt,Iyer:1994ys}. Adopting the standard higher dimensional perspective, the constant $\phi_{0}$ corresponds to the entropy of the higher dimensional black hole in its extremal limit, while $\phi$ tracks deviations away from exact extremality. The $\chi$ contribution, meanwhile, is purely semi-classical, arising solely from the quantum matter action (\ref{eq:polyact2}).  As shown in \cite{Pedraza:2021cvx}, the total entropy (\ref{eq:entSgen}) is the generalized entropy $S_{\text{gen}}$ \cite{Bekenstein:1974ax}, i.e., the sum of the (classical) gravitational entropy and the von Neumann entropy of quantum matter exterior to a horizon.\footnote{In \cite{Myers:1994sg} the Wald functional was used to write down the semi-classical entropy of black hole solutions to the RST model, however, the formula was not recognized to be the generalized entropy.} This realization follows from the fact that, on-shell, the auxiliary field $\chi$ is proportional to the von Neumann entropy of a $\text{CFT}_{2}$ in vacuum reduced to an interval in a curved background. Explicitly, in conformal gauge $ds^{2}=-e^{2\rho}dx^{+}dx^{-}$, the von Neumann entropy is (see, e.g., Eq. (64) of \cite{Fiola:1994ir}) 
\beq S_{\text{vN}}=\frac{c}{6}(\rho(x^{+}_{1},x^{-}_{1})+\rho(x^{+}_{2},x^{-}_{2}))+\frac{c}{6}\log[\delta^{-1}_{1}\delta^{-1}_{2}(x^{+}_{2}-x^{+}_{1})(x^{-}_{2}-x^{-}_{1})]\;,\label{eq:vNent2d}\eeq
with cutoffs $\delta_{1,2}$ at the endpoints of interval $\partial\Sigma=[(x^{+}_{1},x^{-}_{1}),(x^{+}_{2},x^{-}_{2})]$ regulating the logarithmic UV divergence due to entanglement between modes of arbitrarily short-wavelength on a flat background. Meanwhile, $\chi$ has the solution $\chi=-\rho+\psi$, for harmonic function $\psi$ obeying $\Box\psi=0$. In \cite{Pedraza:2021cvx} the short-distance cutoffs $\delta_{1,2}$ arise as integration constants when solving for $\psi$.

As an induced gravity scenario, the holographic braneworld provides a realization of generalized entropy being equivalent to entanglement entropy \cite{Susskind:1994sm,Jacobson:1994iw,Frolov:1996aj,Frolov:1997up}.
To see this in the instance of a flat (Randall-Sundrum) brane, recall that the brane essentially acts like a UV cutoff. We can extend the spacetime behind the brane all the way to an asymptotic boundary on which we imagine a CFT with UV cutoff. The cutoff is given by the location of the brane, i.e., $\sqrt{\epsilon(x)}=e^{-(\phi_0 - \phi)}$ (here we extended the Fefferman-Graham relation (\ref{eq:phi0eps2}) by allowing for fluctuations about $\phi_{0}$).  The von Neumann entropy of a subregion $\Sigma$ for such a CFT in a flat background is
\beq
\begin{split} 
S_{\text{vN}}&=\frac{c}{6}\log\left[e^{(\phi_{0}+\phi(x_{1}))}e^{(\phi_{0}+\phi(x_{2}))}(x_{2}^{+}-x_{1}^{+})(x_{2}^{-}-x_{1}^{-})\right]\\
&=\frac{\phi_{0}+\phi(x_{1})}{4G_{N}}+\frac{\phi_{0}+\phi(x_{2})}{4G_{N}}+\frac{c}{6}\log\left[(x_{2}^{+}-x_{1}^{+})(x_{2}^{-}-x_{1}^{-})\right]\;,
\end{split}
\label{eq:vNinduc}\eeq
where in the second line we used $c=\frac{3}{2G_{N}}$. 
Comparing to \cref{eq:entSgen} with \eqref{eq:vNent2d}, we recognize von Neumann entropy equals the generalized entropy for a cutoff set by the AdS bulk length scale as is expected for the brane theory \cite{Emparan:2006ni,Chen:2020uac,Neuenfeld:2023svs}.
Our findings are consistent with previous (undeformed) holographic braneworld scenarios \cite{Hawking:2000da,Fursaev:2000ym,Emparan:2006ni, Neuenfeld:2023svs}.

\section{Boundary CFT interpretation} \label{sec:BCFTpov}
In the case of an AdS bulk and brane tension smaller than the critical value, ETW branes serve as bottom-up models for holographic BCFTs \cite{Takayanagi:2011zk,Fujita:2011fp}. In this section we will discuss the interpretation of a non-trivial dilaton profile (or equivalently a small deformation of the brane) from the BCFT point of view. We will restrict ourselves to linearized perturbations, i.e., the case of JT gravity on the brane.
Moreover, we will consider the case of the vacuum of a BCFT$_2$ on a half space, whose dual  bulk geometry is described by the $\text{AdS}_{3}$ Poincar\'e patch, see \eqref{eq:poincare_coords} and the discussion in \cref{sec:boundary_term}. 
The induced metric on the brane defined by the trajectory \eqref{eq:standard_solution_poincare} is given by 
\begin{align}
    \label{eq:ads2_poincare}
    ds^2_{\mathcal{Q}} = \frac{1}{1-T^2}\frac{-dt^2 + dy^2}{y^2},
\end{align} 
where $z\equiv\sqrt{1-T^2} y$. Clearly the brane geometry is that of $\text{AdS}_{2}$, and the constant prefactor $(1-T^2)^{-1}$ shows the $\text{AdS}_{2}$ length on the brane depends on the value of the brane tension.

Let us now modify the brane location (\ref{eq:standard_solution_poincare}) to create the ansatz for a deformed brane (recall \cref{fig:fluctuations}),
\begin{align}
    \label{eq:deformed_solution_poincare}
 X(t,z) = \frac{T}{\sqrt{1 - T^2}} \; z + \delta X(t,z),
\end{align}
which we subject to the brane equations of motion and only keep terms up to linear order in $\delta X$.  Again using the coordinate $y = \frac{z}{\sqrt{1 - T^2}}$ on the brane, the expansion of the brane equations of motion (\ref{eq:brane_eom}) produce the following three differential equations (see Appendix \ref{app:branedefsJT}) 
\begin{align}
    \partial_y \delta X(t, y) + y \partial^2_t \delta X(t,y) = 0, \qquad \partial_t  \partial_y \delta X(t,y) = 0,  \qquad \partial_y \delta  X(t, y) - y \partial^2_y \delta X(t,y) = 0,
\label{eq:diffeomsbrane}\end{align}
whose general solution takes the form 
\begin{align}
    \label{eq:on_shell_fluctuations}
    \delta X(t,y) = a + b\, t + c\, (t^2 - y^2),
\end{align}
with arbitrary constants $a,b,c$.

Let us compare (\ref{eq:on_shell_fluctuations}) to solutions of the JT equations of motion on the background (\ref{eq:ads2_poincare})
\begin{align}
    \nabla_i \nabla_j \phi - (1-T_{0}^2) g_{ij} \phi = 0.
\end{align}
The tension-dependent prefactor of the last term effectively cancels the tension-dependence of the metric (\ref{eq:ads2_poincare}). Thus, the general solution is completely independent of $T$ and is given by
\begin{align}
    \label{eq:general_sol_jt}
    \phi(t,y) = \frac{a' + b'\, t + c'\, (t^2 - y^2)}{y},
\end{align}
again with three arbitrary constants $a',b',c'$. 
It is not surprising that, at leading order, we reproduce the JT equations of motion from the equations of motion for brane deformations, given that the action for brane deformations is the JT action. In fact, from \eqref{eq:new_traj} we immediately see that $\phi = \frac{\delta X(t,y)}{y}$. We can thus identify $a = a'$, and similarly for $b$ and $c$.
Due to our choice of coordinates, it is now easy to obtain the location of the BCFT boundary from the dilaton through
\begin{align}
    \label{eq:boundary_deformation}
    \lim_{y \to 0} y \phi(t,y) = \delta X(t,0) = a + b\, t + c\, t^2.
\end{align}
We see a non-trivial dilaton profile indicates a deformation of the location of the BCFT boundary from its zeroth order location $x=0$.

To understand the nature of this deformation, we will naively apply the extrapolate dictionary of AdS/CFT. It states that for a boundary operator with dimension $\Delta$ there is a bulk field whose solution to the bulk equations of motions near the boundary has an expansion of the form
\begin{align}
    \phi(t,y) \sim y^{d-\Delta} \phi_-(t) + \dots  + y^\Delta \phi_+ (t) + \dots. 
\end{align}
If we assume that the dilaton field has a dual boundary operator on the codimension-one boundary of the dual CFT (i.e., $d=1$), \eqref{eq:general_sol_jt} implies that we must choose $\Delta = 2$. Expectation values of this operator are encoded by the boundary limit of the so-called normalizable piece $\phi_+(t)$ and the $\Delta = 2$ operator is sourced in the dual theory by the (rescaled) boundary limit of the non-normalizable mode $\phi_-(t)$.

In terms of the dilaton on the brane, the non-normalizable bulk mode can be identified with the classical solution of $\phi(t,y)$. Its rescaled boundary limit $\phi_-(t) = \delta X(t,0)$, obtained in \eqref{eq:boundary_deformation}, then acts as a source for an operator of dimension $\Delta = 2$. In fact, two-dimensional BCFTs contain a natural boundary operator of dimension $2$ sourced by deformations of the boundary -- the displacement operator $\mathcal D = T^{nn}$, i.e., the limit of the stress-energy tensor towards the boundary with both indices pointing in the normal direction \cite{McAvity:1993ue,Billo:2016cpy}. This leads us to the main point of this section.

We propose a non-trivial dilaton profile on the brane is dual to a BCFT$_2$ that has been deformed with the displacement operator $\mathcal D$. Here we will demonstrate how the dilaton configurations in (\ref{eq:general_sol_jt}) are generated from a BCFT perspective. We leave 
a detailed investigation of the relation between the boundary displacement operator and dilaton theories on the brane to future work.

A natural guess for a three-parameter family of deformations with non-trival action on the boundary is given by the action of the three generators of the Lorentzian $\operatorname{SL}(2,\mathbb R) \times \operatorname{SL}(2,\mathbb R)$ global conformal symmetry which are broken by the presence of the boundary. The geometric action of infinitesimal global transformations is given by conformal killing vectors that are the $n=0,\pm 1$ generators of two Witt-algebras
\begin{align}
    L_n = -z^{n+1} \partial_z, \qquad \bar L_n = - \bar z^{n+1} \partial_{\bar z}, \qquad \text{for }n=\pm 1, 0,
\end{align}
with
\begin{align}
    \label{eq:conformal_generators}
    [L_n, L_m] = (n-m) L_{n+m}, \qquad [\bar L_n, \bar L_m] = (n-m) \bar L_{n+m}.
\end{align}
In our case $z = t+x$ and $\bar z = t-x$. The boundary of the BCFT at $x=0$ breaks certain linear combinations of the generators in (\ref{eq:conformal_generators}). To identify the broken generators we define the location of the brane with an embedding function $f(z,\bar z) = \frac 1 2 (z-\bar z) = x$ such that the boundary is located at $0 = f(z, \bar z)$. It is then useful to consider the following linear combinations of the generators $L_n$,
\begin{align}
    \mathcal L_n = L_n + \bar L_n, && \bar {\mathcal L}_n = L_n - \bar L_n.
\end{align}
One can easily verify that the generators $\mathcal L_n$ preserve the location of the boundary, i.e., $\mathcal L_n f(z,\bar z)|_{z-\bar z = 0} = 0$. On the other hand, a straightforward computation shows
\begin{align}
\delta f(z,\bar z) = \left(a \, \bar {\mathcal L}_{-1} + b\, \bar {\mathcal L}_{0} + c\, \bar {\mathcal L}_{1}\right) f(z,\bar z) = - a - b\, t - c\, (t^2 + x^2).
\end{align}
The location of the boundary after we acted with the broken generators is given by $f(z,\bar z) + \delta f(z, \bar z) = 0$, which at leading order exactly reproduces the deformation obtained from the solution of the dilaton equations of motion, (\ref{eq:boundary_deformation}).

\section{The Ryu-Takayanagi formula with JT gravity on the brane} \label{sec:RTderivation}

An important quantity of interest in BCFTs is the boundary entropy associated with a choice of boundary conditions \cite{Affleck:1991tk}. This boundary entropy can be defined as a constant contribution to the entanglement entropy of a strip-like region which includes the boundary. In the case of holographic BCFTs it was shown by Takayanagi that such additional contributions can be reproduced from a holographic computation using the RT prescription \cite{Ryu:2006bv,Hubeny:2007xt} if the standard homology constraint of the RT formula is modified such that RT surfaces are allowed to end on ETW branes \cite{Takayanagi:2011zk}. More precisely, the RT surface has to obey Neumann boundary conditions at the brane such that its endpoint is dynamically determined by the extremization procedure. 

If the theory on the brane includes JT gravity, the authors of \cite{Almheiri:2019hni} proposed to modify the RT formula to include an endpoint contribution proportional to the dilaton whenever the RT surface touches the brane, i.e., 
\begin{align}
    \label{eq:area_modified}
    S_\text{RT} =\underset{\sigma_A}{\min \operatorname{ext}} \left( \frac{\text{Area}(\sigma_A)}{4 G_N} + \frac{\phi(\sigma_A \cap \mathcal Q)}{4 G_N} \right),
\end{align}
where $\sigma_A$ is a bulk surface and $\mathcal Q$ is the brane with dilaton $\phi$. The additional term proportional to the dilaton does not affect the extremization of the bulk of $\sigma_A$, however, it changes the result of the extremization over the endpoint of $\sigma_A$ at the brane, see \cite{Chen:2020uac}. In \cite{Chen:2020uac,Chen:2020hmv} this prescription was extended to the case where a localized gravitational action is added to the brane (possibly including additional higher derivative terms) and a partial proof for particularly symmetric configurations was given. As far as we know, however, there is no general proof that (\ref{eq:area_modified}) is the correct prescription if JT gravity is added to the brane. 

Above, we argued that JT gravity on the brane can be understood as an effective theory for brane deformations in the semi-classical limit. Now, we will demonstrate that in this case (\ref{eq:area_modified}) follows from Takayanagi's prescription for RT surfaces in the presence of ETW branes.

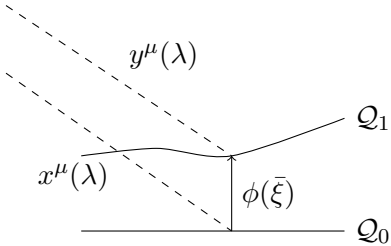
\begin{figure}[t]
    \centering
    \begin{tikzpicture}
    \draw (0,0) -- (3.5,0) node [right] {$\mathcal Q_0$};
    \draw plot [smooth] coordinates {(0,1) (1,1.1) (2,1) (3.5,1.5)} node [right] {$\mathcal Q_1$};
    \draw[->] (2,0) -- node [right] {$\phi(\bar \xi)$} (2,1);

    \draw [dashed] (-1,3) to node [above right] {$y^\mu(\lambda)$} (2,1);
    \draw [dashed] (-1,2.1) to node [below left] {$x^\mu(\lambda)$} (2,0);
    \end{tikzpicture}
    \caption{The deformed $\mathcal Q_1$ and undeformed $\mathcal Q_0$ branes. The dashed lines show bulk geodesics which end at $\bar \xi$ on either $\mathcal Q_0$ or $\mathcal Q_1$. For the shown configuration they are not close to extremal surfaces.}
    \label{fig:expansion_brane}
\end{figure}

Let us again consider the situation discussed in Section \ref{sec:dilagravaction}. That is, we assume we can choose normal coordinates, (\ref{eq:metric_ansatz}) around a reference brane $\mathcal Q_0$ with embedding $\theta_{0}(\xi) = 0$. We consider another brane $\mathcal Q_1$ whose embedding can, at least locally, be treated as a deformation around $\mathcal Q_0$ (see Figure \ref{fig:expansion_brane}),
\begin{align}
    \theta_{1}(\xi) - \theta_{0}(\xi) = \phi(\xi).
\end{align}
The deformation $\phi$ plays the role of the dilaton in the brane theory. In order to obtain statements about the JT limit of the theory on the brane, we will restrict ourselves to expressions up to linear order in the dilaton $\phi \ll 1$ which will thus play the role of our expansion parameter in the following.

Now, consider a bulk extremal surface $y^\mu(\lambda)$ which connects a point $y^\mu(0)$ at the asymptotic boundary to some point $y^\mu(1) = (\theta(\vec {\bar \xi}), \bar \xi_0, \bar \xi_1)$ on $\mathcal Q_1$. For now, we do not require that $y^\mu$ is extremized with respect to variations of its endpoint $y(1)^\mu$.

Our first task is to relate the ordinary area functional of this surface to that of a bulk extremal surface $x^\mu(\lambda)$ connecting $x^\mu(0) = y^\mu(0)$ to $x^\mu(1) = (0, \bar \xi_0, \bar \xi_1)$. In the following we will assume the bulk geometry varies sufficiently smoothly such that, since the chosen endpoints are only $\mathcal O(\phi)$ apart, the bulk surfaces are also perturbatively close. That is, we can find a parametrization where
\begin{align}
\label{eq:expansion_extremal_surface}
    y^\mu(\lambda) = x^\mu(\lambda) + \Delta x^\mu(\lambda),
\end{align}
with $\Delta x^\mu(0) = (0,0,0)$, $\Delta x^\mu(1) = (\phi(\bar \xi), 0,0)$ and where both $\Delta x^\mu(\lambda) = \mathcal O(\phi)$ and $\frac d {d\lambda} {\Delta x}^\mu(\lambda) = \mathcal O(\phi)$. Moreover, to make contact with the RT formula, we assume both $x^\mu(\lambda)$ and $y^\mu(\lambda)$ are globally the smallest extremal surfaces connecting their respective endpoints. It is possible for RT surfaces to discontinuously jump, even if we vary their endpoints only slightly and so this assumption can fail. We therefore exclude configurations where this happens.\footnote{This opens the possibility of large failures of (\ref{eq:area_modified}) for special boundary subregions. We leave an investigation of these fine-tuned situations to future work.}

In the absence of the pathologies just described, the area of the surface $y^\mu$ is given by
\begin{align}
    \begin{split}
    \label{eq:area_expanded}
    \operatorname{Area}[y^\mu] = \int_0^1 \sqrt{h[y]} d\lambda = \int_0^1 \sqrt{h[x]} d\lambda + \int_0^1 d\lambda \left(\dots\right) \Delta x^\mu(\lambda) + \frac{ G_{\mu\nu} \Delta x^\mu(1) \dot x^\nu(1)}{\sqrt{G_{\mu\nu} \dot x^\mu \dot x^\nu}}.
    \end{split}
\end{align}
The first term on the right-hand side is simply $\operatorname{Area}[x^\mu]$ and the second term vanishes since $x^\mu$ is an extremal surface in the bulk. The last term is a boundary term which survives which can be written as
\begin{align}
\frac{G_{\mu\nu} \Delta x^\mu(1) \dot x^\nu(1)}{\sqrt{G_{\mu\nu} \dot x^\mu \dot x^\nu}} = \phi(\bar \xi) n_\mu t^\mu,
\end{align}
where $n^\mu$ is the normal to $\mathcal Q_0$ and $t^\mu = \frac{\dot x^\nu(1)}{\sqrt{G_{\mu\nu} \dot x^\mu \dot x^\nu}}$ is the tangent to the extremal surface $x^\mu$ at $\lambda=1$. In total,
\begin{align}
    \label{eq:area_modified_2}
    \operatorname{Area}[y^\mu] = \operatorname{Area}[x^\mu] + \phi(\bar \xi) n_\mu t^\mu.
\end{align}
This expression does not agree with the area functional used in (\ref{eq:area_modified}) because for a general choice of endpoints $\vec{\bar \xi}$ we have that $n_\mu t^\mu - 1 = \mathcal O(1)$. However, recall that (\ref{eq:area_modified}) also involves extremization over the endpoint. Thus, in order to prove (\ref{eq:area_modified}) is indeed the correct RT formula, it suffices to show the area function used in (\ref{eq:area_modified}) agrees with (\ref{eq:area_modified_2}) in the vicinity of extremal surfaces to leading order in $\phi$.

To see that this is indeed the case, we vary (\ref{eq:area_modified_2}) with respect to the endpoint of the extremal surface,
\begin{align}
\label{eq:variation_area_dilaton}
\begin{split}
\delta \operatorname{Area}[y^\mu] & = \delta \operatorname{Area}[x^\mu] + \delta \phi\, t^\mu n_\mu + \phi\,  (\delta t)^\mu n_\mu    \\
& = \delta \operatorname{Area}[x^\mu] + \delta \phi \, t^\mu n_\mu  + \mathcal O(\phi^2).
\end{split}
\end{align}
The last term is of higher order since $\delta t^\mu = \mathcal O(\phi)$. Moreover, if we define $\tilde t_\mu$ the tangent to $y^\mu$ at $\mathcal Q_1$ and $\tilde n_\mu$ the normal to $\mathcal Q_1$ at the intersection of $y^\mu$ and the brane, we have
\begin{align}
    \tilde n_\mu \tilde t^\mu = 1.
\end{align}
These vectors can be related to $n^\mu$ and $t^\mu$ via
\begin{align}
    \tilde n = \frac{\partial_\theta - \partial_i \phi \partial_{\xi^i}}{\sqrt{1 + \partial_i \phi \partial^i \phi}} = n + \mathcal O(\phi), && \tilde t^\mu = \frac{\dot y^\nu(1)}{\sqrt{G_{\mu\nu} \dot y^\mu(1) \dot y^\nu(1)}} = t^\mu + \mathcal O(\phi),
\end{align}
such that $n_\mu t^\mu = 1 + \mathcal O (\phi)$. This demonstrates
\begin{align}
    \label{eq:proof_extremal_surface}
    \delta \operatorname{Area}[y^\mu] = \delta \operatorname{Area}[x^\mu] + \delta \phi + \mathcal O(\phi^2) \qquad \text{(near an extremal surface $y^\mu$).}
\end{align}

Situations can arise in which $y^\mu$ is a member of a family of extremal surfaces, i.e., there is direction in deformation space such that $y^\mu$ stays extremal. In this case it is obvious from the area variation (\ref{eq:proof_extremal_surface}) that neglecting the $\mathcal O(\phi^2)$ breaks the degeneracy and incorrectly singles out a particular extremal surface. In such fine-tuned examples we must thus be careful to remember extremal surfaces can exist anywhere where 
\begin{align}
    \label{eq:approximate_saddlepoint}
    \delta \operatorname{Area}[x^\mu] + \delta \phi = \mathcal O(\phi^2).
\end{align}
However, the generic case is that $y^\mu$ is an isolated extremal surface, i.e., all deformations of $y^\mu$ move away from extremality. In this case we can neglect the higher orders in (\ref{eq:approximate_saddlepoint}) and it follows from (\ref{eq:proof_extremal_surface}) that the existence of an extremal surface $y^\mu$ ending on $\mathcal Q_1$ implies the existence of a surface $x^\mu$ ending on $\mathcal Q_0$ which extremizes the modified area functional in (\ref{eq:area_modified}). Furthermore, we can integrate (\ref{eq:proof_extremal_surface}) to show the area of the extremal surface $y^\mu$ agrees with the area of the surface $x^\mu$ measured with (\ref{eq:area_modified}). Thus, we conclude  the RT formula in the presence of a perturbatively deformed ETW brane is reproduced by the modified prescription (\ref{eq:area_modified}) when the deformation is modeled by JT gravity on an undeformed brane.

\section{Summary and outlook} \label{sec:disc}

In this article we considered shape deformations of two-dimensional end-of-the-world branes embedded in any three-dimensional ambient background. The action for the induced theory on the brane is a specific semi-classical dilaton gravity and the dilaton encodes the brane deformations. When we restrict to a bulk $\text{AdS}_{3}$ spacetime, which we take to have a holographic dual, the dilaton theory controlling the shape of the brane flows to Liouville gravity in the infrared. We showed that for small fluctutations the brane theory reduces to (A)dS or flat-JT gravity coupled to a holographic large-$c$ CFT, depending on the brane slicing. Specializing to the case of an $\text{AdS}_{2}$ brane, we argued changing the boundary conditions of JT gravity on the brane corresponds to a deformation of the $\text{BCFT}_{2}$ with the displacement operator, which from the viewpoint of the defect theory is an irrelevant deformation. Lastly, in the presence of a non-trivial dilaton profile on the brane, we proved the RT prescription for computing entanglement entropy for holographic BCFTs is modified by a contact term, as in \cite{Almheiri:2019hni}.

Our set-up and analysis open up multiple avenues worth exploring. First, thus far we  restricted ourselves to a three-dimensional bulk spacetime, leading to an induced two-dimensional dilaton gravity on the brane. Nothing in our construction, however, prevents us from working in arbitrary dimension. Indeed, as detailed in Appendix \ref{app:perturbative_computations}, for a deformed $d$-dimensional brane embedded in bulk $\text{AdS}_{d+1}$, we find the dilaton-gravity action up to second order in deformations $\phi$ is (for $\Delta T=0$ and neglecting a boundary term)
\beq 
\begin{split}
\Delta I_{\total}&=\frac{1}{16\pi G_{N}}\int d^{d}\xi\sqrt{-g}\biggr[F(\phi)\mathcal{R}+V(\phi)+a_{d}(\partial_{i}\phi)^{2}\biggr]\;,
\end{split}
\eeq
where $F(\phi)\equiv\phi+\frac{(d-2)}{2}\phi^{2}$ and $V(\phi)\equiv b_{d}\phi+c_{d}\phi^{2}$,
with $a_{d}>0$ and $b_{d}$ and $c_{d}$ are some coefficients.
Thus, the deformations are characterized by a generalized Brans-Dicke-type gravity. The zeroth order contribution, meanwhile, is the usual higher-curvature theory of gravity coupled to a CFT with a UV cutoff (see, e.g., \cite{Emparan:1999pm,Bueno:2022log}). It would be worth looking for black hole or cosmological solutions to such theories. The $d=3$ case would be particularly interesting as there already exist analytic black hole solutions which exactly account for semi-classical backreaction  (without brane deformations) \cite{Emparan:2020znc,Emparan:2022ijy,Panella:2023lsi}. 

Second, we focused on pure three-dimensional Einstein gravity in the bulk, taking the bulk geometry to be empty $\text{AdS}_{3}$. An extension of our work would be to consider a bulk BTZ black hole \cite{Banados:1992wn,Banados:1992gq} or AdS$_{3}$ black string, for which the bulk horizon induces a black hole horizon on the brane, thus allowing for a study of JT black holes (cf. \cite{Geng:2022dua}). Alternatively, our set-up offers a doubly-holographic study of JT black holes dual to accelerated boundary trajectories, i.e., no bulk black hole/string is required and the bath region can be at zero temperature. In this context the horizon on the brane is induced by acceleration (Rindler) horizons, analogous to \cite{Jensen:2013ora,Sonner:2013mba}. Instead of an energy transfer between a black hole and a thermal bath (the half-space CFT), energy would be injected via `kinks' in the boundary. We leave this for future work.

Third, it is well known, as a two-dimensional toy model, AdS-JT gravity reduces to a Schwarzian theory on AdS$_{2}$ boundary \cite{Kitaev_SYK,Maldacena:2016upp}. At the level of the Euclidean path integral,  the remaining degrees of freedom upon integrating out the dilaton -- the Schwarzian mode -- characterize the freedom to remove a patch of a hyperbolic disk with fixed boundary length. In the case of the induced AdS-JT gravity we find (including the correct GHY boundary term), it is not obvious how the Schwarzian theory emerges. In particular, from the bulk perspective, it is not clear what the freedom to cut out a patch of a disk corresponds to. It would be worth understanding how the Schwarzian dynamics arises in our construction.

Further, more generally we found the induced theory on the brane is characterized by a class of two-dimensional dilaton models with a high degree of symmetry. There exist, however, dilaton gravities with less symmetry, e.g., the deformed JT models examined in \cite{Lemos:1993py,Sybesma:2022nby}, which fall in the category of an action of the type $I=\int d^{2}x\sqrt{-g}\left[\Phi \mathcal{R}+\Lambda\Phi^{\alpha}\right]$, for $\alpha\neq1$ and $\Lambda$ some real number.  It would be interesting to see what modifications to the bulk theory could lead to such theories with and understand the symmetry breaking mechanism from the bulk and BCFT perspectives.

Our work also leaves some open technical questions. 
For example, we found that at leading order in the cutoff the induced brane theory is \emph{classical} Liouville gravity, in that $Q=b^{-1}=\sqrt{\frac{c}{6}}$. In quantum Liouville theory, while the two-dimensional action takes the same form, the parameter $Q$ receives quantum corrections, $Q\to b^{-1}+b$, such that $b\to0$ can be understood as a semi-classical limit. This correction follows from modifying the path integration measure over the Liouville field \cite{David:1988hj,Distler:1988jt}.
While we did not carefully discuss the path integral measure, it should be interesting to investigate the precise form of the measure descending from the three-dimensional bulk path integral, as this would provide a means to compute the central charge of the Liouville theory we find at leading order.

Moreover, in this article we primarily focused on the physics as seen from the brane perspective. We argued a non-trivial dilaton profile on the brane is dual to a BCFT$_2$ that has been deformed by the displacement operator $\mathcal D$, and provided some evidence for this proposal. It would be worth exploring this duality in more detail. Most importantly, we have only considered this proposal at the level of a JT gravity approximation to the full theory. An obvious extension would be to better understand which boundary conditions, e.g., Fateev-Zamolodchikov-Zamolodchikov-Teschner (FZZT) D1-branes \cite{Fateev:2000ik,Teschner:2000md} or Zamolodchikov-Zamolodchikov (ZZ) D0-branes \cite{Zamolodchikov:2001ah}, have to be imposed for the full leading order Liouville gravity on the brane and their dual BCFT interpretation.

Lastly, let us end with an intriguing observation. In Section \ref{sec:entropy} we showed how the generalized entropy in the brane perspective equals the entanglement entropy of a two-dimensional CFT with a UV cutoff. Notably, the dilaton contribution to the entropy arises from the UV cutoff, $\sqrt{\epsilon(x)}=e^{-(\phi_{0}+\phi(x))}$. As argued in Section \ref{sec:dynamical_dilaton}, the dilaton is a dynamical field to be integrated over in the path integral (\ref{eq:bulkPIZ}). Combining these observations, this suggests a UV regulated CFT in which the regulator is treated as a dynamical field displays some characteristics of a gravitational theory. It would be interesting to investigate the meaning of integrating over a cutoff, and study its implications. Moreover, this observation suggests a sharp difference between braneworld holography and $T\bar T$-deformed CFTs \cite{Smirnov:2016lqw,McGough:2016lol} whose precise relation is yet to be understood.

\noindent\section*{Acknowledgments}
We are grateful to Nele Callebaut, Roberto Emparan, Antonia Micol Frassino, Ben Freivogel, Hao Geng, Jani Kastikainen, Thomas Mertens, Rodolfo Panerai, Juan Pedraza, Shan-Ming Ruan, and Matteo Selle for useful correspondence. We also thank Antonia Micol Frassino for initial collaboration. DN acknowledges support from the Heising-Simons Foundation “Observational Signatures of Quantum Gravity” collaboration grant 2021-2817. AS is supported by STFC consolidated grant ST/X000753/1. 
WS is supported by the Simons Foundation through an ``INI-Simons Postdoctoral Fellowship in Mathematics''. WS acknowledges support from the Icelandic Research Fund via the Grant of Excellence titled “Quantum Fields and Quantum Geometry”, and the University of Iceland Research Fund. WS thanks Juan Pedraza for his hospitality.

\appendix

\section{Conventions and useful relations} \label{app:conventions}

Here we state our conventions (which largely follow \cite{Poisson:2009pwt}) and provide some useful formulae used in the main text. 

\subsection*{Background geometry} Let $\mathcal{M}$ be a $d+1$ dimensional spacetime endowed with metric $G_{\mu\nu}$ with coordinates $x^{\mu}$ on $\mathcal{M}$. We take a `mostly plus' convention for Lorentzian signature. The Riemann curvature tensor on $\mathcal{M}$ with respect to $G_{\mu\nu}$ is given by 
\beq R^{\rho}_{\;\sigma\mu\nu}=\partial_{\mu}\Gamma^{\rho}_{\;\nu\sigma}-\partial_{\nu}\Gamma^{\rho}_{\;\mu\sigma}+\Gamma^{\rho}_{\;\mu\lambda}\Gamma^{\lambda}_{\;\nu\sigma}-\Gamma^{\rho}_{\;\nu\lambda}\Gamma^{\lambda}_{\;\mu\sigma}\;.\label{eq:Riemanngencon}\eeq
Equivalently, in terms of the Levi-Civita connection $\nabla_{\mu}$, $R^{\rho}_{\;\sigma\mu\nu}=-[\nabla_{\mu},\nabla_{\nu}]V_{\nu}$ and $R^{\rho}_{\;\sigma\mu\nu}V^{\sigma}=[\nabla_{\mu},\nabla_{\nu}]V^{\rho}$ for vectors $V$. Our convention is such that in Euclidean space the intrinsic Ricci scalar of the 2-sphere is positive.

\subsection*{Hypersurface geometry} 

 Let $\Sigma$ denote a timelike or spacelike $d$-dimensional hypersurface embedded in $\mathcal{M}$. We will define our hypersurface as  a restriction on coordinates $x^{\alpha}$, i.e., introduce a scalar function $\Phi(x^{\alpha})$ which obeys the constraint $\Phi(x^{\alpha})=0$.\footnote{When $\Phi(x^{\alpha})$ is spacelike, $\Phi$ is taken to increase toward the future across the hypersurface, while when $\Sigma$ is timelike $\Phi$ increases outward.} We can generate a unit normal $n_{\alpha}$ via $n_{\alpha}=\epsilon \mathcal{N}\partial_{\alpha}\Phi$
with $\epsilon\equiv n^{2}=n_{\alpha}n^{\alpha}=\pm1$, where $\epsilon=+1$ indicates the hypersurface is timelike, $\epsilon=-1$ has $\Sigma$ spacelike, and $\mathcal{N}$ is a normalization, $\mathcal{N}=|G^{\alpha\beta}\partial_{\alpha}\Phi\partial_{\beta}\Phi|^{-1/2}$. Further, we take $n^{\alpha}$ to be outwards pointing.
The induced metric $h_{ij}$ and its inverse on $\Sigma$ are defined by
\beq h_{ij}\equiv G_{\alpha\beta}e^{\alpha}_{i}e^{\beta}_{j}\;,\quad  e^{i}_{\alpha}\equiv h^{ij}G_{\alpha\beta}e^{\beta}_{j}\;,\eeq 
for vectors  $e^{\alpha}_{i}\equiv\frac{dx^{\alpha}}{d\xi^{i}}$ tangent to curves contained in $\Sigma$ and coordinates $\xi^{i}$ intrinsic to $\Sigma$.
By definition, $n_{\alpha}e^{\alpha}_{i}=0$. In terms of the background $G_{\alpha\beta}$,
\beq G_{\alpha\beta}=\epsilon n_{\alpha}n_{\beta}+h_{ij}e^{i}_{\alpha}e^{j}_{\beta}=\epsilon n_{\alpha}n_{\beta}+h_{\alpha\beta}\;,\quad h_{\alpha\beta}\equiv h_{ij}e^{i}_{\alpha}e^{j}_{\beta}\;,\label{eq:completeness2}\eeq
where $h_{\alpha\beta}$ is the projector onto hypersurfaces orthogonal to $n_{\mu}$. Similarly, $h^{\alpha\beta}=G^{\alpha\beta}-\epsilon n^{\alpha}n^{\beta}$.
Clearly, $h^{\alpha\beta}n_{\alpha}=h^{\alpha\beta}n_{\beta}=0$.

A directed surface element on $\Sigma$ is given by 
$d\Sigma_{\alpha}=\epsilon n_{\alpha}d\Sigma$, where $d\Sigma\equiv|h|^{1/2}d^{d}\xi$, such that Stokes' theorem reads for vector field $A^{\alpha}$
\beq \int_{\mathcal{V}}\sqrt{-G}d^{d+1}x\nabla_{\alpha}A^{\alpha}=\oint_{\partial\mathcal{V}}d\Sigma_{\alpha}A^{\alpha}\;,\eeq
where $\partial\mathcal{V}$ is the boundary of a $d+1$-dimensional region $\mathcal{V}$.

\subsubsection*{Extrinsic curvature}

 We define the extrinsic curvature $K_{ij}$ as
\beq K_{ij}\equiv -(\nabla_{\beta}n_{\alpha})e^{\alpha}_{i}e^{\beta}_{j}\;.\eeq
The trace of the extrinsic curvature (or `mean' curvature) is given by 
\beq K=h^{ij}K_{ij}=-\nabla_{\alpha}n^{\alpha}\;.\eeq
The minus sign is included such that a circle in Euclidean space has positive $K$. Equivalently, we may write the extrinsic curvature with respect to bulk coordinates via 
\beq K_{\mu\nu}=e^{i}_{\mu}e^{j}_{\nu}K_{ij}=-h_{\mu}^{\alpha}h_{\nu}^{\beta}\nabla_{\alpha}n_{\beta}\;.\eeq
Using the decomposition for $G_{\mu\nu}$ (\ref{eq:completeness2}), we may write 
\beq K_{\mu\nu}=-\nabla_{\mu}n_{\nu}+\epsilon n_{\mu}a_{\nu}\;,\label{eq:Kmunuacc}\eeq
where $a_{\mu}$ is the acceleration for the integral curves of the unit normal $n_{\mu}$, $a_{\mu}=n^{\nu}\nabla_{\nu}n_{\mu}=-\epsilon\partial_{\mu}\log\mathcal{N}$.
Observe $n^{\mu}a_{\mu}=0$, such that $K_{\mu\nu}n^{\nu}=n^{\nu}\nabla_{\mu}n_{\nu}=0$, and $a_{\mu}\nabla_{\alpha}n^{\mu}=-n^{\mu}\nabla_{\alpha}a_{\mu}$.\footnote{It is also useful to know $n^{\mu}K_{\mu\nu}=n^{\nu}K_{\mu\nu}=0$, $a^{\mu}K_{\mu\nu}=-a^{\mu}\nabla_{\mu}n_{\nu}$, and $a^{\nu}K_{\mu\nu}=-a^{\nu}\nabla_{\mu}n_{\nu}+\epsilon a^{2}n_{\mu}$.} Clearly, $K=h^{\mu\nu}K_{\mu\nu}=G^{\mu\nu}K_{\mu\nu}$.

\subsection*{Action and Neumann boundary condition}

With these conventions, the (Lorentzian) Einstein-Hilbert action supplemented with a GHY boundary term is 
\beq I=\frac{1}{16\pi G_{N}}\int_{\mathcal{M}}d^{d+1}x\sqrt{-G}(R-2\Lambda)-\frac{1}{8\pi G_{N}}\int_{\partial\mathcal{M}}d^{d}\xi\sqrt{|h|}\epsilon K\;,\eeq
where $G_{N}$ is Newton's constant in $d+1$-dimensions. 
The GHY term makes the variational problem well-posed. Indeed, a standard calculation shows that the metric variation of the Einstein-Hilbert term is 
\beq
\begin{split}
16\pi G_{N}\delta I_{\text{EH}}&=\oint_{\partial\mathcal{M}} d^{d}\xi\sqrt{|h|}\epsilon(n_{\rho}G^{\sigma\nu}\delta \Gamma^{\rho}_{\;\nu\sigma}-n^{\rho}\delta\Gamma^{\nu}_{\;\nu\rho})\\
&=
\oint_{\partial\mathcal{M}}d^{d}\xi\sqrt{|h|}\epsilon(K_{\mu\nu}-Kh_{\mu\nu})\delta h^{\mu\nu}+2\delta\left(\oint_{\partial\mathcal{M}}d^{d}\xi\sqrt{|h|}\epsilon K\right)\;,
\end{split}
\eeq
where we have imposed the bulk Einstein equations and assumed the manifold $\mathcal{M}$ is void of codimension-2 corners. Thus, 
\beq \delta I=-\frac{1}{16\pi G_{N}}\oint_{\partial\mathcal{M}}d^{d}\xi\sqrt{|h|}\epsilon(K_{\mu\nu}-Kh_{\mu\nu})\delta h^{\mu\nu}\;.\eeq
The action is then stationary when one either imposes Dirichlet boundary conditions, $\delta h^{\mu\nu}|_{\partial\mathcal{M}}=0$, or Neumann boundary conditions
\beq (K_{\mu\nu}-K h_{\mu\nu})|_{\partial\mathcal{M}}=0\;.\label{eq:NeumannBC}\eeq
Introducing a brane of tension $T$ at $\partial\mathcal{M}$, characterized by a brane action,
\beq I_{T}=-\frac{T}{8\pi G_{N}}\int_{\partial\mathcal{M}}d^{d}\xi\sqrt{|h|}\;,\eeq
 the total action $I+I_{T}$ will be stationary provided the bulk Einstein equations hold and the Neumann boundary condition (\ref{eq:NeumannBC}) is modified to
\beq \epsilon K_{\mu\nu}|_{\partial\mathcal{M}}=(\epsilon K+T)h_{\mu\nu}|_{\partial\mathcal{M}}\;.\label{eq:braneeomNBC}\eeq
 We refer to (\ref{eq:braneeomNBC}) as the brane equation of motion. Taking the trace, the tension is\footnote{Due to our conventions, we differ from \cite{Takayanagi:2011zk,Fujita:2011fp} by an overall minus sign.} 
\beq T=-\frac{\epsilon(d-1)}{d}K\;,\eeq
such that the brane equation of motion (\ref{eq:braneeomNBC}) becomes
\beq \epsilon K_{\mu\nu}|_{\partial\mathcal{M}}=-\frac{T}{(d-1)}h_{\mu\nu}|_{\partial\mathcal{M}}\;.\eeq
For $d=2$ and $\epsilon=+1$, this reduces to (\ref{eq:brane_eom}).

\subsection*{Useful curvature relations}

As described in the main text, the ambient background geometry describing an ETW brane located at $\theta=\phi_{0}$ may be expressed in normal coordinates 
\beq ds^{2}=G_{\mu\nu}dx^{\mu}dx^{\nu}=(d\theta^{2}+\tilde{g}_{ij}(\theta,\xi)d\xi^{i}d\xi^{j})\;,\label{eq:GNCapp}\eeq
with $\tilde{g}_{ij}(\theta,\xi)=f^{2}(\theta)g_{ij}(\xi)$. The Christoffel symbols are easily attained to be
\beq
\begin{split}
&\Gamma^{\theta}_{\;\theta \theta}[G]=\Gamma^{\theta}_{\;\theta i}[G]=\Gamma^{k}_{\;\theta\theta}[G]=0\;,\quad \Gamma^{\theta}_{\;ij}[G]=-\frac{1}{2}\partial_{\theta}\tilde{g}_{ij}=-\frac{f'}{f}\tilde{g}_{ij}\;,\\
&\Gamma^{k}_{\;\theta j}[G]=\frac{1}{2}\tilde{g}^{ik}\partial_{\theta}\tilde{g}_{ij}=\delta^{k}_{\;j}\frac{f'}{f}\;,\\
&\Gamma^{k}_{\;ij}[G]=\Gamma^{k}_{\;ij}[\tilde{g}]=\Gamma^{k}_{\;ij}[g]=\frac{1}{2}g^{kl}(\partial_{i}g_{jl}+\partial_{j}g_{il}-\partial_{l}g_{ij})\;.
\end{split}
\label{eq:christoffelgenGNC}\eeq
Here $f'\equiv \partial_{\theta}f$ and we used $\tilde{g}^{ij}=f^{-2}(\theta)g^{ij}$. Thence, the non-vanishing components of the Riemann curvature (\ref{eq:Riemanngencon}) are 
\beq
\begin{split}
R^{\theta}_{\;j\theta k}[G]&=\partial_{\theta}\Gamma^{\theta}_{\;jk}-\Gamma^{\theta}_{\;ik}\Gamma^{i}_{\;\theta j}=-\frac{1}{2}\partial^{2}_{\theta}\tilde{g}_{jk}+\frac{f'}{2f}\partial_{\theta}\tilde{g}_{jk}\\
&=-\frac{f''}{f}\tilde{g}_{jk}\;,
\end{split}
\eeq
\beq
\begin{split}
R^{i}_{\;jkl}[G]&=R^{i}_{\;jkl}[\tilde{g}]+\Gamma^{i}_{\;k\theta}\Gamma^{\theta}_{\;lj}-\Gamma^{i}_{\;l\theta}\Gamma^{\theta}_{\;kj}\\
&=R^{i}_{\;jkl}[\tilde{g}]+\left(\frac{f'}{f}\right)^{2}\left(\delta^{i}_{\;l}\tilde{g}_{jk}-\delta^{i}_{\;k}\tilde{g}_{lj}\right)\;.
\end{split}
\eeq
The non-zero components of the Ricci tensor are 
\beq
\begin{split}
R_{\theta\theta}[G]&=-\partial_{\theta}\Gamma^{i}_{\;i\theta}-\Gamma^{j}_{\;\theta i}\Gamma^{i}_{\;j\theta}=-d\frac{f''}{f}\;,    
\end{split}
\eeq
\beq 
\begin{split}
R_{kl}[G]&=R_{kl}[\tilde{g}]+\partial_{\theta}\Gamma^{\theta}_{\;kl}+\Gamma^{i}_{\;i\theta}\Gamma^{\theta}_{\;kl}-\Gamma^{i}_{\;l\theta}\Gamma^{\theta}_{\;ik}-\Gamma^{\theta}_{\;il}\Gamma^{i}_{\;\theta k}\\
&=R_{kl}[\tilde{g}]-\frac{1}{2}\partial_{\theta}^{2}\tilde{g}_{kl}-(d-2)\left(\frac{f'}{f}\right)^{2}\tilde{g}_{kl}\\
&=R_{kl}[\tilde{g}]-\frac{1}{f^{2}}[(f')^{2}+ff'']\tilde{g}_{kl}-(d-2)\left(\frac{f'}{f}\right)^{2}\tilde{g}_{kl}\;.
\end{split}
\eeq
Finally, the Ricci scalar $R[G]=G^{\mu\nu}R_{\mu\nu}[G]$ is 
\beq R[G]=R[\tilde{g}]-\frac{d}{f^{2}}[(f')^{2}+2ff'']-d(d-2)\left(\frac{f'}{f}\right)^{2}\;.\eeq

We will be primarily interested in the case the bulk is an Einstein spacetime, such that $R_{\mu\nu}[G]=\frac{2\Lambda}{(d-1)}G_{\mu\nu}$, with $\Lambda=\frac{d(d-1)\kappa}{2}$, where $\kappa=-1,+1$, or $0$ for $\text{AdS}_{d+1},\text{dS}_{d+1}$, or $\text{Mink}_{d+1}$, respectively, and $L=1$. Therefore, from the $\theta\theta$-component,
\beq -\frac{d f''}{f}=\frac{2\Lambda}{(d-1)}\Rightarrow f''=-\kappa f\;.\label{eq:bulkeomEin}\eeq
For bulk $\text{AdS}_{d+1}$ ($\kappa=-1$), solutions of interest include $f(\theta)\propto e^{\theta/L}, \cosh(\theta/L)$ and $\sinh(\theta/L)$, corresponding to flat, AdS or dS slicings of the hypersurface, respectively.

\section{Perturbative expansions} \label{app:perturbative_computations}

Here we provide computational details for the perturbative expansions used to derive the second order action characterizing the brane deformations in Section \ref{sec:dilagravaction}. 

\subsection{Geometry}

\subsection*{First order contribution}

Recall the $(d+1)$-dimensional geometry (\ref{eq:metric_ansatz}) around an end-of-the-world brane located at $\theta=\phi_0$ is 
\begin{align}
\begin{split}
\label{eq:metric_ansatzapp}
ds^2  &=d\theta^2 + \tilde g_{ij}(\theta, \xi) d\xi^i d\xi^j \\
 &=d\theta^2 + \left( \tilde g_{ij}(\phi_0, \xi) + \delta \theta \; \partial_\theta \tilde g_{ij}(\phi_0, \xi) + \frac 1 2 \delta \theta^2\; \partial^2_\theta \tilde g_{ij}(\phi_0, \xi) + \dots \right)  d\xi^i d\xi^j,
\end{split}
\end{align}
where $\theta\in(-\infty,\infty)$ and $\{\xi^{i}\}$ refer to coordinates parametrizing the $d$-dimensional brane. Let $\Phi\equiv \theta-\phi_{0}=0$ denote the hypersurface equation of the undeformed brane, with unit normal
\beq n_{\mu}=\frac{\partial_{\mu}\Phi}{\sqrt{G^{\mu\nu}\partial_{\mu}\Phi\partial_{\nu}\Phi}}\;,\eeq
whose only non-vanishing component is $n_{\theta}=1$. The extrinsic curvature is
\beq K_{ij}=-(\nabla_{\beta}n_{\alpha})e^{\alpha}_{i}e^{\beta}_{j}=-\nabla_{i}n_{j}=-\frac{1}{2}\partial_{\theta}\tilde{g}_{ij}(\theta,\xi)\;,\label{eq:Kijexapp}\eeq
where we used Christoffel symbols (\ref{eq:christoffelgenGNC}). To leading order in the expansion (\ref{eq:metric_ansatzapp}), 
\beq K_{ij}^{(0)}=-\frac{1}{2}\partial_{\theta}\tilde{g}_{ij}(\phi_{0},\xi)+\mathcal{O}(\delta\theta)\;.\label{eq:extcurv0}\eeq
Consequently, 
\beq \tilde{g}_{ij}(\theta,\xi)=g_{ij}(\xi)-2\delta\theta K_{ij}^{(0)}+\mathcal{O}(\delta\theta^{2})\;,\label{eq:leadingordgtilde}\eeq
where $\tilde{g}_{ij}(\phi_{0},\xi)=g_{ij}(\xi)$. Assuming the undeformed brane of tension $T_{0}$ satisfies  $K^{(0)}_{ij}=-T_{0}h^{(0)}_{ij}=-T_{0}g_{ij}$ and setting $\delta\theta=\phi(\xi)$, it follows that the induced metric $\tilde{g}_{ij}$ on the brane at constant $\theta$ to linear order is
\beq \tilde{g}_{ij}=g_{ij}(1+2T_{0}\phi)+\mathcal{O}(\phi^{2})\;.\label{eq:gtildeij}\eeq
Moreover, at this order, $\tilde{g}^{ij}=g^{ij}(1+2T_{0}\phi)^{-1}$ and $\sqrt{\tilde{g}}=\sqrt{g}(1+2T_{0}\phi)^{d/2}$. 

By our analysis, it is clear the $\theta\theta$-bulk Einstein equations yields a second-order differential equation in $\theta$, with both initial conditions $\tilde{g}_{ij}(\theta=\phi_{0},\xi)$ and $\partial_{\theta}\tilde{g}_{ij}(\theta=\phi_{0},\xi)$ being proportional to $g_{ij}(\xi)$. Thus, all higher-order terms in the Taylor expansion (\ref{eq:metric_ansatzapp}) will be proportional to $g_{ij}(\xi)$, and we may replace the series expansion with a function $f^{2}(\theta)$ obeying $f(\theta=\phi_{0})=1$, such that the metric (\ref{eq:metric_ansatzapp}) takes the form $ds^{2}=d\theta^{2}+f^{2}(\theta)g_{ij}(\xi)d\xi^{i}d\xi^{j}$. Consequently, the extrinsic curvature (\ref{eq:extcurv0}) becomes $K^{(0)}_{ij}=-f'(\phi_{0})g_{ij}$, with trace $K^{(0)}=g^{ij}K_{ij}^{(0)}=-df'(\phi_{0})$, where we used $f(\phi_{0})=1$. From this it is useful to know
\beq\label{eq:Kabove} \tilde{g}_{ij}=g_{ij}(1+2\phi f'(\phi_{0}))+\mathcal{O}(\phi^{2})=g_{ij}\left(1-\frac{2\phi}{d}K^{(0)}\right)+\mathcal{O}(\phi^{2})\;.\eeq

\subsection*{Higher-order contributions}

\noindent \textbf{Normal vector and extrinsic curvature.} Now place a deformed brane with tension $T$ at $\theta = \phi_0 + \phi(\xi)$, for small $\phi(\xi)$ relative to $\phi_{0}$. The hypersurface equation is $\Phi\equiv\theta-\phi_{0}-\phi(\xi)=0$, with unit normal
\beq n_{\mu}=\frac{[\delta^{\theta}_{\mu},-\delta^{k}_{\mu}\partial_{k}\phi]}{\sqrt{1+\tilde{g}^{ij}\partial_{i}\phi\partial_{j}\phi}}=\left[\delta^{\theta}_{\mu}\left(1-\frac{1}{2}\tilde{g}^{ij}\partial_{i}\phi\partial_{j}\phi\right),-\delta^{k}_{\mu}\partial_{k}\phi\right]+\mathcal{O}(\phi^{3})\;.\eeq
We have then have, neglecting terms higher than quadratic order
\beq 
\begin{split}
&\nabla_{\theta}n_{\theta}=
-\frac{K^{(0)}}{d}g^{ij}\partial_{i}\phi\partial_{j}\phi\;,\\
&\nabla_{k}n_{\theta}
=(\partial_{k}\phi)\frac{f'(\theta)}{f(\theta)}-\frac{1}{2}[(\partial_{k}\tilde{g}^{ij})\partial_{i}\phi\partial_{j}\phi+2\tilde{g}^{ij}(\partial_{j}\phi)(\partial_{i}\partial_{k}\phi)]\;,\\
&\nabla_{\theta}n_{k}
=(\partial_{k}\phi)\frac{f'(\theta)}{f(\theta)}\;,\\
&\nabla_{k}n_{i}
 =\tilde{\nabla}_{k}n_{i}+\frac{f'(\theta)}{f(\theta)}\tilde{g}_{ik}+\frac{1}{2d}K^{(0)}g_{ik}g^{jl}\partial_{j}\phi\partial_{l}\phi\;.
\end{split}
\eeq
Here $\tilde{\nabla}_{k}n_{i}=\partial_{k}n_{i}-\Gamma^{j}_{\;ik}[\tilde{g}]n_{j}$, the covariant derivative compatible with $\tilde{g}_{ij}$. Further,
\beq \nabla_{i}n^{i}
=-\tilde{\Box}\phi-\tilde{K}(\theta)\left(1-\frac{1}{2}\tilde{g}^{ij}\partial_{i}\phi\partial_{j}\phi\right)\;,\eeq
with $\tilde{K}(\theta)=-\frac{1}{2}\tilde{g}^{ik}\partial_{\theta}\tilde{g}_{ik}=-df^{-1}(\theta)f'(\theta)$ and $\tilde{\Box}=\tilde{g}^{ik}\tilde{\nabla}_{i}\tilde{\nabla}_{k}$. With these expressions in hand, it is easy to show the acceleration $a_{\mu}=n^{\nu}\nabla_{\nu}n_{\mu}$ is of higher-order. Specifically, 
\beq a_{\theta}\sim g^{ij}\partial_{i}\phi\partial_{j}\phi\,,\quad a_{k}\sim -\tilde{g}^{ij}(\partial_{j}\phi)\tilde{\nabla}_{i}n_{k}\;,\label{eq:athetaak}\eeq
entering at second order. Since $a_{\mu}$ is at least of second order in deformations $\phi$, we will ignore such contributions in our analysis.

Let us now work out the extrinsic curvature $K_{\mu\nu}=-\nabla_{\mu}n_{\nu}+n_{\mu}a_{\nu}$. As we show below, we need only $K_{\mu\nu}$ to linear order in $\phi$.  Since $a_{\mu}$ is at least of second order, it will be safe to approximate $K_{\mu\nu}\approx -\nabla_{\mu}n_{\nu}$. Then, it is easy to work out $K_{\theta\theta}\sim\mathcal{O}(\phi^{2})$, while
\beq K_{ij}=-\tilde{g}_{ij}\frac{f'(\theta)}{f(\theta)}-\tilde{\nabla}_{i}n_{j}+\mathcal{O}(\phi^{2})\approx -g_{ij}[T_{0}+(1+T_{0}^{2})\phi]+D_{i}D_{j}\phi+\mathcal{O}(\phi^{2})\;.\label{eq:Kijpertapp}\eeq
Since it will prove useful momentarily, note
\beq K^{ij}K_{ij}\approx d[T_{0}^{2}+2T_{0}(1-T_{0}^{2})\phi]-2T_{0}\Box\phi+\mathcal{O}(\phi^{2})\;,\label{eq:KijKij}\eeq
where $K^{ij}=\tilde{g}^{il}\tilde{g}^{jk}K_{lk}\approx-g^{ij}[T_{0}+(1-3T_{0}^{2})\phi]+D^{i}D^{j}\phi$.

\vspace{2mm}

\noindent \textbf{Trace of extrinsic curvature.}  We have enough to express the trace of the extrinsic curvature $K$ to second order in deformations $\phi$. Restricting ourselves to the case when the bulk is vacuum $\text{AdS}_{d+1}$, such that $f''(\theta)=f(\theta)$   is obeyed  (\ref{eq:bulkeomEin}), then for $\theta=\phi_{0}+\phi$,
\beq \tilde{K}(\theta)=-df^{-1}(\theta)f'(\theta)\approx -dT_{0}-d(1-T_{0}^{2})\phi+dT_{0}(1-T_{0}^{2})\phi^{2}+\mathcal{O}(\phi^{3})\;,\eeq
where we used $f(\phi_{0})=1=f''(\phi_{0})$, $f'(\phi_{0})=f'''(\phi_{0})=-\frac{K^{(0)}}{d}$, and $K^{(0)}=-dT_{0}$. Moreover, since $\Gamma^{i}_{\;jk}[\tilde{g}]=\Gamma^{i}_{\;jk}[g]$ (\ref{eq:christoffelgenGNC}), then $\tilde{\nabla}_{k}n_{i}=D_{k}n_{i}$, where $D_{k}$ is the covariant derivative compatible with metric $g_{ij}$. Thus, 
\beq \tilde{\Box}\phi=\tilde{g}^{ik}D_{k}D_{i}\phi=f^{-2}(\theta)\Box\phi\approx \Box\phi-2T_{0}\phi \Box\phi+\mathcal{O}(\phi^{3})\;,\eeq
with $\Box\phi=g^{ij}D_{i}D_{j}\phi$. Altogether,
\beq 
\begin{split}
K&\approx -dT_{0}+\Box\phi-d(1-T_{0}^{2})\phi-2T_{0}\phi\Box\phi+dT_{0}(1-T_{0}^{2})\phi^{2}+\left(\frac{d}{2}-1\right)T_{0}\phi_{i}^{2}+\mathcal{O}(\phi^{3})\;.    
\end{split}
\label{eq:K2ndord}\eeq
where we introduced notation $g^{ij}\partial_{i}\phi\partial_{j}\phi\equiv\phi_{i}^{2}$. Notice the last term vanishes when $d=2$. 

From here it is straightforward to work out the normal derivative $\partial_{n}K=n^{\alpha}\partial_{\alpha}K$. As explained in the main text and below, we are only interested in this quantity to linear order in deformations, hence we need only consider $\partial_{\theta}K$, 
\beq \partial_{n}K\approx \partial_{\theta}(\tilde{\Box}\phi+\tilde{K}(\theta))+\mathcal{O}(\phi^{2})=-d(1-T_{0}^{2})+2dT_{0}(1-T_{0}^{2})\phi-2T_{0}\Box\phi+\mathcal{O}(\phi^{2})\;,\label{eq:normderK}\eeq
where as above we used $f(\phi_{0})=1$ and $f'(\phi_{0})=T_{0}$. 

Lastly, below we will need $K^{2}$ at most to linear order, i.e., 
\beq K^{2}\approx d^{2}T_{0}^{2}-2dT_{0}\Box\phi+2d^{2}T_{0}(1-T_{0}^{2})\phi+\mathcal{O}(\phi^{2})\;.\label{eq:Ksq}\eeq

\vspace{2mm}

\noindent \textbf{Deformed brane metric determinant.} The induced metric $\gamma_{ij}$ on the deformed brane at $\theta=\phi_{0}+\phi(\xi)$ has line element
\beq ds^{2}_{\gamma}=f^{2}(\phi_{0}+\phi)g_{ij}(\xi)d\xi^{i}d\xi^{j}+(\partial_{i}\phi)(\partial_{j}\phi)d\xi^{i}d\xi^{j}\;.\label{eq:inducedmetdefBapp}\eeq
Then, to second order in deformations, 
\beq 
\gamma_{ij}=
g_{ij}+[2T_{0}\phi+(1+T_{0}^{2})\phi^{2}+\mathcal{O}(\phi^{3})]g_{ij}+(\partial_{i}\phi)(\partial_{j}\phi)\;.\label{eq:gammaijapp}\eeq
We thus expand $\gamma_{ij}$ perturbatively as $\gamma_{ij}=g_{ij}+q_{ij}$ for small $q_{ij}$, 
\beq q_{ij}\equiv [2T_{0}\phi+(1+T_{0}^{2})\phi^{2}]g_{ij}+\phi_{i}\phi_{j}\;,\eeq
with $\phi_{i}\equiv\partial_{i}\phi$.  The metric determinant $\sqrt{\gamma}$ is then
\beq \sqrt{\gamma}=\sqrt{g}\left[1+\frac{1}{2}q^{i}_{\;i}+\frac{1}{8}(q^{i}_{\;i}q^{j}_{\;j}-2q^{ij}q_{ij})\right]\;,\eeq
where indices on $q_{ij}$ are raised using $g^{ij}$. 
Explicitly, 
\beq \sqrt{\gamma}=\sqrt{g}\left[1+dT_{0}\phi+\frac{1}{2}\phi_{i}^{2}+\frac{d\phi^{2}}{2}\left(1+(d-1)T_{0}^{2}\right)\right]\;,\label{eq:metdetdefbrane}\eeq
where to the order of interest, $q^{ij}q_{ij}\approx 4dT_{0}^{2}\phi^{2}$ and $q^{i}_{\;i}q^{j}_{\;j}\approx 4d^{2}T_{0}^{2}\phi^{2}$.

\vspace{2mm}

\noindent \textbf{A comment on relating brane tensions.} The deformed brane of tension $T$ obeys the brane equation of motion such that 
\beq K_{ij}=-T\gamma_{ij}\;.\eeq
Substituting in the extrinsic curvature (\ref{eq:Kijpertapp}) and induced metric  (\ref{eq:gammaijapp}), we find
\beq 
\begin{split}
 0&=g_{ij}(T_{0}-T)+g_{ij}(1+T_{0}^{2})\phi-D_{i}D_{j}\phi-2TT_{0}\phi g_{ij}+\mathcal{O}(\phi^{2})\;.
\end{split}
\label{eq:reltensions}\eeq
Notice when the deformation is turned off this implies the deformed and undeformed brane tensions coincide, $T_{0}=T$. Further, taking the trace with $g^{ij}$ and rearranging yields
$T=T_{0}+\left(1-T_{0}^{2}\right)\phi-\frac{1}{d}\Box\phi+\mathcal{O}(\phi^{2})$. Equally,  $\Box\phi=d(T_{0}-T)+d(1-T_{0}^{2})\phi+\mathcal{O}(\phi^{2})$.

\subsection{Second order action}

Assuming the bulk $(d+1)$-dimensional spacetime $\mathcal{M}$ is governed by Einstein gravity, the action characterizing the embedding of a deformed brane of tension $T$ is (dropping the boundary and corner terms)
\begin{align}
     \hspace{-4mm} I_\total = \frac{1}{16\pi G_N} \int d^d\xi \int^{\phi_0 + \phi} \hspace{-2mm} d\theta \sqrt{-G}\left(R[G] - 2 \Lambda\right) - \frac{1}{8 \pi G_N} \int \hspace{-1mm} d^d \xi \sqrt{-\gamma} (K + T)\Large|_{\theta = \phi_0 + \phi}.
\label{eq:Ibulkgenapp}\end{align}
As described in the main text, we divide this into terms which depend on deformations $\phi(\xi)$, and those which do not. Namely, 
\begin{align}
    &I^{(0)}_{\total} =  \frac{1}{16\pi G_N} \int d^d\xi \int^{\phi_0 } d\theta \sqrt{-G}\left(R[G] - 2 \Lambda\right)\;,\label{eq:I1app}\\
    &\Delta I_\total =  \frac{1}{16\pi G_N} \int d^d\xi \int_{\phi_0}^{\phi_0 + \phi(\xi)} d\theta \sqrt{-G}\left(R[G] - 2 \Lambda\right)  - \frac{1}{8 \pi G_N} \int d^d \xi \sqrt{-\gamma} (K + T)\big|_{\theta = \phi_0 + \phi} \;,
    \label{eq:I2app}
\end{align}
 where thus we far we have only rearranged the $\theta$-integral. 
 
 Focusing on $\Delta I_{\total}$, now invoke the fundamental theorem of calculus. Specifically, recall for any continuous and real-valued integrable function $h$ defined on a closed interval, we can define a uniformly continuous function $F(\theta_{1})=\int_{b}^{\theta_{1}}d\theta h(\theta)$ for any $\theta$ in said closed interval. It follows, to quadratic order,
\beq
\begin{split}
\int_{\theta_{1}}^{\theta_{1}+\delta\theta}d\theta h(\theta)&=F(\theta_{1}+\delta\theta)-F(\theta_{1})\approx \delta\theta h(\theta_{1})+\frac{1}{2!}(\delta\theta)^{2}h'(\theta_{1})+\mathcal{O}(\delta\theta^{3})\\
&\approx\delta\theta h\left(\theta_{1}+\frac{1}{2}\delta\theta\right)+\mathcal{O}(\delta\theta^{3})\;.
\end{split}
\label{eq:higher_orders}\eeq
Combined with  $\sqrt{-G} = \sqrt{-\tilde g}$, we rewrite the action (\ref{eq:I2app}) as 
\begin{align}
    \begin{split}
        \label{eq:action_bulk_2}
        \hspace{-4mm}\Delta I_\total ={}&  \frac{1}{16\pi G_N} \int\hspace{-1mm} d^d\xi  \left[\sqrt{-\tilde g}\phi\left(R[G] - 2 \Lambda\right)\big|_{\theta = \phi_0 + \frac 1 2 \phi} - 2 \sqrt{-\gamma} (K + T)\big|_{\theta = \phi_0 + \phi} \right]+...
        \end{split}
\end{align}
where the ellipsis indicates terms of order $\mathcal{O}(\phi^{3})$ and beyond. 

Next, we invoke the Gauss-Codazzi equation to express the bulk Ricci scalar $R[G]$ in terms of $d$-dimensional curvatures,
\begin{align}
    \label{eq:gauss_codazziapp}
    R = \mathcal{R} - K^2 - K^{\mu\nu} K_{\mu\nu} + 2 n^\mu \partial_\mu K + 2 D_\mu a^\mu - 2 a^2.
\end{align}
Recall the acceleration $a^{\mu}$ is of second order, thus we can neglect the last two terms of (\ref{eq:gauss_codazziapp}) since they will contribute terms of order $\mathcal O(\phi^3)$ in (\ref{eq:action_bulk_2}). Altogether,
\beq
\begin{split}
\Delta I_{\total}&=\frac{1}{16\pi G_{N}}\int d^{d}\xi\biggr[\sqrt{-\tilde{g}}\phi\left(\mathcal{R}[\tilde{g}]-K^{2}-K^{\mu\nu}K_{\mu\nu}+2n^{\mu}\partial_{\mu}K-2\Lambda\right)\big|_{\theta=\phi_{0}+\frac{1}{2}\phi}\\
&-2\sqrt{-\gamma}(K+T)\big|_{\theta=\phi_{0}+\phi}\biggr]\;.
\end{split}
\label{eq:DeltaIbulkgenapp}\eeq

Since we are only interested in the action at second order, to proceed, we need only expand $\tilde g_{ij}(\theta,\xi)$ and $n^\mu \partial_\mu K$ to first order in deformations, and $\gamma$ and $K$ to second order. 
 Substituting in our above perturbative expansions (\ref{eq:leadingordgtilde}), (\ref{eq:K2ndord}), (\ref{eq:normderK}), and (\ref{eq:metdetdefbrane}), the quantity (\ref{eq:DeltaIbulkgenapp}) has terms which contribute at zeroth, first, and second order in deformations. 

 \vspace{2mm}
 
 \noindent \textbf{Zeroth order.} The zeroth order contribution comes solely from the second line of (\ref{eq:DeltaIbulkgenapp}),
\beq -\frac{1}{8\pi G_{N}}\int d^{d}\xi\sqrt{-g}(-dT_{0}+T)=-\frac{1}{8\pi G_{N}}\int d^{d}\xi\sqrt{-g}(K^{(0)}+T)\;.\eeq
We therefore add this to the $I^{(0)}_{\total}$ contribution (\ref{eq:I1app}), such that 
\beq I^{(0)}_{\total}\to \frac{1}{16\pi G_N} \int d^d\xi \int^{\phi_0 } d\theta \sqrt{-G}\left(R[G] - 2 \Lambda\right)-\frac{1}{8\pi G_{N}}\int d^{d}\xi\sqrt{-g}(K^{(0)}+T)\;.\eeq

\vspace{2mm}

\noindent \textbf{Linear order.} It is straightforward to determine linear order contribution, denoted, $\Delta I^{(1)}_{\total}$. First note, at this order,
\beq -\frac{\sqrt{-\gamma}}{8\pi G_{N}}(K+T)\approx -\frac{\sqrt{-g}}{8\pi G_{N}}\left(\Box\phi-d(1-T_{0}^{2})\phi-d^{2}T_{0}^{2}\phi+dT_{0}T\phi\right)\;.\eeq
Then, it follows from (\ref{eq:DeltaIbulkgenapp})
\beq 
\begin{split}
 \Delta I^{(1)}_{\total}&=\frac{1}{16\pi G_{N}}\int \hspace{-1mm} d^{d}\xi\sqrt{-g}\phi\left[\mathcal{R}[g]+d((d-1)-T_{0}^{2})+d^{2}T_{0}^{2}-2dT_{0}T-2\phi^{-1}\Box\phi\right]\;. 
\end{split}
\eeq
where we used the zeroth order contributions of (\ref{eq:KijKij}) and (\ref{eq:Ksq}), $\sqrt{-\tilde{g}}\approx \sqrt{-g}$, and $2\Lambda=-d(d-1)$ (assuming the bulk is $\text{AdS}_{d+1}$).

\noindent \textbf{Second order.} The second order contributions to the action require more effort. First note at order $\mathcal{O}(\phi^{2})$, the second line of (\ref{eq:DeltaIbulkgenapp}) is 
\beq
\begin{split}
&-\frac{1}{8\pi G_{N}}\int d^{d}\xi\sqrt{-g}\biggr\{(d-2)T_{0}\phi\Box\phi-d(d-1)T_{0}(1-T_{0}^{2})\phi^{2}+\left(\frac{T}{2}-T_{0}\right)\phi_{i}^{2}\\
&-\frac{d\phi^{2}}{2}(1+(d-1)T_{0}^{2})(dT_{0}-T)\biggr\}\\
&=\frac{1}{16\pi G_{N}}\int d^{d}\xi\sqrt{-g}\biggr\{-2(d-2)T_{0}\phi\Box\phi+T_{0}\phi_{i}^{2}+d(d-1)T_{0}[3+(d-3)T_{0}^{2}]\phi^{2}\\
&+(T_{0}-T)\phi_{i}^{2}+(T_{0}-T)d(1+(d-1)T_{0}^{2})\phi^{2}\biggr\}\;,
\end{split}
\label{eq:bdry2ndord}\eeq
where to get to the second equality we used $(dT_{0}-T)=(d-1)T_{0}+(T_{0}-T)$.

Next, we must evaluate the first line of (\ref{eq:DeltaIbulkgenapp}) up to second order. It is helpful to separately consider the contributions
\beq \frac{1}{16\pi G_{N}}\int d^{d}\xi \sqrt{-\tilde{g}}\phi \mathcal{R}[\tilde{g}]\big|_{\phi_{0}+\frac{\phi}{2}}\;,\label{eq:intmedcalc2}\eeq
and 
\beq -\frac{1}{16\pi G_{N}}\int d^{d}\xi\sqrt{-\tilde{g}}\phi(K^{2}+K^{\mu\nu}K_{\mu\nu}-2n^{\mu}\partial_{\mu}K+2\Lambda)\big|_{\phi_{0}+\frac{\phi}{2}}\;.\label{eq:intmedcalc3}\eeq
Focus on (\ref{eq:intmedcalc2}) first. Recall the expansion (\ref{eq:gtildeij}) of $\tilde{g}_{ij}|_{\phi_{0}+\frac{\phi}{2}}\approx (1+T_{0}\phi)g_{ij}\approx e^{T_{0}\phi}g_{ij}$. Consequently, 
\beq \mathcal{R}[\tilde{g}]=e^{-T_{0}\phi}\left[\mathcal{R}[g]-T_{0}(d-1)\Box\phi-\frac{T_{0}^{2}}{4}(d-2)(d-1)\phi_{i}^{2}\right]\;,\eeq
and $\sqrt{-\tilde{g}}\approx \sqrt{-g}e^{dT_{0}\phi/2}$. Thus, (\ref{eq:intmedcalc2}) becomes
\beq \frac{1}{16\pi G_{N}}\int d^{d}\xi\sqrt{-g}e^{T_{0}\phi(\frac{d}{2}-1)}\phi[\mathcal{R}[g]-T_{0}(d-1)\Box\phi]+\mathcal{O}(\phi^{3})\;,\eeq
yielding the second order contribution
\beq \frac{1}{16\pi G_{N}}\int d^{d}\xi\sqrt{-g}\left[T_{0}\left(\frac{d}{2}-1\right)\phi^{2}\mathcal{R}[g]-T_{0}(d-1)\phi\Box\phi\right]\;,\label{eq:2ndordbulk1}\eeq
for $\phi$ small. 

Now turn to (\ref{eq:intmedcalc3}). At second order we have\footnote{Since we are evaluating about $\theta=\phi_{0}+\frac{1}{2}\phi$, we replace $\phi$ in $K_{ij}^{2}$ (\ref{eq:KijKij}), $K^{2}$ (\ref{eq:Ksq}), and $\partial_{n}K$ (\ref{eq:normderK}) with $\phi/2$. Likewise, $\sqrt{-\tilde{g}}|_{\phi_{0}+\frac{1}{2}\phi}=\sqrt{-g}(1+T_{0}\phi)^{d/2}\approx \sqrt{-g}\left(1+\frac{d}{2}T_{0}\phi\right)$.}
\beq
\begin{split}
\frac{1}{16\pi G_{N}}\int d^{d}\xi\sqrt{-g}\biggr\{(d-1)T_{0}\phi\Box\phi-d(d-1)T_{0}(1-T_{0}^{2})\phi^{2}+\frac{d^{2}T_{0}}{2}((d-3)-(d-1)T_{0}^{2})\phi^{2}\biggr\}    
\end{split}
\label{eq:2ndordbulk2}\eeq
Adding this to (\ref{eq:2ndordbulk1}), we find that the second order contribution to the first line of (\ref{eq:DeltaIbulkgenapp}) is 
\beq 
\begin{split}
 \frac{1}{16\pi G_{N}}\int d^{d}\xi\sqrt{-g}\biggr\{&\frac{dT_{0}}{2}\left[d^{2}-5d+2-(d-2)(d-1)T_{0}^{2}\right]\phi^{2}+\frac{T_{0}}{2}(d-2)\phi^{2}\mathcal{R}[g]\biggr\}\;,
\end{split}
\label{eq:intmedcalc4}\eeq
where we see the $\phi\Box\phi$ term has cancelled.

Altogether, the second order contribution, denoted $\Delta I_{\total}^{(2)}$, is given by the sum of (\ref{eq:bdry2ndord}),  and (\ref{eq:intmedcalc4}), 
\beq 
\begin{split}
\Delta I^{(2)}_{\total}&=\frac{1}{16\pi G_{N}}\int d^{d}\xi\sqrt{-g}\biggr\{-2(d-2)T_{0}\phi\Box\phi+\frac{T_{0}}{2}(d-2)\phi^{2}\mathcal{R}[g]+T_{0}c_{d}\phi^{2}\\
&+T_{0}\phi_{i}^{2}+(T_{0}-T)\phi_{i}^{2}+(T_{0}-T)d(1+(d-1)T_{0}^{2})\phi^{2}\biggr\}\;,
\end{split}
\eeq
where we introduced coefficient $c_{d}\equiv \frac{d}{2}[d^{2}+d-4+(d-4)(d-1)T_{0}^{2}]$. We can rewrite this as
\beq 
\begin{split}
 \Delta I^{(2)}_{\total}&=\frac{T_{0}}{8\pi G_{N}}\int d^{d}\xi\sqrt{-g}\biggr\{\frac{1}{2}\phi_{i}^{2}+\frac{c_{d}}{2}\phi^{2}-(d-2)\phi\Box\phi+\frac{(d-2)}{4}\phi^{2}\mathcal{R}[g]\biggr\}\\
 &+\frac{(T_{0}-T)}{8\pi G_{N}}\int d^{d}\xi\sqrt{-g}\left(\frac{1}{2}\phi_{i}^{2}+\frac{d}{2}(1+(d-1)T_{0}^{2})\phi^{2}\right)\;.
\end{split}
\eeq

\subsection*{3D bulk with generic cosmological constant} \label{app:kappa}

Here we specialize to a three-dimensional bulk with an arbitrary cosmological constant.

For the $(2+1)$-dimensional geometry (\ref{eq:metric_ansatz}) around an end-of-the-world brane at $\theta=\phi_0+\phi(\xi)$, 
the bulk Einstein equations for generic cosmological constant $\Lambda=\kappa$ give
\begin{equation}
	0
	=
	\left.
	R_{\theta\theta}
	-2\kappa g_{\theta\theta}
	\right|_{\Phi=0}
	=
	\kappa f(\phi_{0})+f''(\phi_{0})
	+
	\mathcal{O}(\phi^{1})
	\,.
\end{equation}
Given $f(\phi_{0})=1$ and following the same steps as above, we find
\begin{align}
    	\tilde g_{ij} 
    &= 
    	g_{ij} (1 + 2 T_{0} \phi) + \mathcal O(\phi^2),    \label{eq:sum11}
    \\
    n^\mu \partial_\mu K 
    &= 
    2 \left(T^{2}_{0} + \kappa \right) 
    - 
    2  \left( 
    	2T_{0}^{3}+3 T_{0}\kappa +f'''(\phi_{0})
    \right)\phi
    - 
    2 T_{0} \Box \phi  + \mathcal O(\phi^2),\\
     K 
    &= 
    - 2 T_{0}
    + 
    \Box \phi 
    +
    2  \left(T_{0}^2+\kappa \right)  \phi
    -
    2T_{0}\phi\box\phi
    +
    \left(
    	2 T_{0}^{3}+3T_{0}\kappa+f'''(\phi_{0})
    \right)\phi^{2}
    + \mathcal O(\phi^3),\\
    \sqrt{-\det \gamma} 
    &  =  
    \sqrt{-\det g} \left(1 + 2 T_{0} \phi + \frac {1} 2 g^{ij} \partial_i \phi \partial_j \phi + \left(T^{2}_{0} -\kappa \right) \phi^2   \right)  + \mathcal O(\phi^3),
	\\
   	\mathcal{R}[\tilde{g}]
	&=
	\mathcal{R}[g]
	-
	2 T_{0}(
		\Box \phi 
		+
		\mathcal{R}[g] \phi
	)
	+
	\mathcal{O}(\phi^{2})
	\,,
	\\
	K_{\mu\nu}K^{\mu\nu}
	&=
	2 T_{0}^{2}
	-
	2T_{0}(
		\Box \phi
		+
		2
		(
			\kappa
			+
			T_{0}^{2}
		)
		\phi
		)
	+
	\mathcal{O}(\phi^{2})\label{eq:sum21}
	\,,
\end{align}
where $\mathcal{R}$ denotes the two-dimensional Ricci scalar.

Recall the action characterizing the embedding of a deformed brane of tension $T$,
\beq
\begin{split}
\Delta I_{\total}&=\frac{1}{16\pi G_{N}}\int d^{2}\xi\biggr[\sqrt{-\tilde{g}}\phi\left(\mathcal{R}[\tilde{g}]-K^{2}-K^{\mu\nu}K_{\mu\nu}+2n^{\mu}\partial_{\mu}K-2\Lambda\right)\big|_{\theta=\phi_{0}+\frac{1}{2}\phi}\\
&-2\sqrt{-\gamma}(K+T)\big|_{\theta=\phi_{0}+\phi}+2\sqrt{-g}(K+T)\big|_{\theta=\phi_{0}}\biggr]\;,
\end{split}
\eeq
where we used \eqref{eq:DeltaIbulkgenapp}.
Substituting in \eqref{eq:sum11} -- \eqref{eq:sum21}, we recover the first and second order contributions to the action characterizing the brane deformations presented in the main text, \eqref{eq:defactfinal}.
Note that in the first line one needs to perform the replacement $\phi\to\phi/2$ to the derived identities in order to use them.

\section{Brane deformations and classical JT}\label{app:branedefsJT}

Here we summarize the details relating brane deformations parallel to the brane and the equations of motion of classical JT gravity, as described in Section \ref{sec:BCFTpov}.

Consider empty $\text{AdS}_{3}$ in Poincar\'e patch coordinates
\begin{align}
    \label{eq:poincare_coordsapp}
    ds^2_{\text{AdS}_{3}} = G_{\mu\nu}dx^{\mu}dx^{\nu}= \frac{-dt^2 + dx^2 + dz^2}{z^2}.
\end{align}
We place an ETW brane $\mathcal{Q}$ of tension $T$ at $\theta=\phi_{0}$, or, equivalently,  trajectory
\begin{align}
    \label{eq:standard_solution_poincareapp}
    x = \frac{T}{\sqrt{1 - T^2}}\; z.
\end{align}
The induced metric $h_{ij}^{\mathcal{Q}}$ on the brane is
\beq h_{ij}^{\mathcal{Q}}=\frac{1}{(1-T^{2})y^{2}}\begin{pmatrix} -1&0\\0&1\end{pmatrix}\;,\label{eq:hijQbrane}\eeq
where we defined $z=\sqrt{1-T^2} y$.

Let us now perturb the location of the brane via a small deformation $\delta X(t,z)$, such that the deformed brane is positioned at
\beq x=\frac{T}{\sqrt{1-T^{2}}}z+\delta X(t,z)\;.\label{eq:deformedbrane}\eeq
The unit normal vector to the hypersurface (\ref{eq:deformedbrane}) is
\beq n_{\mu}=\mathcal{N}\left((\partial_{t}\delta X)\delta^{t}_{\mu}-\delta^{x}_{\mu}+\left(\frac{T}{\sqrt{1-T^{2}}}+\partial_{z}\delta X\right)\delta^{z}_{\mu}\right)\;,\eeq
with normalization
\beq \mathcal{N}=-\frac{L}{z}\left[1+\frac{T^{2}}{(1-T^{2})}+\frac{2T}{\sqrt{1-T^{2}}}(\partial_{z}\delta X)+(\partial_{z}\delta X)^{2}-(\partial_{t}\delta X)^{2}\right]^{-1/2}\;.\eeq
We will only be interested in leading order in $\delta X$. Using the projector $h_{\mu\nu}=G_{\mu\nu}-n_{\mu}n_{\nu}$ together with $y=\frac{z}{\sqrt{1-T^{2}}}$ and $x=Ty+\delta X(t,y)$, such that $\{e^{\mu}_{i}\}=\{\partial_{i} x^{\mu}\}=\{e^{t}_{t}=1,\; e^{z}_{y}=\sqrt{1-T^{2}},\; e^{x}_{t}=\partial_{t}\delta X,\; e^{x}_{y}=T+\partial_{y}\delta X\}$,
the induced metric $h_{ij}=e^{\mu}_{i}e^{\nu}_{j}h_{\mu\nu}$ of the deformed brane is
\beq h_{ij}=\frac{1}{(1-T^{2})y^{2}}\begin{pmatrix}-1&T\partial_{t}\delta X\\ T\partial_{t}\delta X&1+2T\partial_{y}\delta X\end{pmatrix}\;.\eeq
Setting $\delta X=0$, we recover the induced metric of the undeformed brane (\ref{eq:hijQbrane}). 
 
To leading order in $\delta X$, the extrinsic curvature $K_{ij}=-e^{\mu}_{i}e^{\nu}_{j}\nabla_{\mu}n_{\nu}$ has non-vanishing components
\beq 
\begin{split}
&K_{tt}=\left(\frac{T}{(1-T^{2})y^{2}}+\frac{1}{y^{2}}\partial_{y}\delta X+\frac{1}{y}\partial^{2}_{t}\delta X\right)\;,\\
&K_{ty}=
-\frac{T^{2}}{(1-T^{2})y^{2}}\partial_{t}\delta X+\frac{1}{y}\partial_{t}\partial_{y}\delta X\;,\\
&K_{yy}=-\frac{T}{(1-T^{2})y^{2}}(1+2T\partial_{y}\delta X)+\frac{1}{(1-T^{2})y^{2}}(\partial_{y}\delta X-y\partial_{y}^{2}\delta X)\;,
\end{split}
\eeq
where we used $\partial_{z}\delta X=(1-T^{2})^{-1/2}\partial_{y}\delta X$. Imposing $K_{ij}=-T h_{ij}$ we find
\beq
\begin{split}
&K_{tt}=Th_{tt}\Longrightarrow \partial_{y}\delta X+y\partial^{2}_{t}\delta X=0\;,\\
&K_{yt}=Th_{yt}\Longrightarrow \partial_{y}\partial_{t}\delta X=0\;,\\
&K_{yy}=Th_{yy}\Longrightarrow\partial_{y}\delta X-y\partial_{y}^{2}\delta X=0\;,
\end{split}
\eeq
recovering (\ref{eq:diffeomsbrane}) from the main text.

\bibliography{dilabranerefs}

\end{document}